\documentclass[11pt]{article}
\usepackage{amsmath, amsthm, amssymb}
\usepackage[left=1in, bottom=1in, right=1in, top=1in]{geometry}
\usepackage[authoryear, square]{natbib}

\usepackage[normalem]{ulem}
\usepackage{hyperref}
\usepackage{dsfont}
\usepackage{xcolor}
\usepackage{graphicx}
\usepackage{wrapfig}
\usepackage{subfig}
\usepackage{floatrow}
\usepackage{tikz}
\usetikzlibrary{arrows.meta}
\usepackage{authblk} %
\let\svtikzpicture\tikzpicture
\def\tikzpicture{\no\textbf{}indent\svtikzpicture}
\graphicspath{ {fig2/} }
\usepackage{booktabs}
\usepackage{float}
\usepackage{enumitem}
\usepackage{multirow}
\usepackage{array}
\usepackage{tabu}
\usepackage{comment}
\usepackage{algorithm} 
\usepackage{algpseudocode}
\usepackage{caption}
\usepackage[mathscr]{eucal}
\usepackage{etoc}
\AtBeginDocument{\etocdepthtag.toc{main}}

\usetikzlibrary{arrows.meta}
\let\svtikzpicture\tikzpicture
\def\tikzpicture{\no\textbf{}indent\svtikzpicture}
\graphicspath{ {fig2/} }

\allowdisplaybreaks
\numberwithin{equation}{section}

\newcommand{\iid}{\text{iid}}

\newcommand{\pfend}{$\hfill\square$}
\newcommand{\wrt}{\text{with respect to }}

\newcommand{\e}{\mathbf{e}}
\newcommand{\Ec}{\mathcal{E}}
\newcommand{\zb}{\mathbf{z}}
\newcommand{\NN}{\mathbb{N}}
\newcommand{\ZZ}{\mathbb{Z}}
\newcommand{\PP}{\mathbb{P}}
\newcommand{\EE}{\mathbb{E}}
\newcommand{\RR}{\mathbb{R}}
\newcommand{\BB}{\mathbb{B}}
\newcommand{\CC}{\mathbb{C}}

\newcommand{\X}{\mathcal{X}}
\newcommand{\Zcal}{\mathcal{Z}}
\newcommand{\Cscr}{\mathscr{C}}
\newcommand{\Kscr}{\mathscr{K}}
\newcommand{\Sscr}{\mathscr{S}}

\newcommand{\nor}{\mathcal{N}}

\newcommand{\dtilde}[1]{\tilde{\tilde{#1}}}
\renewcommand{\i}{\mathrm{i}}
\newcommand{\re}{\mathrm{Re}}
\newcommand{\im}{\mathrm{Im}}
\newcommand{\dd}{\text{d}}
\newcommand{\scl}{\text{sc}}
\newcommand{\dg}{\text{diag}}
\newcommand{\thres}{\Delta_f}
\newcommand{\thresr}{\Delta_R}
\newcommand{\threse}{\Delta^\mathrm{(est)}_a}
\newcommand{\threseall}{\Delta^\mathrm{(est)}}
\newcommand{\thresa}{\Delta^\mathrm{(ap)}}
\newcommand{\Dt}{\mathcal{D}_\tau}
\newcommand{\kf}{\kappa_{f}}
\newcommand{\cmplt}{^\texttt{c}}

\newcommand{\kt}{\kappa_{\Theta}}
\newcommand{\thetabf}{\boldsymbol{\theta}}
\newcommand{\pdhermit}{\mathcal{H}_{+}^p}
\newcommand{\pbf}{\mathbf{p}}
\newcommand{\wbf}{\mathbf{w}}
\newcommand{\sbf}{\mathbf{s}}
\def\bigO#1{\mathcal{O}(#1)}

\newcommand{\wmj}{\mathcal{W}_m(j)}

\newcommand{\dka}{d_k^{(a)}}
\newcommand{\dkb}{d_k^{(b)}}
\newcommand{\dkma}{d_k^{-(a)}}
\newcommand{\eka}{\varepsilon_k^{(a)}}
\newcommand{\ea}{\varepsilon^{(a)}}
\newcommand{\bka}{\beta_k^{(a)}}
\newcommand{\bjahat}{\hat\beta_j^{(a)}}
\newcommand{\bjastar}{\overline{\beta}_j^{(a)}}
\newcommand{\bjatilde}{{\tilde\beta_j^{(a)}}}
\newcommand{\vka}{v_k^{(a)}}
\newcommand{\vkr}{\re(v_k)}
\newcommand{\vki}{\im(v_k)}
\newcommand{\thetajstar}{\Theta_j^*}
\newcommand{\thetakstar}{\Theta_k^*}
\newcommand{\thetajhattau}{\hat\Theta_j^{(\tau)}}

\newcommand{\tauahatsq}{\hat\tau_a^2}

\newcommand{\muja}{\mu_j^{(a)}}
\newcommand{\are}{\phi_\mathrm{RE}}
\newcommand{\tnf}{{\cal T}_n}

\newcommand{\vmnf}{{\cal V}_{m,n}}

\newcommand{\Snk}{\Sigma_{n,k}}
\newcommand{\Onk}{\Omega_{n,k}}
\newcommand{\Za}{\Zcal^{(a)}}
\newcommand{\Zma}{\Zcal^{-(a)}}

\newcommand{\tscr}{\mathscr{T}(M, \eta)}
\newcommand{\Bcal}{\mathcal{B}}

\newcommand{\vecop}[1]{\mathrm{vec}\left(#1\right)}
\newcommand{\peq}{\overset{\pi}{=}}
\newcommand{\M}{\mathcal{M}}
\newcommand{\Mb}{\mathbb{M}^{2\times 2}}
\newcommand{\rank}{\mathrm{rank}}
\newcommand{\lf}{{\mathcal{L}}}

\newcommand{\thickhline}{
    \noalign {\ifnum 0=`}\fi \hrule height 1pt
    \futurelet \reserved@a \@xhline
}

\newcommand{\vertiii}[1]{{\left\vert\kern-0.25ex\left\vert\kern-0.25ex\left\vert #1 
    \right\vert\kern-0.25ex\right\vert\kern-0.25ex\right\vert}}

\newcommand{\maxnorm}[1]{{\left\vert\kern-0.25ex\left\vert #1 \right\vert\kern-0.25ex\right\vert}_{\max}}
\newcommand{\infnorm}[1]{{\left\vert\kern-0.25ex\left\vert #1 \right\vert\kern-0.25ex\right\vert}_{\infty}}
\DeclareMathOperator*{\esssup}{ess\,sup}

\DeclareMathOperator*{\argmin}{arg\,min}

\DeclareMathOperator*{\cov}{Cov}
\DeclareMathOperator*{\var}{Var}

\DeclareMathOperator*{\trace}{trace}

\theoremstyle{plain}
\newtheorem{theorem}{Theorem}[section]
\newtheorem{assumption}{Assumption}[section]
\newtheorem{lemma}{Lemma}[section]
\newtheorem{proposition}{Proposition}[section]
\newtheorem{corollary}{Corollary}[section]
\newtheorem{definition}{Definition}[section]
\newtheorem{remark}{Remark}[section]

\newtheorem{example}{Example}[section]

\title{Regularized Estimation of Sparse Spectral Precision Matrices}
\date{ }
\author[1]{Navonil Deb \footnote{Email: \href{mailto:nd329@cornell.edu}{\texttt{nd329@cornell.edu}}}}
\author[2]{Amy Kuceyeski \footnote{Email: \href{mailto:amk2012@med.cornell.edu}{\texttt{amk2012@med.cornell.edu}}}}
\author[1]{Sumanta Basu \footnote{Corresponding author. Email: \href{mailto:sumbose@cornell.edu}{\texttt{sumbose@cornell.edu}}}}

\affil[1]{Department of Statistics and Data Science, Cornell University}
\affil[2]{Department of Radiology and Brain and Mind Research Institute, Weill Cornell Medicine}

\begin{document}
\maketitle

\begin{abstract}
Estimation of a sparse spectral precision matrix, the inverse of a spectral density matrix, is a canonical problem in frequency-domain analysis of high-dimensional time series (HDTS), with applications in neurosciences and environmental sciences. 
Existing estimators use off-the-shelf optimizers for complex variables that limit scalability, uniform (non-adaptive) penalization that is not tailored to handle heterogeneity across time series components, and lack a formal non-asymptotic theory that systematically analyzes approximation and estimation errors in high-dimension. 
In this work, develop fast pathwise coordinate descent (CD) algorithms and non-asymptotic theory for a complex graphical lasso (CGLASSO) and an adaptive version CAGLASSO, that adapts penalization to the underlying scale of variability. For fast algorithms, we devise a \textit{realification procedure} based on ring isomorphism, a notion from abstract algebra, that can be used for other high-dimensional optimization problems over complex variables. 
Our non-asymptotic analysis shows that consistency is possible in high-dimension under suitable sparsity assumptions. A key step is to separately bound the approximation and estimation error arising from treating the finite-sample discrete Fourier Transforms (DFTs) as i.i.d. complex-valued data, an issue well-addressed in classical time series but relatively less explored in HDTS literature. We demonstrate the performance of our proposed estimators in several simulated data sets and a real data application from neuroscience.
\end{abstract}

\section{Introduction}\label{sec:intro}

Spectral precision matrix, the inverse of a spectral density matrix, is a key quantity in frequency-domain multivariate time series analysis. Estimation of spectral precision matrix is needed to compute \textit{partial coherency}, the frequency-domain analogue of partial correlation, and for building graphical models of stationary processes that capture conditional relationships among time series components across time lags \citep{dahlhaus2000graphical, davis2016sparse}. These tools have proven effective in diverse disciplines-- including neuroscience, economics, and environmental sciences \citep{granger1969investigating, bowyer2016coherence, fiecas2019spectral}. Of particular relevance, conditional dependence graphs captured by spectral precision matrix are essential for understanding functional connectivity (FC) assessed by neuroimaging methods \citep{baek2021detecting, ombao2022spectral, wodeyar2022structural}. In this work, we develop methods for estimating the spectral precision matrix at a given frequency that characterizes the frequency-specific conditional dependence among the components of the time series. 

Formally, we consider $ X_t = \big(X_t^{(1)},\ldots,\allowbreak X_t^{(p)}\big)^\top$ for $t\in \ZZ$, a $p$-dimensional weakly stationary and centered real-valued time series with autocovariance function
\begin{equation}\label{eq:autocovariance}
    \Gamma(h) := \cov(X_t, X_{t- h})= \EE\left[X_t X_{t-h}^\top\right],\quad h \in \ZZ.
\end{equation}
The spectral density and spectral precision matrices at frequency $\omega \in [-\pi, \pi]$  are defined as 
\begin{equation}\label{eq:spec_density_precision}
    f(\omega) := \frac{1}{2\pi}\sum_{h= -\infty}^\infty \Gamma(h) e^{-\i h \omega}, \quad \Theta(\omega) := [f(\omega)]^{-1}.
\end{equation}
It is customary to estimate these quantities at the Fourier frequencies $\omega_j = 2\pi j/n$, $j \in F_n := \left\{ -\left\lfloor(n-1)/2\right\rfloor, \ldots, \left\lfloor n/2 \right\rfloor \right\}$. 
Given $n$ consecutive observations $X_1, \ldots, X_n$, the \textit{discrete Fourier transform} (DFT) at frequency $\omega_j$ is defined as 
\begin{equation}\label{eq:dft}
    d_j:= d(\omega_j) = \frac{1}{\sqrt{2\pi n}} \sum_{t=1}^n X_t e^{-\i t \omega_j}.
\end{equation}
Under suitable regularity conditions, DFTs in nearby $\pm m$ Fourier frequencies, i.e. $\{d_k\}_{k=j-m}^{j+m}$, can be approximated as i.i.d. complex Gaussian with mean zero and covariance matrix $f(\omega_j)$ (see Section \ref{subsec:classical_estimators} for details), and this opens up several routes to estimate $\Theta(\omega_j)$ as the precision matrix of $2m+1$ i.i.d. complex-valued observations $\{d_k\}_{k = j-m}^{j+m}$. For example in classical setting with fixed $p$ and $n \rightarrow \infty$, a natural estimator of $\Theta(\omega)$ is the inverse of the averaged periodogram $\sum_{k=-m}^m d_j d_j^\dagger/(2m+1)$ \citep{brillinger2001time, priestley1988spectral, brockwell1991time}. In high-dimensional setting and under sparsity assumptions, penalized estimation of $\Theta(\omega)$ can be carried out by extending precision matrix estimators for real-valued i.i.d. samples to handle complex-valued data. For instance, \citet{fiecas2019spectral} propose an estimator based on constrained $\ell_1$-minimization for matrix estimation (CLIME) algorithm \citep{cai2011constrained}. \citet{dallakyan2022time, tugnait2022sparse, baek2023local} build estimators based on \textit{graphical lasso} (GLASSO) for real-valued i.i.d. data \citep{friedman2008sparse, mazumder2012graphical}. 
\subsection{Related works and knowledge gaps}

Despite the recent progress, several computational, methodological and theoretical questions remain open. We start with the issue of scalable computation. Typical fMRI data sets involve $p \in [50, 800]$ brain regions and $n \approx 1000$ time points. Estimating $\Theta(\omega_j)$ for $\bigO{n}$ different frequencies, and tuning over both sparsity penalty as well as bandwidth $\bigO{m}$, becomes computationally demanding. Scalable optimization algorithms over complex variables are direly needed. In contrast to fast pathwise \textit{coordinate descent} (CD) algorithms that are state-of-the-art for high-dimensional estimators such as the lasso or GLASSO, existing estimators broadly use two strategies: (a) separately penalizing real and imaginary components $\Theta(\omega)$ that may deteriorate estimation accuracy by artificially increasing problem dimension \citep{fiecas2010functional}, and (b) use off-the-shelf complex variable optimization procedure such as the alternating direction methods of multipliers (ADMM) \citep{dallakyan2022time, baek2023local}, which are known to be slower than pathwise CD algorithms for real-valued data \citep{gu2018admm} because they are not tailored to exploit underlying sparsity structure of the problem. An effective extension to pathwise CD for complex lasso is not obvious since the lasso penalty on complex numbers is a group lasso penalty on reals, for which closed form CD updates do not exist in general.

On the methodological front, existing estimators  \textit{uniformly} penalize off-diagonals of the spectral precision matrix. This results in inaccurate estimation under heterogeneity, i.e. when the diagonal entries of $\Theta(\omega)$ are substantially different from each other. In this case, estimators of $(\Theta(\omega))_{a,b}, a \neq b$, vary on different scales for different $a, b$. In the context of large covariance and precision matrix estimation with i.i.d. real-valued data, the issue of scale heterogeneity has been addressed through adaptive regularization schemes \citep{cai2011adaptive, cai2016estimating}. This issue of scale adaptation is more complicated for complex-valued random variables because their variances are not scalars, but instead captured by $2 \times 2$ matrices. Standardizing variables before estimating $\Theta(\omega)$, as is common in practice \citep{jankova2018inference}, does not address this problem since the appropriate scales of the penalties should be based on functions of partial variances of the DFT components that capture the heterogeneity of the diagonals of $\Theta(\omega)$, rather than the marginal variances.

Finally, non-asymptotic theoretical guarantees for such estimators are currently lacking in the literature. Deriving estimation error bounds in high dimensions is crucial for proving statistical consistency and characterizing how these bounds depend on model parameters. \citet{ravikumar2011high} establish non-asymptotic error bounds for the GLASSO estimator for real-valued i.i.d. samples. \citet{weylandt2020computational} proves similar results for GLASSO with complex-valued i.i.d. samples. However, these guarantees cannot be directly applied to a GLASSO estimator constructed with $\{d_k\}_{k = j-m}^{j+m}$ since they are not independent and $\Snk$, the covariance matrix of $d_k$  (defined in \eqref{eq:Snk}), varies across $k \in \{j-m,\ldots, j+m\}$. \citet{baek2023local} provide theoretical results for the GLASSO estimator obtained from a \textit{local Whittle likelihood} only near frequency zero. The spectral precision matrix estimators proposed by \citet{jung2015graphical, dallakyan2022time} are not accompanied by theoretical guarantees. Additionally, partial variances required for the scales of adaptive penalization can be obtained from \textit{nodewise regression} (NWR) of the DFTs. In Gaussian graphical models with a sparse precision matrix $\Theta$ and data matrix $\X := [X^{(a)}: \ldots : X^{(p)}]$ having i.i.d. Gaussian rows, regressing $X^{(a)}$ for $a\in [p]$ on its the other columns of $\X$ yields the NWR coefficient $\beta \in \RR^{p-1}$ with entries $\beta_b, b \in [p] \setminus \{a\}$ such that $\beta_b = 0$ if and only if $\Theta_{b,a} = 0$ \citep{meinshausen2006high} -- leading to exactly sparse regression coefficients and thereby allowing direct application of high dimensional theory of sparse regression \citep{wainwright2019high, van2014asymptotically}. However when linearly projecting one coordinate of $d_k$ on its other coordinates, the regression coefficients are not sparse, and both the regression errors and the rows of the design matrix are heteroskedastic (see Section \ref{subsubsec:nwr_pop} for details) due to dependent and heteroskedastic structure of $d_k$'s across frequencies -- rendering the theory of high-dimensional sparse regression not directly applicable to NWR of DFTs. For proving non-asymptotic guarantees for the standard estimators in high dimension, one needs to quantify the error of approximating the DFTs as i.i.d. data.

\subsection{Our contributions} We propose the \textit{Complex Graphical Lasso} (CGLASSO) \eqref{eq:graphical-lasso}, an estimator of $\Theta(\omega_j)$ at arbitrary Fourier frequency $\omega_j$, and present a CD algorithm (Algorithm \ref{alg:cglasso}). The key algorithmic contribution is a \textit{ring isomorphism} (Section~\ref{sec:realification}) between complex numbers and real $2 \times 2$ orthogonal matrices, which preserves the canonical operations of complex matrices within an equivalent ring of real matrices and is applicable to a broader range of optimization problems. Leveraging this isomorphic structure, we implement \textit{Complex Lasso} (CLASSO) \eqref{eq:classo} in the inner optimization layer of CGLASSO. Under realification, the CLASSO objective is equivalent to a group lasso problem with orthogonal predictors, enabling closed-form CD updates and improving computational efficiency. Along the regularization path, CGLASSO exploits pathwise CD with warm start and active set screening, achieving substantial computational gains over state-of-the-art methods. Empirical results in Section \ref{subsec:runtime_improvement} show that CGLASSO performs significantly faster (20-100 folds) than ADMM and the performance is scalable with the dimension.

Methodologically, incorporating joint penalty for real and imaginary parts of $\Theta(\omega_j)$ gives room for more accurate estimation (Section \ref{subsec:rmse_improvement}) over separate penalties \citep{fiecas2010functional}. Maximizing the joint likelihood of the full data improves model selection accuracy (Section \ref{subsec:model_selection_improvement}) relative to only NWR-based approaches \citep{krampe2025frequency}. Building on CGLASSO and NWR of DFTs, we develop \textit{Complex Adaptive Graphical Lasso} (CAGLASSO) \eqref{eq:caglasso}, a variant of CGLASSO with penalties on the off-diagonals adapted to the estimated NWR partial variances (defined in \eqref{eq:nwr_partial_variance2}) that capture the scales of conditional variability. Empirical results in Section \ref{subsec:adaptation_improvement} show that partial variance–adapted penalties yield more accurate estimates than either uniform or marginal variance–adapted penalties when $\Theta(\omega_j)$ contains highly heterogeneous entries.

On the theoretical front, we derive non-asymptotic error bounds for the CGLASSO and CAGLASSO estimators. Our analysis shows that the error bounds for GLASSO with i.i.d. samples continue to hold for CGLASSO, with additional terms due to an \textit{approximation bias} $\vmnf$ (defined in \eqref{eq:approximation_bias}), which contains the finite sample truncation bias of $d_k$ and the smoothing bias due to averaging the periodograms across nearby frequencies (Theorem \ref{thm:consistency_cglasso}). These additional terms arise because the DFTs are not i.i.d., hence treating them as independent samples in our estimation algorithms entails an approximation cost captured by $\vmnf$. Additionally for deriving error bounds for the NWR partial variance estimators that serve as the penalty weights in CAGLASSO, we establish concentration bounds on the estimation error and prediction error of NWR of DFTs (Theorem \ref{thm:caglasso_consistency}). To this end, we adapt the framework of \textit{approximately linear truth} for sparse regression \citep[Section 6.2.3]{buhlmann2011statistics} and define a sparse oracle $\bjastar$ for $a\in [p], j \in F_n$ in \eqref{eq:oracle_beta} for linearly approximating the NWR means. Using perturbation bounds for inverted matrices, we show that the oracle approximation error can be explicitly controlled by $\vmnf^2$ (Proposition \ref{prop:nwr_approx_oracle}) only under a summability condition (Assumption \ref{asn:summable}) -- which is a non-trivial work. In contrast to the NWR results for i.i.d. real-valued samples \citep{van2014asymptotically}, our derivations allow for heteroskedasticity of the errors. A further technical contribution lies in developing concentration inequalities for averages of quadratic forms of dependent DFTs, which are essential for proving the \textit{restricted eigenvalue} (RE) (Proposition \ref{prop:re}) and \textit{deviation conditions} (Proposition \ref{prop:deviation}) underlying Theorem \ref{thm:consistency_cglasso}. This dependence precludes the use of standard concentration bounds for covariance matrix of i.i.d. samples and requires \textit{Hanson–Wright}–type concentration inequalities for complex-valued matrices. By carefully tracing the approximation errors, we obtain consistency results for the estimated partial variances in DFT-NWR (Corollary \ref{cor:consistency_partial_variance}), which serve as adaptive penalty weights, leading to the final non-asymptotic error bound for the CAGLASSO estimator (Theorem \ref{thm:caglasso_consistency}).

\subsection{Organization of the paper}

In Section \ref{sec:method}, we set up the background of the necessary frequency domain objects and then we describe our proposed methods CGLASSO and CAGLASSO. We describe the estimation algorithms for CGLASSO and CAGLASSO in Section \ref{sec:algo}. In Section \ref{sec:theory}, we study the theoretical properties of our proposed estimators. We provide simulation studies in Section \ref{sec:simulation}, followed by an application to a fMRI real data in Section \ref{sec:FC_fMRI}. Finally we discuss some future directions in Section \ref{sec:discussion}. The proofs and the supporting materials are deferred in the Appendix \ref{sec:supp_dft_nwr} to \ref{sec:intro_algebra}.

\subsection{Notation} We use $\emptyset$, $\NN, \ZZ, \RR$ and $\CC$ to denote empty set, the set of natural numbers, integers, real numbers and complex numbers respectively. For $x\in \RR$, $\lfloor x\rfloor$ is the greatest integer smaller than or equal to $x$. $\i$ denotes the imaginary number $\sqrt{-1}$. For $z\in \CC$, $\re(z)$ and $\im(z)$ denote the real and imaginary parts of $z$ respectively, and $z^\dagger$ denote its complex conjugate. $|z|$ denotes $\sqrt{\re(z)^2 + \im(z)^2}$ i.e. the complex modulus of $z$. For any positive integer $p$, $\e^{(p)}$ denotes the $p$ dimensional basis vector with $1$ in the $p$\textsuperscript{th} position and 0 elsewhere. For a complex vector $\zb = (z_1,\ldots, z_p)\in \CC^p$, $\re(\zb)$, $\im(\zb)$ and $\zb^\dagger$ denotes its real part, imaginary part and conjugate transpose respectively. For any $q > 0$, $\|\zb\|_q := \left(\sum_i |z_i|^q\right)^{1/q}$ denotes $\ell_q$ norm of $\zb$. For a matrix $A$, $A^\dagger$ denotes its conjugate transpose, $\vecop{A}$ denotes its vector representation and $A_{\cdot j}$ denotes its $j$\textsuperscript{th} column. $\rank(A)$ denotes the rank of $A$ i.e. the dimension of the column space of $A$. For two matrices $A$ and $B$, the Kronecker product of $A$ and $B$ is denoted by $A \otimes B$. If $A$ is a square matrix, $\trace(A)$ and $\det(A)$ denote its trace and determinant respectively. $A^+$ denotes the vector consisting of the diagonal entries of $A$, and $A^-$ denotes the matrix obtained from setting the diagonals of $A$ to zero. $\vertiii{A}_\infty, \vertiii{A}_1, \|A\|, \|A\|_F$ and $\|A\|_\infty$ denote the maximum modulus row sum norm i.e. $\max_{r} \sum_{s}|A_{rs}|$, maximum modulus column sum norm i.e. $\max_{s} \sum_{r}|A_{rs}|$ spectral norm or the square root of the largest eigen value of $A^{\dagger} A$, Frobenius norm $\sqrt{\trace(A^\dagger A)}$ and element-wise maximum modulus value $\max_{r,s}|A_{rs}|$ respectively. $\pdhermit$ denotes the set of all $p\times p$ Hermitian positive definite complex matrices. For two sequences $(x_n)_{n \in \NN}$ and $(y_n)_{n\in \NN}$, we denote $x_n \succsim y_n$ if there is a universal constant $c$, independent of the dimension and model parameters, such that $x_n \ge c y_n$ for all $n \in \NN$. $x_n\precsim y_n$ and $x_n = \bigO{y_n}$ are defined analogously. $x_n \asymp y_n$ denotes that both $x_n \succsim y_n$ and $x_n \precsim y_n$ holds. We also use $c, c', c_i > 0$ to denote universal constants whose values are allowed to change from line to line within a proof.

\section{Background and methods}\label{sec:method}

Before describing our methods and the estimators, we provide a background of the interpretation of $\Theta(\omega)$ in capturing frequency-specific conditional structure of $X_t$, and state the properties of $\{d_k\}_{k= j-m}^{j+m}$, our data in complex domain.

\subsection{Frequency-specific conditional dependence}\label{subsec:frequency_graph}

A zero-mean strictly stationary process $X_t$ with finite autocovariance has a Cram\'{e}r's representation \citep[Theorem 4.6.2]{brillinger2001time} 
$$X_t = \int_{-\pi}^\pi e^{\i t\omega}~ \dd Z(\omega),$$ 
where $Z(\omega) = \left(Z^{(1)}(\omega), \ldots, Z^{(p)}(\omega)\right)^\top, \omega \in [-\pi, \pi]$, is a $p$-dimensional \textit{orthonormal increment process} \citep[Section 2.3.1]{von2019spectral} with zero mean and covariance matrix in differential notation with the following entries \citep[Theorem 4.6.1]{brillinger2001time} for $a, b \in \{1, \ldots, p\}$:
$$\cov(\dd Z^{(a)}(\omega), \dd Z^{(b)} (\omega')) = \EE[\dd Z^{(a)}(\omega) \overline{\dd Z^{(b)}(\omega')}] = (f(\omega))_{a, b}\ \delta(\omega - \omega')\ \dd \omega\ \dd \omega'.$$
where $\delta(.)$ denotes the \textit{Dirac delta} function. Thus $X_t$ can be viewed as a continuous superposition of harmonic components $e^{\i t \omega}$ at all frequencies $\omega \in [-\pi, \pi]$, and $\dd Z(\omega)$ is the random amplitude at frequency $\omega$ \citep{ombao2022spectral} whose marginal dependence structure across the coordinates are captured by $f(\omega)$. Hence similar to real-valued precision matrix for i.i.d. Gaussian data that capture the conditional dependence among the data coordinates, $\Theta(\omega) = [f(\omega)]^{-1}$ can be viewed as a frequency-specific analogue of precision matrix capturing the conditional dependence among the coordinates of $Z(\omega)$ (see Appendix \ref{sec:freq_graphical_model} for a detailed illustration). Since the DFT in \eqref{eq:dft}, an inner product between $X_t$ and the Fourier basis at a given frequency $\omega_j$, is a natural finite sample analogue of $dZ(\omega_j)$, its sample covariance matrix serves as a key instrument for our estimation.

\subsection{Asymptotic distribution of DFT and averaged periodogram}\label{subsec:classical_estimators}

We denote the set of \textit{boundary frequencies} as $B_n := \{-\lfloor(n-1)/2 \rfloor, 0, \lfloor n/2 \rfloor \}$. For a bandwidth $m$, the set of $2m+1$ \textit{nearby} frequencies around $j$ is denoted by $\wmj := \{k \in F_n: |k - j| \le m\}$. Here the indices of Fourier frequencies $j$ are evaluated ``modulo $n$" as $k-j$ may fall outside $F_n$. Following \citet[Theorem~11.7.1]{brockwell1991time}, \citet[Theorem~4.4.1]{brillinger2001time} and under regularity conditions on $X_t$, $d_j$ for $j \in F_n \setminus B_n$ converges in distribution to a mean-zero complex Gaussian random vector with covariance matrix $f(\omega_j)$ as $n \to \infty$. Under regularity conditions on the underlying time series, $f(\omega_k)$ can be approximated by $f(\omega_j)$ if $k$ is near $j$ (Lemma \ref{lem:lip_smooth} in Appendix). Hence we can treat $\{d_k\}_{k\in \wmj}$ as nearly independent i.i.d. complex Gaussian random vectors with mean zero and complex valued covariance matrix $f(\omega_j)$, and construct estimators of $f(\omega_k)$ and $\Theta(\omega_k)$, $k \in F_n$ with analogous methods available for i.i.d. real-valued data involving kernel estimators \citep{brockwell1991time, brillinger2001time, priestley1988spectral, hannan2009multiple}.

The periodogram at frequency $\omega_j$ is defined as $I(\omega_j):= d_j d_j^\dagger$. We denote the DFT matrix centered at $j$ as
\begin{equation}\label{eq:dft_mat}
    \Zcal = \left[\Zcal^{(1)} : \ldots : \Zcal^{(p)}\right] := \begin{bmatrix}
        d_{j-m}^\dagger\\
        \vdots\\
        d_{j+m}^\dagger
    \end{bmatrix}_{(2m+1) \times p}.
\end{equation}
For brevity, $j$ is excluded in the notation of $\Zcal$. The averaged periodogram estimator at $\omega$ is 
\begin{equation}\label{eq:avg_periodogram}
    \hat f(\omega_j) := \frac{1}{2m+1}\sum_{k \in \wmj} I(\omega_k) = \frac{1}{2m+1} \Zcal^\dagger \Zcal.
\end{equation} 
In classical asymptotic regime, $m = \bigO{n^{-4/5}}$ is a standard choice of the bandwidth \citep{beltrato1987determining}. In high dimensional regimes, $m = \bigO{\sqrt{n}}$ \citep{bohm2009shrinkage} ensures that $\hat f(\omega_j)$ is consistent to $f(\omega_j)$. In Section \ref{sec:theory}, we explore regimes of $m$ depending on $n$ and $p$ and other model dependent parmaters for consistent estimation of $\Theta(\omega_j)$.

\subsection{Complex graphical lasso (CGLASSO)}\label{subsec:penalized_whittle_liklihood}

Penalized Whittle likelihood methods have been widely explored in recent works related to estimation of spectral precision matrix under sparsity \citep{jung2015graphical, dallakyan2022time, baek2023local, tugnait2022sparse}. We propose an estimator of $\Theta(\omega_j)$ that builds upon a local approximation of the Whittle likelihood of the underlying multivariate time series, and imposes $\ell_1$-penalty on the off-diagonal entries to accommodate sparsity in high dimension. 

Since $d_k, k \in \wmj$ are approximately independent and asymptotically complex normal for $\omega_j$ sufficiently away from zero i.e. $B_n \cap \wmj = \emptyset$ (Section \ref{subsec:classical_estimators}), we use the DFTs to construct an approximate Gaussian likelihood around $\omega_j$ with $2m+1$ nearby  frequencies. This construction is similar to \textit{Whittle likelihood} of DFTs \citep{whittle1951hypothesis, sykulski2019debiased}. Under smoothness conditions on $f(\cdot)$ (See Lemma \ref{lem:lip_smooth} in Appendix), we approximate $\Theta(\omega_k)$ with $\Theta(\omega_j)$ for $k \in \wmj$. The approximated negative-log likelihood is
\begin{align}
    & \sum_{k \in \wmj} \left(-\log \det \Theta(\omega_j) +  d^\dagger_k \Theta(\omega_j) d_k \right) \nonumber \\
    =~ & -(2m+1)\log\det\Theta(\omega_j) + \trace\left(I(\omega_k)\Theta(\omega_j)\right). \label{eq:penalized_whittle}
\end{align}
The estimator $\hat \Theta_j:= \hat\Theta(\omega_j)$ of $\Theta(\omega_j)$ that maximizes the $\ell_1$-penalized approximated likelihood is defined as the solution of the following optimization problem
\begin{equation}\label{eq:graphical-lasso}
     \min_{\Theta\in \pdhermit} \left\{ - \log \det \Theta +\trace(\hat f(\omega_j)\Theta) + \lambda \|\Theta^-\|_1 \right\},
\end{equation}
where $\lambda > 0$ is the penalty parameter, and $\left\|\Theta^-\right\|_1$ is the $\ell_1$-norm of the vectorized version of $\Theta^-$, i.e. the sum of the absolute off-diagonal elements of $\Theta$. We call the optimization problem \eqref{eq:graphical-lasso} as \textit{complex graphical lasso} (CGLASSO). The CGLASSO estimator is a complex valued analogue to graphical lasso estimator of precision matrix for i.i.d. real valued data \citep{friedman2008sparse, ravikumar2011high}-- the inputs being $\hat f(\omega_j)$ in place of the sample gram matrix.

\begin{remark}[Estimation near boundary frequencies]
    We propose the CGLASSO estimator in \eqref{eq:graphical-lasso} for $\omega_j$ sufficiently away from the boundary frequencies. Later shown in Theorem \ref{thm:consistency_cglasso}, consistency of $\hat\Theta(\omega_j)$ around $\Theta(\omega_j)$ relies on the fact that $\hat f(\omega_j)$ is consistent for $f(\omega_j)$ (Lemma \ref{lem:single_deviation_bound}). Near the boundary frequencies i.e. $B_n \cap \wmj \neq \emptyset$, $\hat f(\omega_j)$ remains consistent for $f(\omega_j)$ and the CGLASSO estimator in \eqref{eq:graphical-lasso} can still be applied to estimate $\Theta(\omega_j)$ for frequencies near the boundary. Since our focus is on estimation rather than inference for $\Theta(\omega_j)$, potential boundary effects due to the repeated DFTs (up to conjugate) around the boundary frequencies are asymptotically negligible for estimating $\Theta(\omega_j)$.
\end{remark}

\subsection{Complex adaptive graphical lasso (CAGLASSO)}\label{subsec:caglasso}

In CGLASSO \eqref{eq:graphical-lasso}, a uniform $\lambda$ across the entries is not ideal when variables are on different scales, leading to heterogeneous variances \citep{jankova2018inference, carter2023partial}. This necessitates careful handling of the scales associated with the DFTs. We now introduce \textit{complex adaptive graphical lasso} (CAGLASSO) that adapts the entry-wise penalties in CGLASSO in \eqref{eq:graphical-lasso} by the estimated partial variances of the DFTs. In adaptive thresholding of covariance matrix \citep{cai2011adaptive}, the choice of threshold can be recovered by estimating the diagonals of the covariance matrix. In contrast, the choice of penalty for estimating precision matrix should analogously depend on the diagonals of the precision matrix that are more difficult to recover. We apply nodewise regression (NWR) of the DFTs for recovering the scales of the diagonals of the spectral precision matrix.

NWR involves regression of the columns of the design matrix one at a time on its other columns \citep{meinshausen2006high, van2014asymptotically, jankova2017honest, jankova2018inference}. For i.i.d. designs, this pseudo-likelihood method recovers the support of the precision matrix. We implement a NWR method on $\Zcal$ to construct data-driven and adaptive weights that are subsequently integrated into the entry-wise adaptive penalty of CGLASSO. This procedure involves solving $p$ separate penalized regression problems. When $p$ is large, each NWR problem is solved using lasso to obtain sparse solutions.

\subsubsection{Estimation of partial variance with NWR}\label{subsubsec:method-nodewise-reg}

For $a \in [p]$, we denote the $a$\textsuperscript{th} column of the DFT matrix $\Zcal$ by $\Za$, and the submatrix containing all the other columns by $\Zma$. The estimated NWR coefficients are $\bjahat = \{(\bjahat)_b: b \in [p] \setminus \{a\} \}$ such that
\begin{equation}\label{eq:nodewise-reg}
    \bjahat = \argmin_{\beta \in \CC^{p-1}}\left\{\frac{1}{2(2m+1)}\big\|\Za - \Zma\beta \big\|_2^2 + \lambda_a \|\beta \|_1\right\},
\end{equation}
where $\|\beta\|_1 = \sum_{b\ne a}|\beta_b|$ and $\lambda_{a} > 0$ is the penalty parameter for recovering the $a$\textsuperscript{th} column. Following \citet{van2014asymptotically, jankova2017honest}, the estimator of the regression noise levels is
\begin{equation}\label{eq:nwr_partial_variance}
     \tauahatsq := \frac{1}{2(2m+1)} \big\| \Za - \Zma \bjahat \big\|_2^2 + \lambda_a \big\|\bjahat\big\|_1.
\end{equation}
$\tauahatsq$ is referred interchangeably as NWR residual variance estimator, regression noise estimators in the literature \citep{van2014asymptotically, jankova2017honest}. For the rest of our paper, we refer to $\tauahatsq$ as the \textit{estimated partial variance}. The Karush-Kuhn-Tucker (KKT) condition of lasso
$$ \frac{1}{2m+1}( \Zma )^\dagger \left(\Za - \Zma \bjahat\right) + \lambda_a \partial\left( \big\| \bjahat \big\|_1 \right) = 0, $$
readily implies
\begin{equation}\label{eq:nwr_partial_variance2}
    \tauahatsq = \frac{1}{2m+1}\left(\Za - \Zma \bjahat\right)^\dagger\Za.
\end{equation}
Similar to NWR with real valued i.i.d. data, we can construct a pseudo-estimator for the columns of $\Theta(\omega_j)$ using $\bjahat$ for $a \in [p]$ and subsequently recover the sparsity structure of $\Theta(\omega_j)$ \citep[Section 2.1.1]{van2014asymptotically}. We use the estimated partial variances $\hat\tau_a^2,~ a \in [p]$ to construct weights for penalizing each entry of $\Theta(\omega)$.

\subsubsection{Estimating $\Theta(\omega_j)$ with partial variance adapted penalty} \label{subsubsec:graphical_lasso_adaptive}

The CAGLASSO estimator $\hat \Theta_j^{(\tau)}$  is obtained from the following optimization problem:
\begin{equation}\label{eq:caglasso}
 \min_{\Theta\in \pdhermit} \left\{ - \log \det \Theta +\trace(\hat f(\omega_j)\Theta) + \lambda \sum_{a\ne b} \hat \tau_a \hat\tau_b |\Theta_{a,b}| \right\},
\end{equation}
Since $\Theta(\omega_j)$ captures the conditional dependence structure in a fixed frequency (Section \ref{subsec:frequency_graph}), scaling the penalties by partial variances yields more accurate entry-wise estimate of $\Theta(\omega_j)$. If the diagonals of $\Theta(\omega_j)$ are known, using penalties proportional to $[(\Theta(\omega_j))_{a,a}(\Theta(\omega_j))_{b,b}]^{-1/2}$ is equivalent to applying a uniform $\ell_1$-penalty on the scales of partial coherencies. As shown later in Corollary \ref{cor:consistency_partial_variance}, $\hat\tau_a^2$ concentrates around $1/(\Theta(\omega_j))_{a,a}$, justifying our choice of the adaptive penalization. 

Solving  \eqref{eq:caglasso} is equivalent to scaling each $(a,b)$\textsuperscript{th} entry of $\hat f(\omega_j)$ with $\hat \tau_a \hat \tau_b$ and using the scaled averaged periodogram as the input to CGLASSO, and rescaling the output thereafter. The detailed steps are described in Algorithm \ref{alg:caglasso}.

\vspace{5pt}

\begin{figure}[t]
    \centering
    \subfloat[]{\includegraphics[trim=0 10 20 20, clip, width=0.33\linewidth]{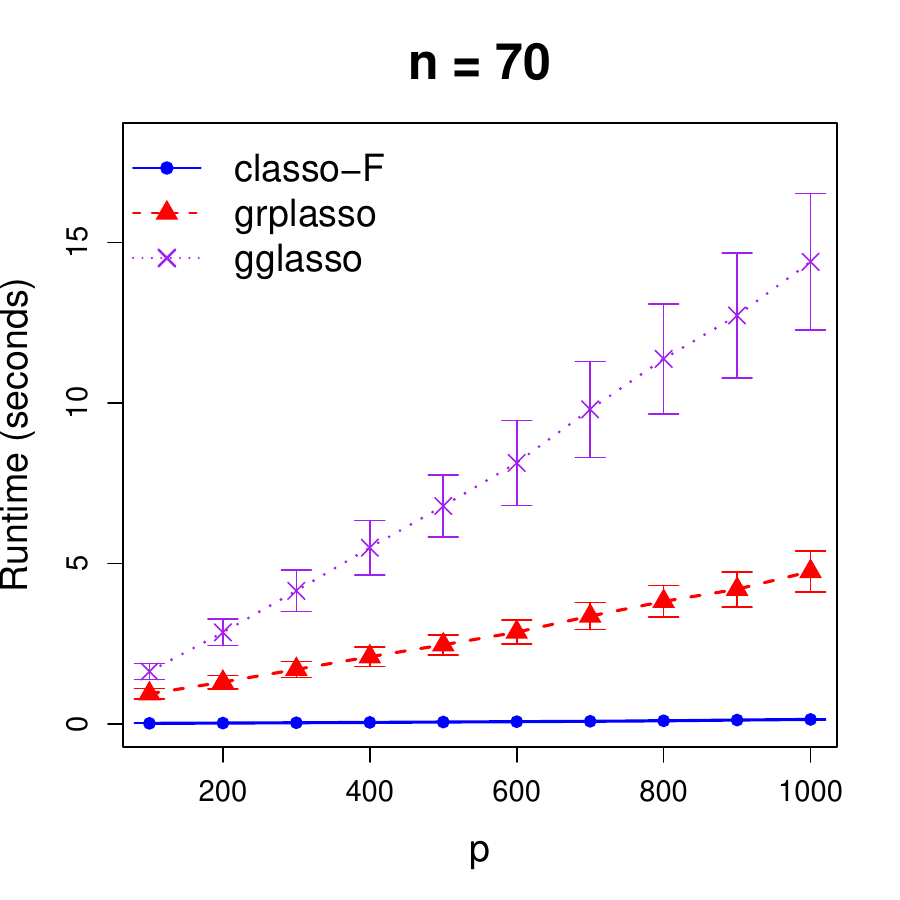}}
    \subfloat[]{\includegraphics[trim=0 10 20 20, clip, width=0.33\linewidth]{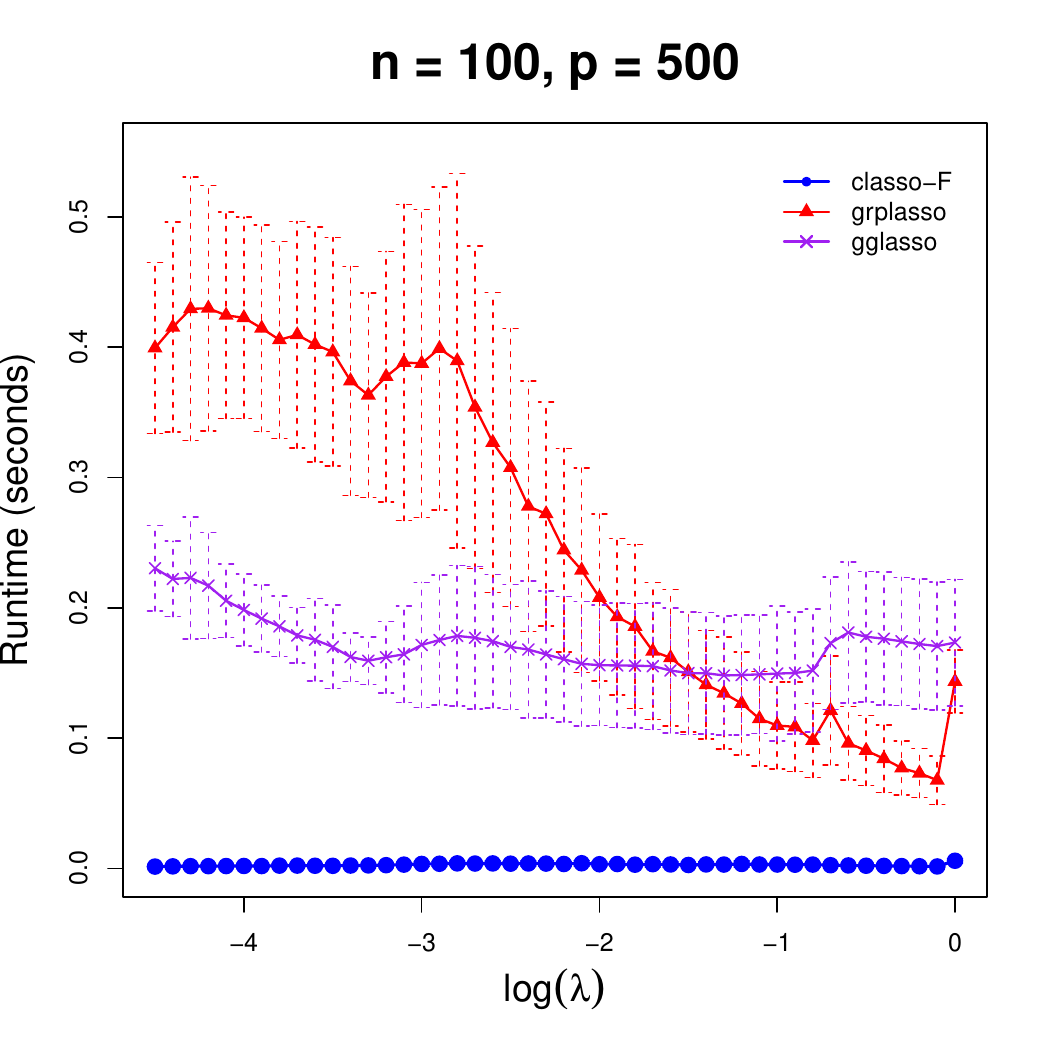}}
    \subfloat[]{\includegraphics[trim=0 0 0 0, clip, width=0.33\linewidth]{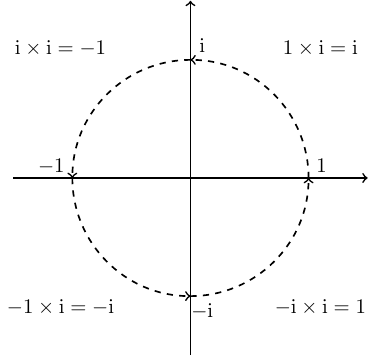}}
    \caption{\textbf{Runtime Improvement of CLASSO leveraging the within-group orthogonal structure of the predictors in the equivalent group lasso problem} (see Remark \ref{rem:classo} for details). Panel (a) shows the total runtime for a regularization path (median across 50 trials) versus dimension $p$ for CLASSO (blue circles), \texttt{grplasso} (red triangles), and \texttt{gglasso} (purple crosses) ($n=70$; details in Appendix \ref{subsec:speec_comp_exp}). Panel (b) shows the median runtime across 50 trials versus $\log \lambda$ for the same methods ($n=100, p = 500$). In panels (a) and (b), vertical bars indicate the median absolute deviation of runtime. CLASSO is substantially faster than the benchmarks across both dimensions and the regularization path. Panel (c) illustrates the isomorphism between complex multiplication and orthogonal rotation that is key to the implementation of CLASSO: multiplication by $\i$ is isomorphic to $\pi/2$ counterclockwise rotation in $\mathbb{R}^2$, mapping $(1,0)\mapsto(0,1)$, $(0,1)\mapsto(-1,0)$, $(-1,0)\mapsto(0,-1)$ and $(0,-1)\mapsto(1,0)$.}
    \label{fig:advantage_of_phi}
\end{figure}

\section{Optimization algorithms}\label{sec:algo}

We describe the algorithms for solving the optimization problems \eqref{eq:graphical-lasso}, \eqref{eq:nodewise-reg} and \eqref{eq:caglasso}. Standard convex optimization solvers e.g. ADMM can solve \eqref{eq:graphical-lasso}, however they do not leverage the problem’s sparsity structure. In high-dimensional regression and Gaussian graphical models (GGMs) literature, coordinate descent (CD) algorithms are widely used due to their computational efficiency, enabled by strategies such as warm starts and active set screening \citep{hastie2015statistical}. However, extending CD to the complex graphical lasso is nontrivial, as the lasso penalty $\|\beta\|_1$ for $\beta \in \CC^p$ corresponds to a group lasso penalty $\sum_{j=1}^p\sqrt{(\re(\beta_j))^2 + (\im(\beta_j))^2}$. Unlike the standard lasso, blockwise CD updates for group lasso generally lack closed-form solutions, requiring a separate optimization subproblem at each iteration \citep[Equation 4.14]{hastie2015statistical}, that considerably slows down the algorithm.

Our key observation is that the specific group lasso regression for complex lasso has an attractive feature. The predictors within a single group are \textit{orthogonal}, allowing  \textit{closed form} updates in blockwise CD iterations \citep[Equation 4.15]{hastie2015statistical}. Using this observation, we first propose CLASSO, a CD algorithm for lasso with complex variables. Then we use CLASSO to build Algorithm \ref{alg:cglasso} for solving the CGLASSO estimator \eqref{eq:graphical-lasso} and Algorithm \ref{alg:caglasso} for the CAGLASO estimator. As shown in Figure \ref{fig:advantage_of_phi}(a) and \ref{fig:advantage_of_phi}(b), leveraging the orthogonal structure among the witin-group predictors offers a substantial improvement of runtime for CLASSO than some popular existing \texttt{R} packages (see Appendix \ref{subsec:speec_comp_exp} for the details).

While this observation of orthogonality alone is sufficient for developing fast algorithms, we argue that this is not a coincidence. It captures a deep connection between complex and real matrices that can be leveraged to systematically \textit{realify} a broader family of complex optimization problems in statistics. We formalize this connection first, and then present the algorithms for solving CLASSO and CGLASSO.  

\subsection{Realifying complex scalars, vectors and matrices}\label{sec:realification}

At the core of our algorithm lies a well-known \textit{field isomorphism} between the set of complex scalars $\mathbb{C}$ and a set of $2 \times 2$ \textit{orthogonal} real matrices. This field isomorphism can be extended from $\mathbb{C}$ to a \textit{ring isomporphism} on complex-valued matrices $\mathbb{C}^{m \times n}$. At a high-level, sum, product and inverse and many other operations of $m \times n$ complex matrices have direct parallel to the same operations in the space of $2m \times 2n$ real-valued matrices. The upshot is that generic complex-variable optimization problems arising in frequency-domain time series can be reduced to well-understood optimization problems over real variables, for which fast pathwise CD algorithm exists. 

As shown in Figure \ref{fig:advantage_of_phi}(c), the identities $\i \times 1 = \i$ and $\i \times \i = -1$ imply that \textit{multiplication by} $\i$ can be viewed as an orthogonal rotation in $\RR^2$ that maps the unit vector $(1,0)$ to $(0,1)$, and maps the unit vector $(0,1)$ to $(-1,0)$. The two linear maps \textit{identity operation} and \textit{orthogonal rotation} correspond to the $2 \times 2$ orthogonal matrices 
\begin{equation*}
I_2 = \begin{bmatrix}
    1 & 0\\
    0 & 1
\end{bmatrix}, \quad
J_2 := \left[ \begin{array}{cc} 0 & -1 \\ 1 & 0 \end{array} \right].
\end{equation*}
respectively. If we correspond $1$ with $I_2$ and $\i$ with $J_2$, any complex number $z$ has a $2\times 2$ matrix representation $\re(z) I_2 + \im(z) J_2$. While these linear maps are well-known in complex analysis, isomorphic nature of this map has not been leveraged to speed up optimization problems in complex domain. We formalize the properties of these linear maps with the following propositions. For a brief introduction to the prerequisite abstract algebraic structures, we refer the readers to Appendix \ref{sec:intro_algebra}. 

\begin{proposition}[Realification of complex scalars]\label{prop:phi1}
Define $\varphi : \CC \rightarrow \RR^{2\times 2}$ as
$$ \varphi(z) := \normalfont{\begin{bmatrix}
    \re(z) & -\im(z)\\
    \im(z) & \re(z)
\end{bmatrix}} .$$
Then $\varphi$ is a field isomorphism between $\CC$ and $\Mb$ with
\begin{align}
    \Mb := \left\{ \begin{bmatrix}
    a & -b\\
    b & a
    \end{bmatrix}: a, b\in \RR \right\}, \label{eq:matrix-class}
\end{align}
implied by the following properties:
\begin{enumerate}[label = (\roman*)]
    \item $\varphi(0) = \mathbf{0}_{2 \times 2}$ and $\varphi(1) = I_{2}$.
    \item  $\varphi(z + z') = \varphi(z) + \varphi(z')$ for $z, z' \in \CC$. In particular, $\varphi(-z) = -\varphi(z)$ for  $z \in \CC$.
    \item $\varphi(z z') = \varphi(z) \varphi(z')$ for all $z, z' \in \CC$. In particular, $\varphi(z^{-1}) = [\varphi(z)]^{-1}$ for $z \in \CC, z \ne 0$.
\end{enumerate}
Additionally for any $z\in \CC$, and $\varphi(z) = [\varphi_1(z) : \varphi_2(z)] $, the following hold:
\begin{enumerate}[label = (\roman*)]
    \item[(a)] $ \varphi_1(z)^\top \varphi_2 (z)  = 0$.
    \item[(b)] $\|\varphi_1(z)\|_2^2 = \|\varphi_2(z)\|_2^2 = \frac12 \|\varphi(z)\|_F^2 = |z|^2$.
    \item[(c)] $\varphi(z^\dagger) = [\varphi(z)]^\top.$
\end{enumerate}
\end{proposition}

The proof is deferred to Appendix \ref{pf:phi1}. Since $\varphi$ is a field isomorphism, $\varphi$ retains the structural properties of $\CC$ in $\Mb$.
The map $\varphi$ doubles the dimension of an input complex number, however the computational complexity of operations with objects from $\Mb$ remains the same as operations involving complex numbers.

The isomorphism $\varphi$ has natural extensions to higher dimensional vector and matrix spaces by replacing every entry with its $2\times 2$ realified image. However the map $\varphi$ will no longer be a field isomorphism since for $m, n\ge 1$, $\CC^n$ and $\CC^{m\times n}$ are rings but not necessarily fields. For any $\zb = (z_1,\ldots, z_n)^\top \in \CC^n$, we extend $\varphi: \CC^n \rightarrow \RR^{2n\times 2}$ as
\begin{equation}\label{eq:phi-extn-1}
    \varphi(\zb) = \begin{bmatrix}
    \varphi(z_1)\\
    \vdots \\
    \varphi(z_n).
\end{bmatrix} \end{equation}
We denote the first column of $\varphi(\zb)$ as $\varphi_1(\zb)$ and the second column as $\varphi_2(\zb)$. Similarly, for any matrix $Z = ((Z_{i,j}))_{i \in [m], j \in [n]} \in \CC^{m\times n}$, the extension $\varphi: \CC^{m\times n}\rightarrow \RR^{2m\times 2n}$ is
\begin{equation}\label{eq:phi-extn-2}
    \varphi(Z) = \begin{bmatrix}
    \varphi(Z_{1,1}) & \cdots & \varphi(Z_{1,n})\\
    \vdots & \ddots & \vdots\\
    \varphi(Z_{m,1}) & \cdots & \varphi(Z_{m,n})
\end{bmatrix}. 
\end{equation}
The extended maps preserve properties (i) and (ii) of Proposition \ref{prop:phi1} in higher dimensions, hence they are \textit{group isomorphisms}. The preservation of the inverse as in property (iii) of Proposition \ref{prop:phi1} may not necessarily hold in higher dimensions -- resulting the extended $\varphi$ maps for vectors and matrices to be a \textit{ring isomorphisms}, formally stated in the next two propositions. $\mathrm{Image}(\varphi)$ denotes the image under the map $\varphi$.

\begin{proposition}[Realification of complex vectors]\label{prop:phi2}
The map $\varphi: \CC^{n}\rightarrow \RR^{2n\times 2}$ in \eqref{eq:phi-extn-1} is a ring isomorphism between $\CC^n$ and $\mathrm{Image}(\varphi)$. Furthermore, the following properties are satisfied:
\begin{itemize}
    \item[(a)] $\varphi_1(\zb)^\top \varphi_2 (\zb) = 0$.
    \item[(b)] $\|\varphi_1(\zb)\|_2^2 = \|\varphi_2(\zb)\|_2^2 = \frac12 \|\varphi(\zb)\|_F^2 = \|\zb\|_2^2$.
\end{itemize}
\end{proposition}

\begin{proposition}[Realification of complex matrices]\label{prop:phi3}
    The map $\varphi: \CC^{m\times n}\rightarrow \RR^{2m\times 2n}$ in \eqref{eq:phi-extn-2} is a ring isomorphism between $\CC^{m\times n}$ and $\mathrm{Image}(\varphi)$. Furthermore, the following properties are satisfied:
    \begin{itemize}
        \item[(a)] $ \varphi_1(Z_{\cdot, j})^\top \varphi_2 (Z_{\cdot, j}) = 0$ for $j \in [n]$.
        \item[(b)] $ \|\varphi(Z)\|_F^2 = 2\|Z\|_F^2$.
        \item[(c)] $\|\varphi(Z)\| = \|Z\|$.
        \item[(d)] $\varphi(Z^\dagger) = [\varphi(Z)]^\top$.
        \item[(e)] Let $\mathrm{GL}_n(\CC)$ be the set of all $n\times n$ invertible matrices with entries in $\CC$. Then
        for any $Z\in \mathrm{GL}_n(\CC)$, $\varphi(Z^{-1}) = [\varphi(Z)]^{-1}$. 
    \end{itemize}
\end{proposition}

The proofs are deferred to Appendix \ref{pf:phi2} and \ref{pf:phi3} respectively.

\subsection{Realifying standard statistical optimization problems}\label{subsec:realifying_stat_opt}

While the realification map $\varphi$ is useful for establishing algebraic properties, an appropriate rearrangement of the rows and columns in its image allows for a more concise description of the algorithms. For any $k \in \NN$, we define a permutation matrix
$$ \Pi_k := \left[ \e_1\ : \e_3 : \ldots : \e_{2k-1} : \e_{2} : \e_4 : \ldots \e_{2k} \right]_{2k \times 2k}. $$
In words, $\Pi_k$ maps the vector $(1, \ldots, 2k)^\top$ to $(1, 3, \ldots, 2k-1, 2, 4, \ldots, 2k)^\top$. Then for $\zb\in \CC^n$ and any $Z\in \CC^{m\times n}$, we denote
\begin{align*}
    \tilde{\zb} := & \Pi_n^\top \varphi(\zb) \e_1 =  \begin{bmatrix}
        \re(\zb) \\ \im(\zb)
    \end{bmatrix},\\
    \tilde{\tilde{Z}} := &        \Pi_m^\top \varphi(Z) \Pi_n = \begin{bmatrix}
        \re(Z) & -\im(Z)\\
        \im(Z) & \re(Z)\\
    \end{bmatrix}.
\end{align*}
The isomorphism $\varphi$ has a broader impact on solving penalized complex-valued optimization problems of statistical interest. Since $\varphi$ preserves a ring isomorphic structure on complex vectors and matrices, the underlying convex objective function can be translated as a real valued objective and the isomorphic structure can be used to develop fast optimization routines. We demonstrate this idea with two lemmata.

\begin{lemma}[OLS with complex variables]\label{lem:ols}
    For a regression problem with response $Y\in \CC^n$, predictor $X\in \CC^{n\times p}$ and coefficient $\beta\in \CC^p$, $\varphi$ admits an equivalence between residual sum of squares as $\|Y-X\beta\|_2^2 = \frac{1}{2}\|\tilde Y - \dtilde{X}\tilde\beta\|_2^2$. Furthermore, this correspondence extends to the normal equations as
    $X^\dagger X\beta = X^\dagger Y \iff \dtilde{X}^\top \dtilde{X} \beta = \dtilde{X}^\top \tilde{Y}$.
\end{lemma}

\begin{lemma}[Complex log-determinant program]\label{lem:log-det}
    Assume that $Z_1\ldots, Z_n$ are i.i.d. observations from a complex-valued normal distribution with mean zero and complex-valued covariance matrix $\Theta^{-1}$ with $ \Theta \in \pdhermit$. Then for the complex-valued sample Gram matrix $\hat \Sigma := \frac{1}{n} \sum_{i=1}^n Z_i Z_i^\dagger$, the convex log determinant barrier function 
    $L_\CC(\Theta) := -\log \det\Theta + \trace(\hat \Sigma\Theta)$ can be translated to the real analogue $L_\RR(\Theta) := -\log \det\dtilde{\Theta} + \trace\bigg(\dtilde{\hat \Sigma}\dtilde{\Theta}\bigg)$, as 
    $L_\CC(\Theta) = \frac{1}{2}L_\RR(\Theta)$.
\end{lemma}

The proofs are deferred to Appendix \ref{pf:ols_complex} and \ref{pf:glasso_complex} respectively. Lemma \ref{lem:ols} and \ref{lem:log-det} establish a concrete link between two complex-valued optimization problems in statistical settings and their real-valued analogue, for which efficient solution techniques are well developed. Thus one can mimic the OLS and GLASSO algorithms in complex domain with the same operations on the realified variables. Leveraging this correspondence, we propose algorithms for solving CLASSO and CGLASSO.

\subsection{CLASSO optimization algorithm}\label{subsec:classo}

\begin{figure}[!t]
\centering
\begin{minipage}[t]{0.48\linewidth}
\begin{algorithm}[H]
\caption{CLASSO with CD} \label{alg:classo}
\begin{algorithmic}
\State \textbf{Input}~~ $X \in \CC^{n \times p}$ with $\|X_j\|_2 = \sqrt{n}$ for all $j \in [p]$; response $Y \in \CC^n$; initial parameter $\hat\beta^{(0)} \in \CC^p$; penalty $\lambda > 0$.
\State \textbf{Initialize}~~ $\hat\beta \gets \hat\beta^{(0)}$;~ $r \gets Y - X \hat\beta$
\Repeat
  \For{$j = 1$ to $p$}
    \State $r^{(j)} \gets r + X_j \hat\beta_j$~~~~// $j$\textsuperscript{th} partial residual
    \State $\hat\beta_j \gets \mathcal{S}_\lambda \left( \frac{1}{n} X_j^\dagger r^{(j)} \right)$
    \State $r \gets r^{(j)} - X_j \hat\beta_j$
  \EndFor
\Until{convergence}
\State \Return $\hat\beta$
\end{algorithmic}
\end{algorithm}
\end{minipage}
\hfill
\begin{minipage}[t]{0.48\linewidth}
\begin{algorithm}[H]
\caption{CLASSO.COV with CD} \label{alg:classo-cov}
\begin{algorithmic}
\State \textbf{Input}~~ $S_{XX} \in \CC^{p \times p}$ with $\operatorname{diag}(S_{XX}) = 1$; $\sbf_{XY} \in \CC^p$; initial $\hat\beta^{(0)} \in \CC^p$; penalty $\lambda > 0$.
\vspace{0.5pt}
\State \textbf{Initialize}~~ $\hat\beta \gets \hat\beta^{(0)}$;~ $\sbf_{Xr} \gets \sbf_{XY} - S_{XX} \hat\beta$
\Repeat
  \For{$j = 1$ to $p$}
    \State $v \gets \sbf_{Xr} + S_{XX} \hat\beta_j$
    \State $\hat\beta_j \gets \mathcal{S}_\lambda(v_j)$
    \State $\sbf_{Xr} \gets v - S_{XX} \hat\beta_j$
  \EndFor
\Until{convergence}
\State \Return $\hat\beta$
\vspace{18pt}

\end{algorithmic}
\end{algorithm}
\end{minipage}
\end{figure}

We use the isomporphism trick with $\varphi$ to build up a pathwise CD algorithm for the complex lasso (CLASSO) problem. The real-complex isomorphic structure enables us to transform the complex lasso into a group lasso problem \citep{yuan2006model} with the reals and imaginary parts of the regression coefficient forming a group. For response $Y = (Y_1,\ldots, Y_n)^\top \in \CC^n$, predictor $X = [X_1 :\ldots: X_p] \in \CC^{n\times p}$ and coefficient $\beta = (\beta_1,\ldots, \beta_p)^\top \in \CC^p$, the CLASSO problem is formulated as follows
\begin{equation}\label{eq:classo}
    \hat \beta = \argmin_{\beta\in \CC^p} \left\{\frac{1}{2n}\|Y - X\beta\|_2^2 + \lambda\|\beta\|_1\right\},
\end{equation}
Using $\varphi$ on \eqref{eq:classo}, the transformed problem with real variables is
\begin{equation}\label{eq:grplasso}
    \hat{\tilde\beta} = \argmin_{\tilde{\beta}_1,\ldots, {\tilde \beta}_p \in \RR^2} \left\{ \frac{1}{2n}\|\tilde{Y} - \sum_{j = 1}^p \tilde{\tilde{X_j}} \tilde{\beta}_j \|_2^2 + \lambda \sum_{j = 1}^p \|\tilde{\beta}_j\|_2\right\}.
\end{equation}
The optimization problem \eqref{eq:grplasso} is a group lasso \citep{yuan2006model} with $p$ groups, each of size 2. 

\begin{remark}[Closed form solution for orthogonality]\label{rem:classo}
The block CD update for \eqref{eq:grplasso}, in general, has no closed form expression. However using Proposition \ref{prop:phi3}(a), the columns of $\dtilde{X}_j$ are orthogonal and hence the block CD has a closed form expression \citep[Equation 4.15]{hastie2015statistical}.
\end{remark}

When the response, the predictors and the coefficients are real valued, standard CD solves lasso by iteratively soft-thresholding the partial residuals \citep{hastie2009elements}. In complex variable setting, \textit{the soft thresholding operator} ${\cal S}_\lambda: \CC^p \rightarrow \CC$ defined as
$$ {\cal S}_{\lambda}(z) := z~ \max\left\{0,\ 1 - \frac{\lambda}{\|z\|_2}\right\}. $$
The CD algorithm for \eqref{eq:classo} is illustrated in Algorithm \ref{alg:classo}, with the details on the closed form updates due to the within-group orthogonal structure of the predictors are deferred to Appendix \ref{subsec:classo_simplification}. Next we address some advantages of using Algorithm \ref{alg:classo} for solving CLASSO.
 
\paragraph*{Computational benefit of orthogonal group structure} We emphasize the role of the orthogonal predictor structure in the group lasso formulation of CLASSO. While the reformulation of CLASSO as a group lasso problem and its connection to standard solvers has been discussed in prior work \citep{maleki2013asymptotic}, the use of within-group predictor orthogonality to enable closed-form updates has, to our knowledge, not been explored. In Appendix~\ref{subsec:speec_comp_exp}, we present a runtime comparison between CLASSO and popular group lasso \texttt{R} packages, demonstrating the computational gains achieved by exploiting this orthogonality.

\paragraph*{Covariance update} CLASSO can be implemented with a covariance updated version CLASSO.COV illustrated in Algorithm \ref{alg:classo-cov} (details in Appendix \ref{subsec:classo_simplification}) where the knowledge of $X^\dagger X$ and $X^\dagger Y$ suffices for carrying out the steps of CLASSO. Covariance update is suitable for graphical lasso where the columns of the covariance-like matrix are inner products between the columns of the DFT matrix.

\paragraph*{Speed improvement strategies} Speed of CLASSO can be further improved by computational strategies offered with the pathwise CD algorithm, such as \textit{warm start} for a regularization path containing a grid of $\lambda$, and \textit{active set screening} [Section 5.10]\citep{hastie2015statistical} for discarding predictors. Details are deferred to Appendix \ref{subsec:speed_improvement_strategies}.

\subsection{CGLASSO optmization algorithm}\label{subsec:cglasso}
We return to the optimization problem \eqref{eq:graphical-lasso}. For brevity, we drop $j$ for the frequency being fixed, and denote $\hat f(\omega_j)$ as $P$.

\begin{algorithm}[!t]
\caption{CGLASSO with CD update} \label{alg:cglasso}
\begin{algorithmic}
\State \textbf{Input}~~ $P = \hat f(\omega_j)$, initial ${\cal B}^{(0)} \in \CC^{(p-1)\times p}$, penalty $\lambda > 0$
\State \textbf{Initialize}~~ $W \gets P$ (fix $\operatorname{diag}(W)$);\quad ${\cal B} \gets {\cal B}^{(0)}$
\Repeat
  \For{$a = 1$ to $p$ (cyclically)}
    \State $W_{1,1} \gets W_{-a,-a}$; \quad $\mathbf{w}_{1,2} \gets W_{-a,a}$; \quad $w_{2,2} \gets W_{a, a}$ ~~~ // Partition submatrices
    \State $\hat\beta \gets \text{CLASSO.COV}(W_{1,1}, \mathbf{w}_{1,2}, {\cal B}_{\cdot,a}, \lambda)$;\quad  ${\cal B}_{\cdot,a} \gets \hat\beta$;\quad $\mathbf{w}_{1,2} \gets W_{1,1} \hat\beta$
  \EndFor
\Until{convergence}
\For{$a = 1$ to $p$} ~~~ // Final cycle
  \State $\hat\theta_{2,2} \gets \frac{1}{w_{2,2} - \mathbf{w}_{1,2}^\dagger \hat\beta}$; \quad $\hat\thetabf_{1,2} \gets -\hat\beta \cdot \hat\theta_{2,2}$
\EndFor
\State \Return $\hat\Theta$
\end{algorithmic}
\end{algorithm}

Using Lemma \ref{lem:existence_uniqueness}, the solution of \eqref{eq:graphical-lasso} is equivalent to solving the subgradient equation $P - \Theta^{-1} + \lambda \Psi = 0$ for $\Theta$, where $\Psi= ((\psi_{a,b}))_{a, b \in [p]}$ with $\psi_{a,a} = 0$, and
\begin{equation}\label{eq:sign}
\psi_{a,b} = 
\begin{cases}
\Theta_{a,b}/|\Theta_{a, b}| & \text{if } \Theta_{a,b} \neq 0,\\
\in \{z\in \CC: |z|\leq 1\} & \text{if } \Theta_{a,b} = 0.
\end{cases}
\end{equation}
for $a\neq b$. \eqref{eq:graphical-lasso} can be solved using block CD \citep{friedman2008sparse}. At each iteration, the current estimates of $W = \Theta^{-1}$ are partitioned as
$$W = \begin{pmatrix}
W_{1,1} & \wbf_{1,2} \\
\wbf_{1,2}^\dagger & w_{2,2}
\end{pmatrix},$$
where $W_{1,1} \in \CC^{(p-1) \times (p-1)}$, $\wbf_{1,2} \in \CC^{p-1}$, and $w_{2,2} \in \CC$. Similar partitions are considered for $\Psi, P$ and $\Theta$. Solving the last row and column $\thetabf_{1,2}$ at each step is equivalent to solving 

\begin{equation}\label{eq:classo_cov_cglasso}
W_{1,1}\beta - \pbf_{1,2} + \lambda \boldsymbol{\psi}_{1,2}=0,
\end{equation}
where $\beta = -\thetabf_{1,2}/\theta_{2,2}$. which can be solved using CLASSO.COV (Algorithm \ref{alg:classo-cov}) with $S_{XX} = W_{1,1},\ \sbf_{XY} = \pbf_{1,2}$ and $\lambda >0$. Based on Algorithm \ref{alg:classo-cov}, we develop a block CD Algorithm \ref{alg:cglasso} for solving CGLASSO problem in \eqref{eq:graphical-lasso}. In Algorithm \ref{alg:cglasso}, we denote $W_{-a,a}$ as the $a$\textsuperscript{th} column of $W$ without the $a$\textsuperscript{th} entry, and $W_{-a, a}$ as the submatrix of $W$ without the $a$\textsuperscript{th} row and $a$\textsuperscript{th} column.

\paragraph*{Warm and warmer start} Now we demonstrate a two-step computational efficiency warm up implemented in Algorithm \ref{alg:cglasso} that enhances its performance over a regularization path.
\begin{enumerate}
    \item[(a)] \textit{Warm start (CLASSO).}~ Similar to the warm start for CLASSO in Section \ref{subsec:classo}, $\lambda_0$ is set reasonably high in order to obtain all zero solutions for the off-diagonal entries. For each smaller $\lambda_t$ along the regularization path, the solution $\hat \Theta(\lambda_{t-1})$ is used as an initial value to compute $\hat \Theta(\lambda_t)$.
    \item[(b)] \textit{Warmer start (CGLASSO).}~ For a fixed $\lambda_t$ and at each iteration $a = 1 \text{ to } p$ cyclically, $\hat\beta$ from the previous cycle i.e. $\mathcal{B}_{\cdot, a}$ is used as initial values for the current cycle in order to make the CLASSO step in each cycle computationally more efficient.
\end{enumerate}

\subsection{Scaled variants of CGLASSO}\label{subsec:scaled_penalty}

We propose the scale-adaptive method CAGLASSO in Section \ref{subsec:caglasso} that adjusts the entry-wise penalties NWR estimated partial variances. 
Additionally, we consider two simpler alternatives strategies to scale the CGLASSO algorithm. We implement these alternative variants in our simulation study in Section \ref{sec:simulation}. We denote $P = \hat f(\omega_j)$, and drop $j$ from the estimated spectral precision matrix.

\subsubsection{CGLASSO with scaling variant-1 (CGLASSO-sc1)} The first approach is similar to \citet[Equation 10]{jankova2018inference}. We first solve a CGLASSO problem based on the sample coherency matrix $ \hat R =  \hat D^{-1}  P \hat D^{-1}$ with $\hat D^2 = \dg(P_{1,1}, \ldots, P_{p,p})$. Next we rescale the output $\hat K$ with $\hat D$. The CGLASSO-sc1 estimator is $\hat\Theta = \hat D^{-1} \hat K \hat D^{-1}$ where
\begin{equation*}
    \hat K \gets \text{CGLASSO}(\hat R, {\cal B}^{(0)}, \lambda),
\end{equation*}
which is equivalent to a graphical lasso with the penalty for $(a,b)$\textsuperscript{th} entry ($a \ne b$) being proportional to $\sqrt{ P_{a, a} P_{b, b}}$ \citep[Equation 9]{jankova2018inference}.

\subsubsection{CGLASSO with scaling variant-2 (CGLASSO-sc2)} In this variant, each invoke of CLASSO.COV within Algorithm \ref{alg:cglasso} takes the normalized inputs. For $a = 1,\ldots, p$, the update for $a$\textsuperscript{th} row and column is:
\begin{align}
    W_{1,1}^{\scl} & \leftarrow \hat D_{1,1}^{-1} W_{1,1} \hat D_{1,1}^{-1};\quad \hat D_{1,1}^2 := \dg(W_{1,1}) = \dg(P_{1,1}); \quad \pbf_{1,2}^{\scl} \leftarrow D_{1,1}^{-1} \pbf_{1,2};  \nonumber \\
    \hat\beta^{\scl} & = \text{CLASSO.COV}(W_{1,1}^{\scl}, \pbf_{1,2}^{\scl}, \mathcal{B}_{\cdot, a}, \lambda); \nonumber \\
    \mathcal{B}_{\cdot, a}  & \leftarrow \hat\beta^{\scl};\quad
    \hat\beta  \leftarrow D_{1,1}^{-1} \hat\beta^{\scl}; \quad
    \mathbf{w}_{1,2} \leftarrow W_{1,1}\hat\beta. \nonumber
\end{align}
The scaling matrix $\hat D_{1,1}$ in CGLASSO-sc2 is similar to the $\hat D$ in CGLASSO-sc1, except that the last diagonal entry $\hat D_{p,p}$ in CGLASSO-sc2 is disregarded in every update due to discarding the last row and column.

\subsection{Computation of CAGLASSO}\label{subsec:caglasso_computation}

\begin{algorithm}[!t]
\caption{CAGLASSO} \label{alg:caglasso}
\begin{algorithmic}
\State \textbf{Input}~~ $P = \hat f(\omega_j)$, NWR penalties $\lambda_a > 0$ for $a \in [p]$, CGLASSO penalty $\lambda > 0$, ${\cal B}^{(0)} = \mathbf{0}_{(p-1)\times p}$
\State \textbf{Initialize} $\tauahatsq \gets 1$ for all $a \in [p]$
\For{$a = 1$ to $p$} ~~~ // Calculating penalty weights with DFT-NWR
  \State $S_{XX} \gets P_{-a, -a};\quad s_{XY} \gets P_{-a,a};\quad s_{YY} \gets P_{a,a}$
  \State $\bjahat \gets \text{CLASSO.COV}(S_{XX}, s_{XY}, \lambda_a)$
  \State $\tauahatsq \gets \frac{1}{2(2m+1)} \left(s_{YY} - 2 \re(s_{XY}^\dagger \bjahat) + (\bjahat)^\dagger s_{XX} \bjahat\right) + \lambda_a \big\|\bjahat\big\|_1$
\EndFor

\State $\hat D_\tau^2 \gets \dg(\hat\tau_1^2, \ldots, \hat \tau_p^2)$
\State $\hat R_\tau \gets \hat D_{\tau}^{-1} P \hat D_{\tau}^{-1} $
\State $\hat K_\tau \gets \text{CGLASSO}(\hat R_\tau, {\cal B}^{(0)}, \lambda) $ ~~~ // Graphical lasso on the scaled averaged periodogram
\State $\hat \Theta^{(\tau)}\gets \hat D_{\tau}^{-1} \hat K_\tau \hat D_{\tau}^{-1} $
\State \Return $\hat\Theta^{(\tau)}$
\end{algorithmic}
\end{algorithm}

The CAGLASSO estimator in \eqref{eq:caglasso} is computed in Algorithm \ref{alg:caglasso} that undergoes two steps-- (a) computing the DFT-NWR partial variances $\tauahatsq$ for $a \in [p]$, and (b) applying CGLASSO on the scaled averaged periodogram, and then rescaling it back for obtaining the final estimator. While computing $\tauahatsq$ for each $a\in [p]$, the penalty parameters $\lambda_a$ can be selected with the minimum value of cross-validation error or Bayesian information criterion (BIC) \citep[Section 7.7 and Section 7.10]{hastie2009elements} of each NWR problem.

\section{Theoretical properties}\label{sec:theory}

Our theoretical results envelope the guarantees for the CGLASSO and CAGLASSO estimators. We first establish non-asymptotic error bound and support recovery guarantee of the the CGLASSO estimator in \eqref{eq:graphical-lasso} for a fixed frequency $\omega_j$. Next we derive estimation consistency of the DFT-NWR estimated coefficient $\bjahat$ in \eqref{eq:nodewise-reg}, and we show prediction consistency of $\Zma \bjahat$ around the true NWR prediction means, that enables us to establish error bounds for $\tauahatsq$. Finally, we prove error bound of the CAGLASSO estimator in \eqref{eq:caglasso}. The true spectral density is denoted by $f^*(\cdot)$ and the true spectral precision matrix as $\Theta^*(\cdot)$. We fix a Fourier frequency $\omega_j$ for $j \in F_n$, and denote the spectral density matrix with $f_j^* \equiv f^*(\omega_j)$, and the spectral precision matrix with $\thetajstar \equiv \Theta^*(\omega_j)$.

Before illustrating the theoretical findings, we state an assumption on $X_t$.

\begin{assumption}[First order summability]\label{asn:summable}
    $\{X_t\}_{t\in [n]}$ is a stable, Gaussian and centered time series with autocovariance function in \eqref{eq:autocovariance} satisfying
    \begin{equation}\label{eq:summable}
        \sum_{h=0}^\infty h \|\Gamma(h)\| < \infty.
    \end{equation}
\end{assumption}

Assumption \ref{asn:summable} is stated for Gaussian time series but can be extended to a broader class of linear processes (see Appendix \ref{sec:consistency_linear} for details). Assumption \ref{asn:summable} restricts the class of underlying time series to have short range dependence, and ensures existence and finiteness of spectral density across all frequencies. A similar assumption also appears in \citet[Assumption 3.1]{sun2018large}, with the sum of entry-wise absolute maximum of $\Gamma(\cdot)$ across all the lags being finite. Our refinement to the first order summability condition restricts the memory of the autocovariance to enjoy the following two properties of the underlying spectral density.

\paragraph*{Bounded spectrum} We define two stability measures as 
\begin{equation}\label{eq:stability_f}
    \vertiii{f^*} := \esssup_{\omega \in [-\pi, \pi]}\|f^*(\omega)\|, ~~~~ \vertiii{\Theta^*} := \esssup_{\omega\in [-\pi, \pi]}\|\Theta^*(\omega)\|.
\end{equation}
Stability measures similar to $\vertiii{f^*}$ and $\vertiii{\Theta^*}$ appear in \citet{basu2015regularized}. Under Assumption \ref{asn:summable}, the spectral density is finite \citep[Theorem 4.3.2]{brockwell1991time}. Hence in the definitions \eqref{eq:stability_f}, the suprema are attained as the maxima. We define a unified \textit{measure of stability} 
$$\M := \max\{\vertiii{f^*}, \vertiii{\Theta^*}\}$$

that captures the strength of the marginal and conditional temporal and cross sectional dependence among the time series components. Assumption \ref{asn:summable} also ensures frequency-wide bounded spectrum for $f^*(\cdot)$ and $\Theta^*(\cdot)$.

\paragraph*{Lipschitz continuity} For meaningful estimation of $\Theta^*(\omega_j)$, the DFTs cannot be allowed to be largely different across frequencies near $\omega_j$. Assumption \ref{asn:summable} restricts the variation of the DFTs across $k \in \wmj$ by implying \textit{Lipschitz continuity} on $f^*(\cdot)$ (Lemma \ref{lem:lip_smooth}) with Lipschitz constant $\lf/4\pi$, where
$$ \lf := 4\sum_{h=0}^\infty h \|\Gamma(h)\|. $$
The smoothness condition can be generalized to \textit{H\"{o}lder continuity} under a higher order summability condition without substantially changing the theoretical results to follow.

We now introduce the approximation bias, the sparsity parameters and the model parameters required for our analysis. Since $j \in F_n$ is fixed, we omit $j$ in the notation of these parameters for brevity.

\paragraph*{Approximation bias} We define a model dependent parameter
\begin{equation}\label{eq:tnf}
     \tnf := \frac{1}{2\pi} \sum_{|h|>n} \| \Gamma(h) \|,
\end{equation}
that captures the strength of the temporal and contemporaneous memory at tail of the autocovariance function \citep{sun2018large}. We define
\begin{equation}\label{eq:approximation_bias}
    \vmnf := \tnf + \frac{m}{n}\lf
\end{equation}
as the \textit{approximation bias} that appears in the error bound of the averaged periodogram $\hat f_j$ and subsequently error bounds and the sample size and bandwidth complexities of our theoretical analysis to follow. $\vmnf$ consists of two sources of biases: 
(a) $\tnf$ captures the \textit{truncation bias} arising from the use of finite-sample DFTs $d_k$, $k \in \wmj$, as data. It appears in the deviation of the DFT covariance matrix $\Snk$ (defined in \eqref{eq:Snk}), a truncated average of the autocovariances for finite samples, from the true spectral density $f_k^*$ which is an average of autocovariance functions across infinite lags (\eqref{eq:Snk_f_diff} in Lemma \ref{lem:dft_cov_approx}), and 
(b) $\frac{m}{n}\lf$ captures the \textit{smoothing bias} induced for calculating the averaged periodogram estimator $\hat f(\omega_j)$ with DFTs $d_k$ whose covariance matrices $\Snk$ vary across $k \in \wmj$. This variation arises due to the variation of the spectral densities $f_k^*$ across frequencies (Lemma \ref{lem:lip_smooth}). These two biases together control the bias of the averaged periodogram i.e. $\EE[\hat f_j] - f_j^*$ (Lemma~\ref{lem:single_deviation_bound}). For a large class of stationary processes, $\tnf \to 0$ with $n \to \infty$ and $\lf$ is asymptotically bounded above \citep[Proposition 3.4]{sun2018large}. Hence the approximation bias is bounded above as $\vmnf \precsim \frac{m}{n}$. In particular, if $m \asymp n^\xi$ for some $\xi \in (0,1)$, $\vmnf \precsim n^{-(1-\xi)}$.

\paragraph*{Sparsity parameters} The edge set of $\thetajstar$ is denoted by $E(\thetajstar) :=  \{(a,b)\in [p]^2: a\ne b, (\thetajstar)_{a,b} \neq 0\}$, and the augmented edge set by $S:= S(\thetajstar) = E(\thetajstar)\cup \{(1, 1),\cdots, (p,p)\}$. The two sparsity parameters required for our analysis are $s := |E(\thetajstar)|,$ i.e. the \textit{number of edges}, and $d := \max_{a \in [p]}\left|\{b \in [p]:\ (\thetajstar)_{a,b} \neq 0\}\right|$, i.e. the \textit{maximum degree}.

\paragraph*{Model parameters} We consider the function $g(\Theta) = \log \det \Theta,\ \Theta \succ0$. Following \citet{boyd2004convex}, the Hessian of $g$ is given by 
\begin{equation}\label{eq:hessian}
    \Upsilon^* := \nabla^2 g(\thetajstar) = {\thetajstar}^{-1}\otimes {\thetajstar}^{-1} = f_j^* \otimes f_j^* \in \CC^{p^2\times p^2},
\end{equation}
and is indexed with the edges such that for $e_1, e_2 \in [p]^2$, the entries $(\Upsilon^*)_{e_1, e_2}$ are of the form $\frac{\partial^2 g}{\partial \Theta_{e_1}\partial \Theta_{e_2}}(\Theta)$ evaluated at $\Theta = \thetajstar$. We define  $\kf := \vertiii{f_j^*}_\infty$ and $\kt := \vertiii{(\Upsilon^*)_{S,S}}_\infty$.

We now state an assumption to be used in our non-asymptotic analysis.

\begin{assumption}[Incoherence] \label{assumption:incoherence}
There exists $\alpha \in (0,1]$ such that 
$$\max_{e\in S\cmplt} \|\Upsilon^*_{e,S} \left(\Upsilon^*_{S,S} \right)^{-1} \|_1 \leq 1-\alpha.$$
\end{assumption}

Assumption \ref{assumption:incoherence} is standard in the literature of GLASSO, ensuring that the non-edge indices of the Hessian has limited influence on the edge indices \citep{zhao2006model, ravikumar2011high}. We additionally denote 
$C_\alpha := 1+ \frac{8}{\alpha}$ that is required in the results.

\subsection{Consistency of CGLASSO}\label{subsec:consistency}

We first state a concentration bound on the departure of the CGLASSO estimator $\hat \Theta_j$ from the true spectral precision matrix $\thetajstar$, and provide the results for edge recovery guarantee. For a constant $A>0$ we denote
\begin{equation}\label{eq:threshold}
\thres := \M\sqrt{\frac{A\log p}{2m+1}} + \vmnf,
\end{equation}
that controls the error bounds of $\hat\Theta_j$ in the result to follow. $\M\sqrt{\log p/m}$ captures the deviation bound associated with the CGLASSO estimator, and $\vmnf$ captures the approximation bias. These two terms are analogous to the variance and the bias for estimating the spectral density $f_j^*$ with the averaged periodogram estimator $\hat f_j$ (See Lemma \ref{lem:single_deviation_bound}).

\begin{theorem}[Consistency of CGLASSO for Gaussian time series]\label{thm:consistency_cglasso}
Let $\{X_t\}_{t\in[n]}$ be a stationary time series satisfying Assumption \ref{asn:summable} and \ref{assumption:incoherence}. Assume that $\kf, \kt \ge 1$, the sample size $n$ satisfies $n \succsim \lf \M^3 \log p$ and $\tnf \le 1/(4\M)$, and $C_\alpha := 1+ \frac{8}{\alpha}$. Then for any $A>0$, $\thres$ as in \eqref{eq:threshold} satisfying $d \thres \leq [6 \kt^2 \kf^3 C_\alpha]^{-1}$, the choice of penalty $\lambda = \frac{8}{\alpha}\thres$ and the bandwidth satisfying $m \succsim \M^2 \log p$ and $ m \le n/(4\M \lf)$, there exist constants $c, c'>0$ such that $\hat\Theta_j$ in \eqref{eq:graphical-lasso} satisfies the following with probability greater than $1 - c \exp(-(c'A - 2)\log p)$:
\begin{enumerate}
\item[(a)] $\big\|\hat\Theta_j - \thetajstar \big\|_\infty \le 2\kt C_\alpha \thres$,
\item[(b)] $E(\hat\Theta_j) \subseteq E(\thetajstar)$, and includes all edges $(a,b)$ with $|(\thetajstar)_{a,b}| > 2\kt C_\alpha \thres$,
\item[(c)] The estimator $\hat\Theta_j$ satisfies
\begin{align*}
&\|\hat \Theta_j - \thetajstar \|_F  \le 2\kt C_{\alpha}\sqrt{s+p}\ \thres, \\
& \|\hat \Theta_j - \thetajstar \| \le 2\kt C_{\alpha}\min\{\sqrt{s+p}, d + 1\}\ \thres. 
\end{align*}
\end{enumerate}
\end{theorem}

Theorem \ref{thm:consistency_cglasso} derives the error bounds for $\hat\Theta_j$ in maximum absolute norm, Fr\"{o}benius norm and operator norm, and edge recovery consistency in terms of $\thres$. If $\M$ is asymptotically bounded away from zero and infinity, the error bounds have $\sqrt{\log p/m}$ deviation bound that is the same as the estimation error bound for GLASSO with $m$ i.i.d. real-valued samples \citep[Corollary 1 and 3]{ravikumar2011high} -- with an additional $\vmnf$ capturing the approximation bias. Theorem \ref{thm:consistency_cglasso} additionally states that entry-wise model selection consistency is achieved when the true entry is larger than a signal depending on $\thres$.

\begin{remark}\label{rem:main}
The condition $d \thres  \le [6\kt^2 \kf^3 C_\alpha]^{-1}$ ensures that both the deviation bound $\M\sqrt{\log p/m}$ and the approximation bias $\vmnf$ are controlled by $[\kt^2 \kf^3 C_\alpha d]^{-1}$. If $X_t$ is a Gaussian time series satisfying standard mixing conditions \citep[Prop 3.4]{sun2018large}, and $m \asymp n^{\xi}$ for $\xi \in (0, 1)$, and $\M, \kf$, $\kt$, $\alpha$ are asymptotically bounded away from zero and infinity, the deviation asymptotically dominates the approximation bias and we obtain $\thres \asymp \sqrt{\log p/m} \precsim 1/d$, i.e. $m \succsim d^2\log p$. The bandwidth complexity in Theorem \ref{thm:consistency_cglasso} becomes the same as the order of sample size for consistently recovering the GLASSO estimator with i.i.d. sub-Gaussian samples \citep[Corollary 1]{ravikumar2011high}. Furthermore, consistency of $\hat\Theta_j$ around $\thetajstar$ is ensured in the regime $d^2 \log p/m \rightarrow 0$.
\end{remark}

\paragraph*{Proof sketch} The proof of Theorem \ref{thm:consistency_cglasso} is deferred to Appendix \ref{pf:consistency_cglasso} and uses a primal-dual witness method (Section \ref{subsubsec:primal_dual_witness}). We extend the proof of consistency for the GLASSO estimator for real-valued i.i.d. data \citep[Theorem 1]{ravikumar2011high} to CGLASSO with averaged periodogram as the input that consists of non-i.i.d. DFTs as observations. Under a carefully chosen single deviation bound on $\|\hat f_j - f_j^*\|_{\infty}$, the primal witness $\tilde \Theta_j$ concentrates around $\thetajstar$ (Lemma \ref{lem:dual_witness_bound}) and the remainder function of the Taylor series of $\Theta \mapsto \Theta^{-1}$ at $\Theta = \thetajstar$, i.e. $R(G) = (\thetajstar + G)^{-1} - {\thetajstar}^{-1}$ is bounded at $G = \tilde\Theta_j - \thetajstar$ (Lemma \ref{lem:remainder_bound}). The single deviation term and the remainder being bounded, the \textit{strict duality condition} of CGLASSO holds i.e. $\hat\Theta_j = \tilde\Theta_j$ (Lemma \ref{lem:strict_duality}), and hence $\hat\Theta_j$ concentrates around $\thetajstar$. The proof of Lemma \ref{lem:dual_witness_bound} involves application of \textit{Brouwers' fixed point theorem} for complex-matrix valued functions, and the proof of Lemma \ref{pf:remainder_bound} involves power series expansion of complex-valued operators.

\subsection{Consistency of CAGLASSO} \label{subsec:caglasso_consistency}

We now establish non-asymptotic error bound on the CAGLASSO estimator $\thetajhattau$ in \eqref{eq:caglasso}. We derive error bounds for the NWR estimated partial variances $\hat\tau_a^2,\ a \in [p]$, and combine with the error bound of the CGLASSO estimator (Theorem \ref{thm:consistency_cglasso}) to obtain the results.

As discussed in Section \ref{subsec:penalized_whittle_liklihood}, $d_k$ has an asymptotic zero-mean complex normal distribution with covariance matrix $f_k^*$. In finite samples, the entries of the covariance matrix of $d_k$, denoted by $\Snk$ in \eqref{eq:Snk}, differ from their true spectral density $f_k^*$ due to presence of truncation bias. In population, a linear projection of one coordinate of $d_k$ on its other coordinates result in NWR coefficients with entries $ -(\Snk^{-1})_{a,b} / (\Snk^{-1})_{a,a}$ -- implying that the NWR coefficients are not necessarily sparse even if $\Theta_k^*$ is sparse. Additionally, $\Snk$ also varies across $k \in \wmj$ -- resulting to a variation of the NWR coefficients across nearby frequencies. Hence the NWR consistency results derived for real-valued i.i.d. data with a single sparse regression coefficient \citep{van2014asymptotically} are not directly applicable for bounding the estimation error of the partial variances $\tauahatsq$ obtained from NWR of the DFTs. To address this issue, we leverage the theory of lasso under an \textit{approximate oracle} that is sparse and linearly approximates the population prediction means of NWR -- with the consistency being attainable up to an oracle approximation error \citep{buhlmann2011statistics}. Under Assumption~\ref{asn:summable}, we explicitly bound this oracle approximation error in terms of the approximation bias $\vmnf$ (Proposition \ref{prop:nwr_approx_oracle}), which in turn allows us to establish estimation consistency and prediction consistency of our proposed NWR estimator in \eqref{eq:nodewise-reg}, and error bounds of $\tauahatsq$ (Theorem \ref{thm:consistency_dft_nwr} and Corollary \ref{cor:consistency_partial_variance} respectively). These results form the basis for deriving the non-asymptotic error bounds of our proposed estimator $\thetajhattau$. We note that \citet{krampe2025frequency} address the connection between partial coherency and the DFT-NWR coefficients in their inference framework by restricting the difference between the average of the covariance matrices $\Snk$ and $f_j^*$ to $o(\log n)$ \citep[Assumption 3]{krampe2025frequency}. Our results, on the other hand, only imposes the summability condition in Assumption \ref{asn:summable} and derives the error bounds explicitly in terms of $\vmnf$.

\subsubsection{NWR of DFT in population}\label{subsubsec:nwr_pop}

We set up the NWR of DFT in population that will be required for deriving our results. Since the DFTs calculated in finite sample setting admits approximation bias, we invoke the population setting of DFT-NWR with least squares regression coefficients that varies across frequencies and are not necessarily sparse.

The covariance matrix of $d_k$ for $k \in F_n$ and its inverse are respectively
\begin{align}
    \Snk :=  \EE[d_k (d_k)^\dagger] = & \EE[I(\omega_k)] = \frac{1}{2\pi} \sum_{|h| < n}\left(1 - \frac{|h|}{n}\right) \Gamma(h) e^{-\i h \omega_k}, \label{eq:Snk}\\
    \Onk := & (\Snk)^{-1} \label{eq:Onk}.
\end{align}
For $a \in [p]$, we use the shorthand $d_k^{-(a)} \in \CC^{p-1}$ to denote the vector $d_k$ without the $a$\textsuperscript{th} entry. If $\dka$ is regressed on on $\dkma$, the population NWR coefficient obtained from least squares is
$$ \bka := \argmin_{\beta \in \CC^{p-1}} \frac{1}{2m+1} \EE\big|\dka - \beta^\dagger \dkma\big|^2, $$
where $\bka = \left((\bka)_1, \ldots, (\bka)_{a-1}, (\bka)_{a+1}, \ldots, (\bka)_{p} \right)^\top$ with entries
\begin{equation}\label{eq:nwr_true_beta}
    \big(\bka\big)_b := - \frac{(\Onk)_{a,b}}{(\Onk)_{a,a}},\quad \text{for } b \ne a.
\end{equation}
The NWR residuals $\eka, k \in \wmj$ are $\eka := \overline{\dka} - \big(\dkma\big)^\dagger \bka$. Equivalently,
\begin{equation}\label{eq:nwr_model}
    \overline{\dka} = \big(\dkma\big)^\dagger \bka + \eka,
\end{equation}
where 
\begin{equation}\label{eq:exp_varnwr_err}
    \EE\big[\eka \big| \dka\big] = 0,\quad \EE\big[\big|\eka\big|^2 \big] = \frac{1}{(\Onk)_{a,a}}.
\end{equation}
We define the vector of the NWR means as
\begin{equation}\label{eq:muja}
    \muja := \left({d_{j-m}^{-(a)}}^\dagger \beta_1^{(a)}, \ldots, {d_{j}^{-(a)}}^\dagger \beta_j^{(a)}, \ldots, {d_{j+m}^{-(a)}}^\dagger \beta_{j+m}^{(a)}\right)^\top,
\end{equation}
and $\ea := \big(\varepsilon_{j - m}^{(a)}, \ldots, \varepsilon_{j}^{(a)}, \ldots, \varepsilon_{j + m}^{(a)}\big)^\top$. The entries \eqref{eq:nwr_model} can be expressed in a vector notation as
\begin{equation}\label{eq:nwr_model_mat}
    \Za = \muja + \ea,
\end{equation}
with $\EE\big[ \ea \big| \muja \big] = 0$.

\subsubsection{Oracle of DFT-NWR}\label{subsubsec:oracle_nwr}

We note that the NWR mean vector $\muja$ in \eqref{eq:muja} is not necessarily a linear combination of the columns of $\Zma$ due to variation of $d_k$ and $\bka$ over frequencies near $\omega_j$. However, it follows from \eqref{eq:nwr_model} and \eqref{eq:exp_varnwr_err} that $\EE[\dka | \dkma] = \Zma \bka$, i.e. each coordinate of $\EE[\dka | \dkma]$ is linear in $\dkma$. If for $ k \in \wmj$, $\dkma$ and $\bka$ are ``close'' to each other, we can approximate $\muja$ in \eqref{eq:nwr_model_mat} by a linear combination of the columns of $\Zma$. We refer to this approximating regression coefficient as the \textit{NWR oracle} that we formally define in \eqref{eq:oracle_beta}, and conduct the analysis of DFT-NWR with this oracle. Consequently, we establish prediction consistency results for DFT-NWR consistency results with respect to this oracle \citep[Section 6.2.3]{buhlmann2011statistics} -- that further enables us to derive error bounds for $\tauahatsq$ in CAGLASSO.

For any $\beta \in \CC^{p-1}$,
\begin{align}
    \frac{\|\Zma \bjahat - \muja\|_2^2}{2m+1} + \lambda_a\|\bjahat\|_1 \le & \frac{2}{m+1} {\ea}^\dagger \Zma (\bjahat - \beta) + \lambda_a\|\beta\|_1 \label{eq:basic_ineq_plus}\\
    & \hspace{10pt} + \frac{\|\Zma \beta - \muja\|_2^2}{2m+1}. \nonumber
\end{align}
The inequality \eqref{eq:basic_ineq_plus} refers to the \textit{basic inequality} of lasso with an additional term $\|\Zma \beta - \muja\|_2^2/(2m+1)$ \citep[Equation 6.6]{buhlmann2011statistics} (See Section \ref{subsec:basic_ineq} for the proof). We denote the support and the number of non-zero entries of the $a$\textsuperscript{th} column of $\thetajstar$ as
\begin{align*}
    S_a := \{b\in [p]\setminus\{a\}: (\thetajstar)_{a,b} \ne 0 \}, \text{ and }
    s_a := |S_a|
\end{align*}
respectively. The DFT-NWR oracle is defined as the minimizer of the additional term in \eqref{eq:basic_ineq_plus} under the support restricted to $S_a$, i.e.
\begin{equation}\label{eq:oracle_beta}
    \bjastar := \argmin_{\beta\in \CC^{p-1}: \beta_{S_a\cmplt} = 0} \frac{\big\| \Zma \beta - \muja \big\|_2^2}{2m+1}.
\end{equation}
In words, $\bjastar$ is a least squares estimator of regressing $\muja$ on $\Zma$ where the support of the solution is restricted to $S_a$.  We refer to the minimum value of the additional error term at $\beta = \bjastar$ as the \textit{oracle approximation error}. We note that $\beta$ on the right side of \eqref{eq:basic_ineq_plus} is arbitrary. The choice of the oracle $\bjastar$ in \eqref{eq:oracle_beta} is not unique to the problem, and solely depends on the choice of the restricting support set \citep[Equation 6.7]{buhlmann2011statistics}. We choose the oracle to match its support with $S_a$ which is the support of $(\thetajstar)_{\cdot a}$, and thereby enforce sparsity for bounding the $\frac{2}{m+1} {\ea}^\dagger \Zma (\bjahat - \beta) + \lambda_a\|\beta\|_1 $ on the right side of the basic inequality \eqref{eq:basic_ineq_plus}. Enforcing sparsity on $\bjastar$ enables us to derive the theoretical results for lasso by carrying the oracle approximation error. We now establish in the next proposition that the oracle approximation error is bounded by a function of the approximation bias $\vmnf$ with high probability for large enough sample size and bandwidth.

\begin{proposition}[Oracle approximation error]\label{prop:nwr_approx_oracle}
    Consider the DFT-NWR setup \eqref{eq:nwr_model} satisfying Assumption \ref{asn:summable}. Assume that $\tnf \le 1/(4\M)$. Then for $a \in [p]$, $A \ge 1$ and the bandwidth $m \le n / (4\M \lf)$, there exists a constant $c > 0$ such that with probability greater than $ 1 -  4 \exp(-c(2m+1)A) $,
    \begin{equation}\label{eq:nwr_approx_oracle}
        \frac{\big\| \Zma \bjastar - \muja \big\|_2^2}{2m+1} \le 18(2A + 3)\M^{11} \vmnf^2.
    \end{equation}
\end{proposition}

The oracle approximation error bound in \eqref{eq:nwr_approx_oracle} has two components -- (a) $\M^{11}$ captures the effect of temporal dependence, and (b) $\vmnf^2$ captures the effect of approximation bias for using finite sample DFTs with variation across nearby frequencies. The upper bound on bandwidth, i.e. $m\le n/(4\M\lf)$ implies that the truncation bias is bounded as $\vmnf \le 1/(2\M)$, which enables us to apply the matrix perturbation bounds in Lemma \ref{lem:spectral_difference_inv}.

\paragraph*{Proof sketch} The detailed proof of Proposition \ref{prop:nwr_approx_oracle} is deferred to Appendix \ref{pf:nwr_approx_oracle}. We denote $\bjatilde$ as the $(p-1)$-dimensional vector in \eqref{eq:bja_tilde}, with entries of the form $(\bjatilde)_b = -(\thetajstar)_{b, a} / (\thetajstar)_{a, a}$, $b \neq a$ and the support being $S_a$. By the definition of $\bjastar$ in \eqref{eq:oracle_beta}, the oracle approximation error is bounded above by 
$\frac{1}{2m+1}\sum_{k \in \wmj} |(\dkma)^\dagger (\bjatilde - \bka)|^2$, that is an average of quadratic forms of $\dkma$ for $k \in \wmj$. We first establish a high probability upper bound $\|\bjatilde - \bka\|_2 \le 6\M^5 \vmnf$ (Lemma \ref{lem:onk_theta_approx}) using a perturbation bound for inverted matrices (\eqref{eq:spectral_difference_inv4} in Lemma \ref{lem:spectral_difference_inv}). We then use a concentration bound on average of quadratic forms of $d_k$ (Lemma \ref{lem:dk_quad_concentration}) that completes the proof.

\begin{remark}[Oracle approximation error for white noise process]
If $\{X_t\}_{t\in [n]}$ is a stationary white noise process with $\Gamma(0) = \Sigma$ and $\Gamma(h) = 0$ for $h \neq 0$, then $\Onk = \thetakstar = 2\pi\Sigma^{-1}$. Hence $\muja = \Zma \bjastar$ with $\big(\bjastar\big)_b = - \frac{(2\pi\Sigma^{-1})_{a,b}}{(2\pi\Sigma^{-1})_{a,a}}$ for $ b \ne a$, i.e. $\bjastar$ exactly recovers the DFT-NWR mean vector $\muja$ with the oracle approximation error being zero. Additionally, $\tnf = 0$ and $\lf = 0$ -- resulting to the left side of \eqref{eq:nwr_approx_oracle} being zero as well as $\vmnf = 0$. Hence the bound \eqref{eq:nwr_approx_oracle} becomes tight.
\end{remark}

\subsubsection{Error bounds for the DFT-NWR estimator}\label{subsubsec:consistency_dft_nwr}

Equipped with the necessary precursory results, we now establish the error bounds for the estimated regression coefficient $\bjahat$ in $\ell_1$-norm. Additionally, we derive upper bound on the prediction error $\|\Zma \bjahat - \muja\|_2^2/(2m+1)$ and error bound on the estimated partial variance $\tauahatsq$. We define two additional terms that appear in the error bounds as follows
\begin{equation}\label{eq:threshold_nwr}
   \threse := \sqrt{\M}\lambda_a,\quad \thresa := \M^{11/2} \vmnf. 
\end{equation}
$\threse$ captures the deviation bound of $\bjahat$ around $\bjastar$ in the regression \eqref{eq:nodewise-reg} and $\thresa$ captures the approximation error with the oracle in \eqref{eq:oracle_beta}. We note that ${\thresa}^2$ appears in the error bounds for the oracle approximation in Proposition \ref{prop:nwr_approx_oracle}.

\begin{theorem}[Consistency of DFT-NWR]\label{thm:consistency_dft_nwr}
    Consider the DFT-NWR setup \eqref{eq:nwr_model} satisfying Assumption \ref{asn:summable}. Assume $a \in [p]$ and the sample size satisfies  
    $n \succsim \M^5 \lf s_a\log p$ and $\tnf \le 1/(4\M)$. Then for $A \ge 1$, $\lambda_a \ge 4\M^2 \sqrt{\frac{3A\log p}{2m+1}}$, and the bandwidth $m \succsim \M^4 s_a \log p$ and $m \le n/(4\M\lf)$, there exist constants $c_i >0 $ such that the solution $\bjahat$ of the NWR problem \eqref{eq:nodewise-reg} satisfies the following with probability greater than $1 - c_0 \exp(-(c_1A-1) \log p)$,
    \begin{align}
        \|\bjahat - \bjastar\|_1 \le & ~48 \sqrt{\M}s_a \threse + 270 \frac{(\thresa)^2}{\lambda_a}, \label{eq:estimation_consistency_nwr}\\
        \frac{\|\Zma \bjahat - \muja\|_2^2}{2m+1} \le &~ 48s_a{\threse}^2 + 270(\thresa)^2. \label{eq:prediction_consistency_nwr}
    \end{align}
\end{theorem}

Theorem \ref{thm:consistency_dft_nwr} provides concentration bound \eqref{eq:estimation_consistency_nwr} of the NWR coefficient $\bjahat$ around the oracle $\bjastar$. In contrast to NWR with i.i.d. samples where the estimated regression coefficients, scaled with the partial variances, concentrate around the respective columns of the precision matrix \citep[Theorem 2.4]{van2014asymptotically}. Here the estimated regression coefficients concentrates around the $s_a$-sparse oracle $\bjastar$. Theorem \ref{thm:consistency_dft_nwr} also provides an upper bound on the DFT-NWR prediction error in \eqref{eq:prediction_consistency_nwr}.

\begin{remark}
\begin{itemize}
\item[(a)] The error bound in \eqref{eq:estimation_consistency_nwr} has two components- (i) $s_a \threse$ is of the same order as the rate of convergence $\sqrt{s_a \log p /m}$ for regression of $m$ i.i.d. samples with $s_a$-sparse regression coefficients \citep[Theorem 7.13]{wainwright2019high}, (ii) $(\thresa)^2 / \lambda$ captures the estimation error due to the oracle approximation.
\item[(b)] Under the conditions of Theorem \ref{thm:consistency_dft_nwr}, \textit{restricted eigenvalue condition} can be shown to hold with high probability (Proposition \ref{prop:re} in Appendix). Hence the prediction error of DFT-NWR has a bound of the form $s_a \log p/m$ that is the same as the fast convergence rate $s\log p / n$ of prediction error in lasso with i.i.d. samples \citep[Theorem 7.20]{wainwright2019high}. 
\item[(c)] The bandwidth lower bound $m \succsim \M^4 s_a \log p$ is similar to the sample size requirement for regression with i.i.d. samples \citep{wainwright2019high} with an additional $\M^4$ that captures the effect of temporal dependence. In the estimation error \eqref{eq:estimation_consistency_nwr}, if $\M$ and $\lf$ are asymptotically bounded away from zero and infinity, and $\tnf\to 0$, we obtain the upper bound on the estimation error $\bjahat$ to be $s_a \sqrt{\log p/m}$ that is the same as the rate of convergence in $\ell_1$-norm for regression with $m$ i.i.d. samples and $s_a$-sparse regression coefficient if the allowable upper bound on the bandwidth is $m \precsim (n^2 s_a \log p)^{1/3}$.
\end{itemize}
\end{remark}

\paragraph*{Proof sketch} The proof of Theorem \ref{thm:consistency_dft_nwr} is deferred to Appendix \ref{pf:consistency_dft_nwr}. Using the basic inequality \eqref{eq:basic_ineq_plus}, we first bound the sum of the prediction error and $\lambda_a \times $ the estimation error. Under Assumption \ref{asn:summable} and the conditions of Theorem \ref{thm:consistency_dft_nwr}, the restricted eigenvalue (RE) condition and the deviation error bound hold with high probability (Proposition \ref{prop:re} and \ref{prop:deviation} respectively). Therefore sum of the prediction error and $\lambda_a \times $ the estimation error can be bounded for $\beta = \bjastar$ using the proof techniques of lasso under approximately linear truth \citep[Theorem 6.2]{buhlmann2011statistics}. The key highlight of the proof is to apply the oracle approximation error bound in Proposition \ref{prop:nwr_approx_oracle} and explicitly track it in the error of DFT-NWR.

We now derive in the following corollary that the estimated residual variance in \eqref{eq:nwr_partial_variance} are consistent to the reciprocals of the diagonals of spectral precision matrix.

\begin{corollary}[Consistency of DFT-NWR residual variance]\label{cor:consistency_partial_variance}
    Consider a DFT-NWR setup satisfying the assumptions of Theorem \ref{thm:consistency_dft_nwr}. Then for the choice of $\lambda_a$ and $m$ in Theorem \ref{thm:consistency_dft_nwr}, there exist constants $c_i > 0$ such that $\tauahatsq$ in \eqref{eq:nwr_partial_variance} satisfies the following with probability greater than $1 - c_0 \exp(-(c_1 A - 1) \log p)$,
    \begin{equation}\label{eq:consistency_partial_variance}
        \left|\tauahatsq - \frac{1}{(\thetajstar)_{a,a}} \right| \le c_2\sqrt{\M}\left(\sqrt{s_a}\threse + \thresa \right).
    \end{equation}
\end{corollary}

The proof is deferred to Appendix \ref{pf:consistency_partial_variance} and uses the prediction consistency result \eqref{eq:prediction_consistency_nwr} in Theorem \ref{thm:consistency_dft_nwr}. In Gaussian graphical models, the diagonals of the precision matrix capture the partial variances of NWR. In DFT-NWR setting, the approximation bias $\vmnf$ controls the departure between the diagonals of $\Onk$ i.e. the population partial variance of DFT-NWR and the diagonals of $\thetajstar$. Corollary \ref{cor:consistency_partial_variance} establishes that under appropriate regime and choices of tuning parameters in DFT-NWR setup, the estimated partial variances concentrate around the reciprocals of diagonals of $\thetajstar$.

\begin{remark}\label{rem:consistency_partial_variance}
For the choices of penalty and bandwidth as in Theorem \ref{thm:consistency_dft_nwr}, the error bound of the variance estimator has two components -- (a) $\M\sqrt{s_a \log p/m}$ is the same as the rate of partial variance estimator in nodewise regression of i.i.d. samples \citep[Theorem 2.4]{van2014asymptotically}, and (b) $\M^6 \vmnf$ controlling the effect of the oracle approximation error on partial variance estimation.
\end{remark}

\subsubsection{Error bounds of the CAGLASSO estimator}
Our next result derives non-asymptotic error bounds on the adaptive estimator $\thetajhattau$ in \eqref{eq:caglasso}. In \eqref{eq:caglasso}, CAGLASSO sets the penalty for any entry $(a,b)\in E(\Theta)$ proportional to $\hat\tau_a \hat \tau_b$. For suitably chosen $\lambda_a$ and $m$, Corollary \ref{cor:consistency_partial_variance} establishes error bounds on the NWR estimated partial variance $\tauahatsq$ around the true partial variance $1/(\thetajstar)_{a,a}$. We combine this result with the results for the CGLASSO estimator in Section \ref{subsec:consistency}, and derive the error bounds for $\hat\Theta_j^{(\tau)}$.

The oracle scaled spectral precision is denoted by $K_\tau := D_\tau \thetajstar D_\tau$, with $D_\tau^2 := \dg(1/(\thetajstar)_{1,1},\ldots, 1/(\thetajstar)_{p,p})$ being the scaling matrix. Additionally, we define $R_\tau := D_\tau^{-1} f_j^* D_\tau^{-1}$. We further define 
\begin{equation}\label{eq:hetero}
    r_S := \left( \frac{\max_{(a, b) \in S\cmplt}(\thetajstar)_{a, a} (\thetajstar)_{b, b}}{\min_{(a, b) \in S}(\thetajstar)_{a, a} (\thetajstar)_{b, b}} \right)^{1/2}
\end{equation}
as the \textit{degree of heterogeneity}. In words, $r_S$ captures the heterogeneity of scales in the incoherence condition of the scaled precision matrix $K_\tau$, and is upper bounded by the condition number of $\dg((\thetajstar)_{1,1},\ldots, (\thetajstar)_{p,p})$. Analogous to $\alpha$, we denote 
\begin{equation}\label{eq:alpha_prime}
    \alpha' := 1 - r_S(1 - \alpha)
\end{equation}
as the incoherence constant for $K_\tau$. Similar to $\threse$ in \eqref{eq:threshold_nwr}, we define
$$ \threseall := \sqrt{\M}\ \max_{a \in [p]}\lambda_a. $$
Analogous to $\thres$,
\begin{equation}\label{eq:thresr}
    \thresr := \sqrt{d} \threseall + \thresa
\end{equation}
controls the error bounds on the scaled averaged periodogram $\hat R_\tau$ (Lemma \ref{lem:scaled_avg_periodogram}) and consequently appears in the error bounds of $\thetajhattau$.

\begin{theorem}[Estimation consistency of CAGLASSO] \label{thm:caglasso_consistency}
Let $\{X_t\}_{t\in[n]}$ be a stationary time series satisfying Assumption \ref{asn:summable} and \ref{assumption:incoherence} with $\alpha \in (1 - r_S^{-1}, 1)$. Assume that  $\kf, \kt \ge 1/\M^2$, and the sample size satisfies $n \succsim \M^5\lf d\log p$ and $\tnf \le 1/(4\M)$. Then for $A \ge 1$, $\thresr$ in \eqref{eq:thresr} satisfying $\thresr \le 1/(2\M^{3/2})$ and $d \thresr \leq [6 \kt^2 \kf^3 \M^7 C_\alpha']^{-1}$, NWR and CAGLASSO penalties $\lambda_a \ge 4\M^2 \sqrt{\frac{3A\log p}{2m+1}}$ for $a\in [p]$ and $\lambda = \frac{32}{\alpha'}\M^{7/2}\thresr$ respectively, and the bandwidth  $m \succsim \M^4 d \log p$ and $ m\le n/(4\M\lf)$, there exist constants $c_i >0 $ such that the solution $\thetajhattau$ of the CAGLASSO problem \eqref{eq:caglasso} satisfies the following with probability greater than $1 - c_0 \exp(-(c_1 A - 2) \log p)$,

\begin{enumerate}
\item[(a)] $\big\|\thetajhattau - \thetajstar \big\|_\infty \le 10\kt \M^{13/2} C_\alpha' \thresr $,
\item[(b)] $E(\thetajhattau) \subset E(\thetajstar)$, and includes all edges $(a,b)$ with $|(\thetajstar)_{a,b}| > 10\kt \M^{13/2} C_\alpha' \thresr$,
\item[(c)] The estimator $\hat\Theta_j^{(\tau)}$ satisfies
\begin{align*}
&\|\thetajhattau - \thetajstar \|_F  \le 10\kt \M^{13/2} C_\alpha' \sqrt{s+p} \thresr, \\
& \|\thetajhattau - \thetajstar \| \le 10\kt \M^{13/2} C_\alpha' \min\{\sqrt{s+p}, d+1\}\thresr. 
\end{align*}
\end{enumerate}
\end{theorem}

The error bounds for the CAGLASSO estimator $\thetajhattau$ shares similarity with that of the CGLASSO estimator in Theorem \ref{thm:consistency_cglasso}. The threshold $\thresr$ is restricted as $d\thresr\le  [6 \kt^2 \kf^3 \M^7 C_\alpha']^{-1}$, ensuring controls on the deviation bound $\sqrt{d}\threseall$ and the approximation bias $\thresa$. In \eqref{eq:alpha_prime}, $\alpha'$ decreases as the degree of heterogeneity $r_S$ increases -- hence the error bound containing $C_\alpha'$ increases with higher degree of heterogeneity.

\begin{remark}\label{rem:caglasso_consistency}
If $X_t$ is a Gaussian time series satisfying standard mixing conditions 
\citep[Proposition 3.4]{sun2018large}, with 
$m \asymp n^{\xi}$ for some $\xi \in (0,1)$, 
and $\kf$, $\kt$, $\alpha$, $r_S$, and $\M$ asymptotically bounded above, 
then the error bound of the CAGLASSO estimator in entry-wise maximum absolute norm is
$\thresr \asymp \sqrt{d \log p / m}$, which is slower than the rate $\sqrt{\log p / m}$ in Theorem~\ref{thm:consistency_cglasso}. Consequently, the condition $\thresr \precsim 1/d$ implies that the bandwidth required for consistency of CAGLASSO is $m \succsim d^3 \log p$, which is more stringent than the bandwidth complexity $m \succsim d^2 \log p$ for CGLASSO. The difference is an artifact of our proof techniques.
\end{remark}

\paragraph*{Proof sketch} The proof of Theorem \ref{thm:caglasso_consistency} is deferred to Appendix \ref{pf:caglasso_consistency}. We apply Corollary \ref{cor:consistency_partial_variance} and Lemma \ref{lem:single_deviation_bound} to show that the scaled averaged periodogram $\hat R_\tau$ concentrates around $R_\tau$ with high probability (Lemma \ref{lem:scaled_avg_periodogram}). Additionally, the incoherence condition holds for the scaled spectral precision matrix $K_\tau$. Hence the proof of Theorem \ref{thm:consistency_cglasso} is carried out for the scaled spectral precision matrix and the corresponding error bounds for $\hat\Theta_j^{\tau}$ are subsequently derived. The key highlight of the proof is to adapt the consistency framework of CGLASSO, including the incoherence condition and parameter representations, to CAGLASSO, with a careful analysis of the oracle approximation error and the estimation error.

\paragraph*{Extension to general linear process} In Appendix \ref{sec:consistency_linear}, our theoretical results for CGLASSO and CAGLASSO stated with Gaussian time series are generalized to observations from a linear process with errors having possibly heavier tails. With heavier tail of the error distributions, we tailor the thresholds $\thres$ in Theorem \ref{thm:consistency_cglasso}, $\thresa$ and $\threse$ in Theorem \ref{thm:consistency_dft_nwr} and $\thresr$ in Theorem \ref{thm:caglasso_consistency} according to the deviation bounds that increases the sample size and the bandwidth complexity for ensuring consistency. The error bounds and sample complexity depends on the threshold $\thres$ controlled by the tail behavior.
\section{Simulation study}\label{sec:simulation}

We compare the performance of CGLASSO and CAGLASSO estimators with benchmark methods through four experiments: (a) runtime comparison with CD, (b) Relative mean squared error (RMSE) comparison with joint penalty on real and imaginary components, (c) model selection comparison with Area under ROC curve (AUROC), and (d) RMSE comparison under adaptive regularization. The CLASSO and CGLASSO algorithms are implemented from our repository: \url{https://github.com/yk748/cxreg}.  Simulations use bandwidth $m = \lfloor\sqrt{n}\rfloor$ \citep{bohm2009shrinkage}. All experiments are conducted on Cornell University’s BioHPC high-performance computing platform for getting robust Monte Carlo results across 50 trials.

\subsection{CD improves runtime over ADMM}\label{subsec:runtime_improvement}

\begin{figure}[!t]
    \centering
    \subfloat{\includegraphics[width=0.3\linewidth]{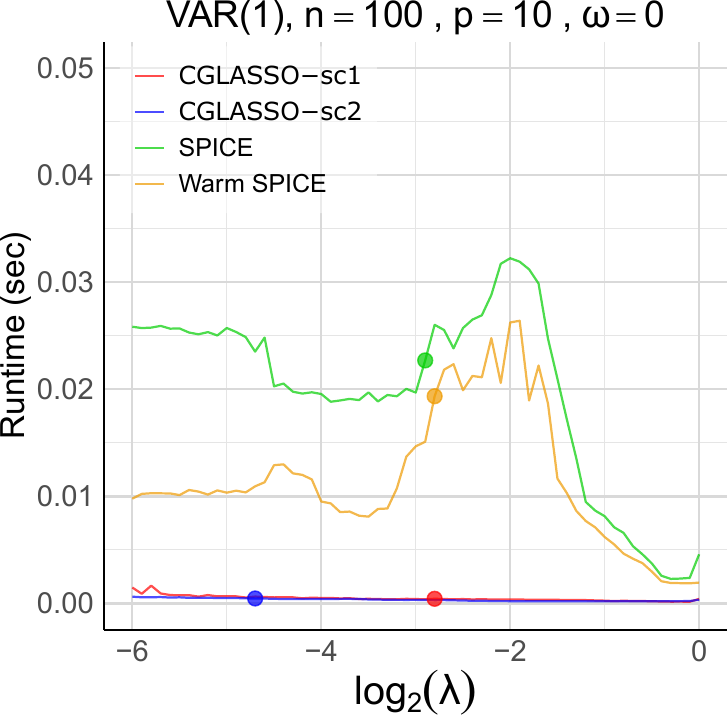}}
    \subfloat{\includegraphics[width=0.3\linewidth]{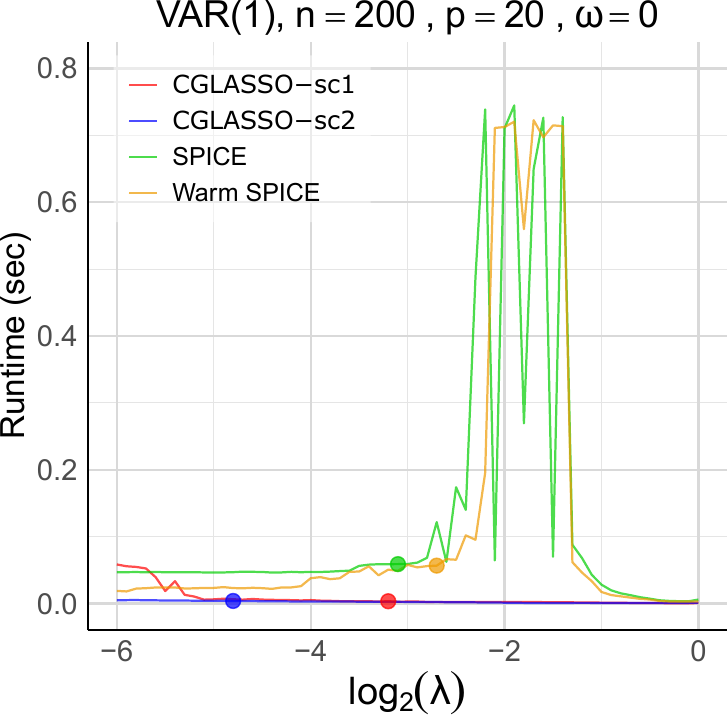}}
    \subfloat{\includegraphics[width=0.3\linewidth]{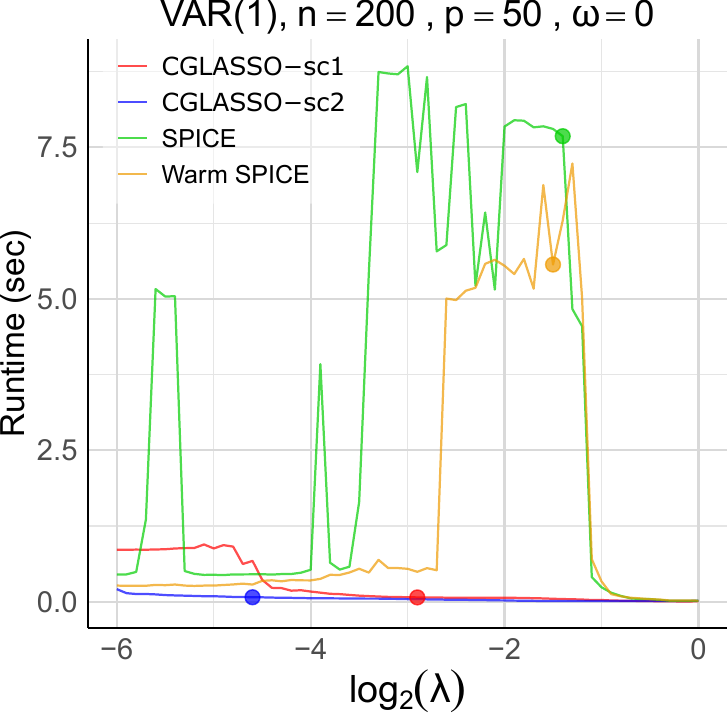}}
    \caption{\textbf{Runtime (median over 50 trials) of CGLASSO-sc1 (red) and 2 (blue) with ADMM-based SPICE (green) and Warm-SPICE (orange).} The underlying DGP is VAR(1) with $n \in \{100, 200\}$, $ p \in \{10, 20, 50\}$ and $\omega = 0$. The dots on the regularization paths indicate the runtime corresponding to $\lambda$ that minimizes the RMSE. CGLASSO-sc1 (red curve) and 2 (blue curve) are faster than SPICE algorithms (green and yellow curves) for various combinations of $n$ and $p$.}
    \label{fig:runtime_final}
\end{figure}

\begin{table}[!t]
\centering
\caption{Runtime comparison (milliseconds; mean and standard deviation) between CGLASSO variants and ADMM-based solvers across different $(n, p)$ settings under a VAR(1) process. Lower values (in bold) means faster computation.}
\label{tab:runtime_results}
\small
\setlength{\tabcolsep}{6pt}
\begin{tabular}{lcccc}
\toprule
$(n, p)$ & CGLASSO-sc1 & CGLASSO-sc2 & SPICE & Warm-SPICE \\
\midrule
(100, 10)   & \textbf{0.37} (0.08)   & 0.45 (0.10)   & 19.22 (6.62)   & 15.39 (6.75) \\
(200, 20)   & \textbf{2.79} (0.65)   & 4.08 (0.62)   & 60.89 (19.06)  & 48.24 (13.59) \\
(200, 50)   & \textbf{73.25} (6.10)  & 79.86 (10.81) & 7762.00 (630.32) & 8195.59 (1897.72) \\
\bottomrule
\end{tabular}
\end{table}

CGLASSO solves the inner lasso problems using CD (CLASSO.COV in Algorithm \ref{alg:classo-cov}), in contrast to ADMM-based solvers that rely on joint \textit{Lagrangian} optimization \citep{baek2021detecting}. We compare the performance of CD that exploits model sparsity with off-the-shelf benchmark ADMM. For each algorithms, the performance is assessed by the runtime (in seconds) to compute the estimators across a regularization path. For each $\lambda$ across the path, we record the Monte Carlo median runtime across 50 trials as a measure of runtime performance.

\paragraph*{Convergence safeguards for small $\lambda$} Despite the simple and efficient implementation, CD is known in the literature to face convergence issues for very small values of $\lambda$ \citep{mazumder2012graphical, friedman2007pathwise}. Nonetheless, the convergence issues are of limited practical concern, as model selection and penalty tuning is governed by a selection criterion that identifies optimal estimators at moderate $\lambda$ where CD yields stable solution. We implement an early stopping rule for the regularization path based on a selection criterion ${\cal C}$, e.g. RMSE, BIC. When the true precision matrix $\Theta$ is known (e.g., in simulation settings), RMSE provides a principled choice for ${\cal C}$. In empirical applications where $\Theta$ is unknown, we instead use BIC as a data-driven proxy. For selection criterion ${\cal C}$ on the estimated precision matrix $\hat\Theta(\lambda)$, if $\lambda^*$ is the global minimizer of ${\cal C}(\hat\Theta(\lambda))$ \wrt $\lambda$ along the regularization path, the path is terminated at the current iteration $t$ and penalty $\lambda_t$ if for some threshold $\epsilon \in (0,1)$, 
$$ {\cal C}(\hat\Theta(\lambda_t)) - {\cal C}(\hat\Theta(\lambda^*)) > \epsilon \left[{\cal C}(\hat\Theta(\lambda_0)) - {\cal C}(\hat\Theta(\lambda^*))\right]. $$
Intuitively, when the selection criterion exhibits a ``U-shaped'' profile along the path, the stopping rule terminates the path once the deviation from the optimum at $\lambda^*$ exceeds $\epsilon$-fraction of the total improvement \wrt\ the initial penalty $\lambda_0$. This strategy is particularly effective in preventing unstable updates for very small values of $\lambda$.

\paragraph*{Experiment Setup} The experiment is conducted on data generated from a $p$-dimensional VAR(1) process $X_t = AX_{t-1} + \varepsilon_t$. $A$ is a tri-diagonal matrix with entries $A_{i,i} = 0.5$, $A_{i,i+1} = -0.3$ and $A_{i,i+2} = 0.1$. The errors $\varepsilon_{i,t}\overset{\iid}{\sim} \nor(0, \sigma_\varepsilon^2)$ with $\sigma_\varepsilon^2 = 0.1$. The simulation is run with $X_0 = \mathbf{0}_{1\times p}$ and a burn-in period 1000. We consider $(n,p) = (100,10),\ (200, 20),\ (200, 50)$, and $\log_2(\lambda) \in [-6, 0]$ with the grid space (in $\log_2$-scale) being 0.1. The spectral precision is estimated at frequency 0. We compare the runtime of CGLASSO-sc1 and CGLASSO-sc2 with that of ADMM based solver for complex graphical lasso. The ADMM based method \citep[Algorithm 2]{baek2023local} uses the \textit{sparse inverse covariance selection} (SICS) algorithm of \citep{scheinberg2010sparse}, based on alternating linearization method (ALM), a variant of ADMM. We refer to the ADMM-based method as the \textit{sparse inverse covariance estimator} (SPICE), and its warm-start variant as Warm-SPICE.

\paragraph*{Results} The median the total runtime for regularization paths are illustrated in Figure \ref{fig:runtime_final} and the values are given in Table \ref{tab:runtime_results}, with mean absolute deviation (MAD) in parenthesis. The runtime of CGLASSO-sc1 and 2 are substantially (20-100 times) smaller than that of the ADMM-based benchmarks SPICE and Warm-SPICE. For larger $p$, the median absolute deviation (MAD) of ADMM is substantially higher compared to CGLASSO-sc1 and 2. The median runtime for both the CGLASSO methods at $\lambda$ that minimizes the RMSE across the regulation path is smaller than the same for SPICE. Therefore, our simulation supports that CD based optimizers enjoy the speed gain over off-the-shelf algorithms by leveraging the orthogonality within groups exhibited by the complex predictors.

\begin{figure}[!t]
    \centering
    \subfloat{\includegraphics[width=0.25\linewidth]{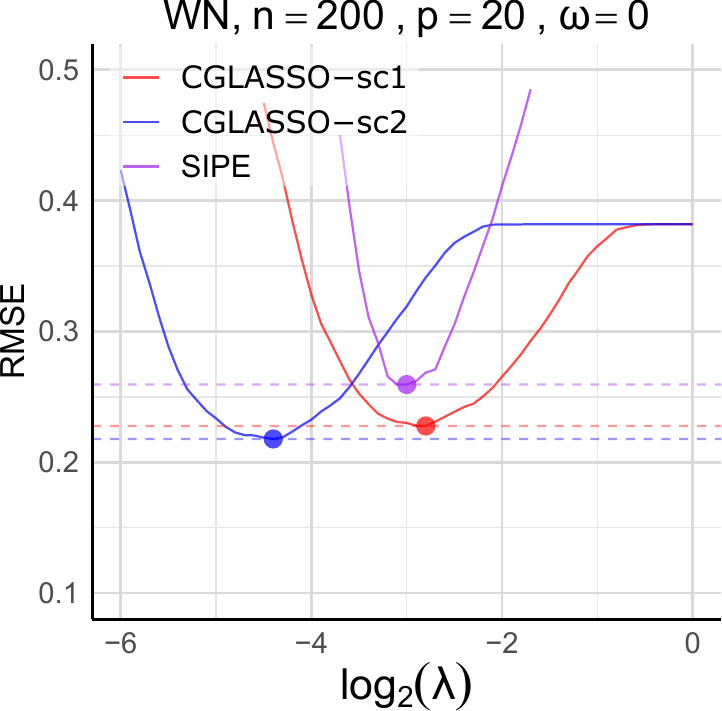}}
    \subfloat{\includegraphics[width=0.25\linewidth]{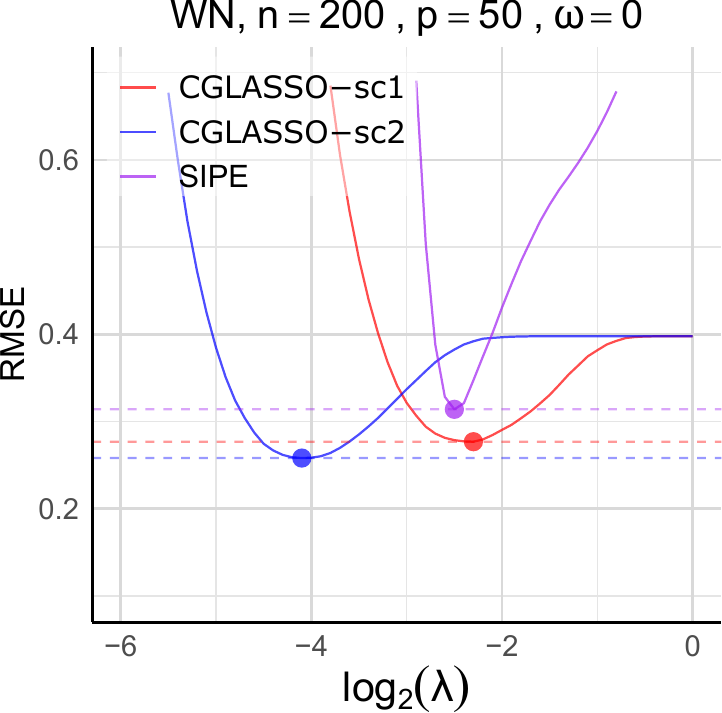}}
    \subfloat{\includegraphics[width=0.25\linewidth]{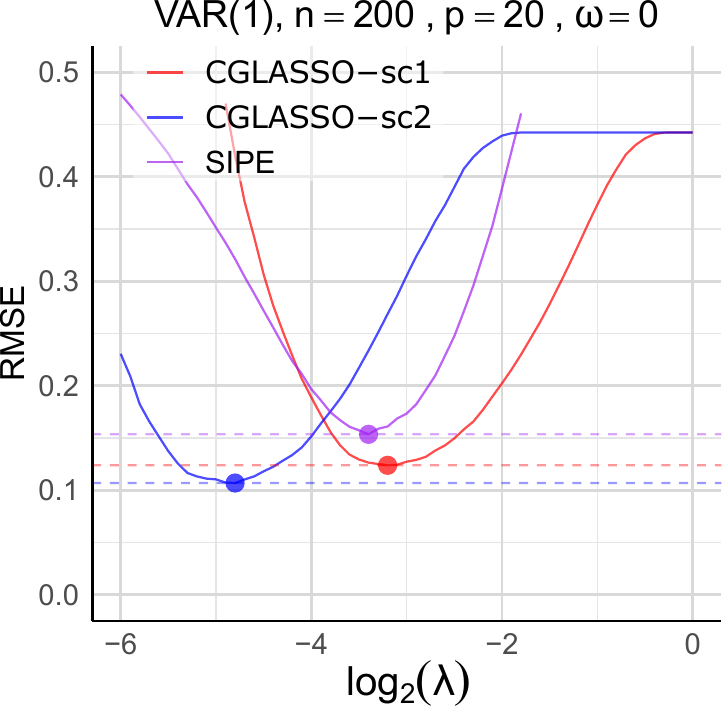}}
    \subfloat{\includegraphics[width=0.25\linewidth]{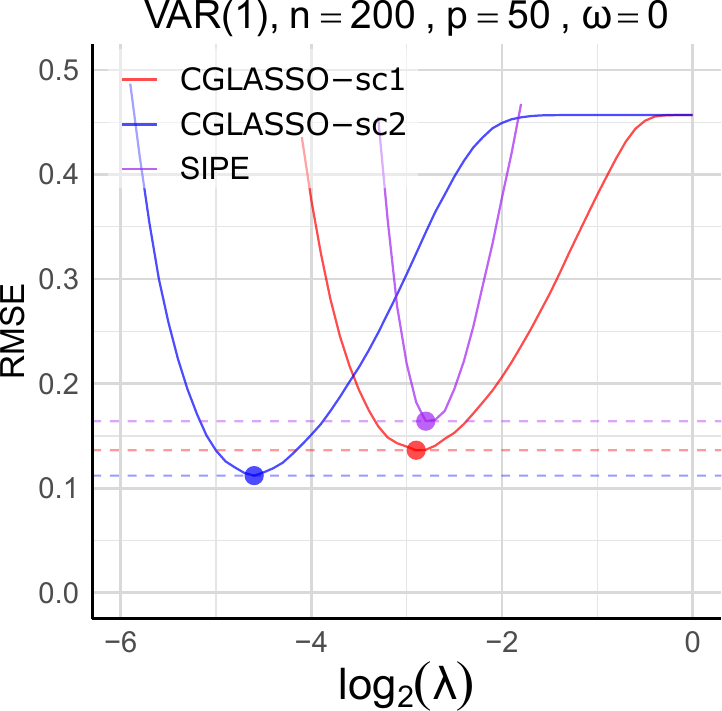}}
    \caption{\textbf{RMSE (median over 50 replicates) of CGLASSO-sc1 (red), CGLASSO-sc2 (blue) and SIPE (purple) across regularization path (different values of $\lambda$)}. CGLASSO-sc1 and 2 have substantially lower minimum RMSE than SIPE (marked by dots on the paths) for different choices of $p$.}
    \label{fig:rmse_final}
\end{figure}
\begin{table}[!t]
\centering
\caption{\textbf{Minimum median RMSE (\%) for CGLASSO and SIPE.} The underlying DGP are white noise and VAR(1) models with $n = 200$, $p \in \{10, 20, 50\}$, and $\omega = 0$. Lower RMSE (in bold) indicates better performance.}
\label{tab:rmse_final}
\small
\setlength{\tabcolsep}{2pt}
\begin{tabular}{llccc@{\hspace{10pt}}ccc}
\toprule
 & & \multicolumn{3}{c@{\hspace{10pt}}}{WN($\Sigma$)} & \multicolumn{3}{c}{VAR(1)} \\
\cmidrule(lr){3-5} \cmidrule(lr){6-8}
$p$ &  & CGLASSO-sc1 & CGLASSO-sc2 & SIPE & CGLASSO-sc1 & CGLASSO-sc2 & SIPE \\
\midrule
10 &  & 18.25 (2.50) & \textbf{16.95} (2.41) & 21.49 (2.48) & 10.93 (2.55) & \textbf{9.63} (2.18)  & 14.78 (3.10) \\
20 &  & 22.77 (2.27) & \textbf{21.78} (1.81) & 25.94 (3.06) & 12.39 (1.98) & \textbf{10.70} (1.43) & 15.35 (1.66) \\
50 &  & \textbf{27.69} (1.00) & 25.82 (0.91) & 31.40 (1.27) & 13.63 (0.81) & \textbf{11.20} (0.71) & 16.41 (1.67) \\
\bottomrule
\end{tabular}
\end{table}

\subsection{Joint penalty improves estimation accuracy over separate penalty}\label{subsec:rmse_improvement}

This experiment empirically shows that  joint penalization of real and imaginary parts in CLASSO and CGLASSO estimates the spectral precision matrix more accurately than estimating the real and imaginary parts separately as in SIPE \citep{fiecas2019spectral}. \citet{maleki2013asymptotic} also reports improved signal recovery measured via phase transitions-- when the non-zero real and imaginary parts are grouped under CLASSO regression.

The estimation error is evaluated with the RMSE in Fr\"{o}benius norm between the estimated spectral precision $\hat \Theta_j$ and the oracle  $\thetajstar$, defined as 
\begin{equation}\label{eq:rmse}
    \text{RMSE}(\hat\Theta_j, \thetajstar) := \frac{\|\hat\Theta_j - \thetajstar \|_F^2}{\|\thetajstar\|_F^2}.
\end{equation}
For VARMA($q, r$) process $X_t = \sum_{h = 1}^q A_h X_{t-h} + \varepsilon_t - \sum_{\ell = 1}^r B_\ell \varepsilon_{t-\ell}$ with $\varepsilon_t \sim WN(0, \Sigma_\varepsilon)$, the oracle is calculated as $\thetajstar = [f(\omega_j)]^{-1}$ where
\begin{equation}\label{eq:spec_density_varma}
    f(\omega_j) = \frac{1}{2\pi} \mathcal{A}^{-1}(e^{-\i \omega_j}) \Bcal (e^{-\i \omega_j}) \Sigma_\varepsilon (\Bcal (e^{-\i \omega_j}))^\dagger (\mathcal{A}^{-1}(e^{-\i \omega_j}))^\dagger,
\end{equation}
with $\mathcal{A}(z) := I_p - \sum_{t=1}^q A_t z^t $ and $\Bcal(z) := I_p - \sum_{t=1}^r B_t z^t $ satisfying $\det(\mathcal{A}(z)) \ne 0$ and $\det(\Bcal(z)) \ne 0$ on $\{z \in \CC : |z| = 1 \}$.

\paragraph*{Experiment Setup}  We consider the following data generating processes (DGP):
\begin{itemize}
    \item[(a)] \textit{Gaussian White noise model} $\text{WN}(0, \Sigma)$.~ $X_t \sim \nor(0, \Sigma)$ i.i.d., $\Sigma$ is a tri-diagonal matrix with $\Sigma_{i,i} = 0.7$, $\Sigma_{i,i-1} = 0.3$, $\Sigma_{i,i-2} = 0.2$. By \eqref{eq:spec_density_varma}, $\thetajstar = 2\pi \Sigma^{-1}$.
    
    \item[(b)] \textit{Vector autoregressive model of order 1} (VAR(1)).~ This DGP is implemented in Section \ref{subsec:runtime_improvement}. The true spectral density is calculated as \eqref{eq:spec_density_varma} and the spectral precision matrix as $\thetajstar = [f(\omega_j)]^{-1}$.
\end{itemize}
The data is generated for $n=200$ and $p \in \{10, 20, 50\}$. We work with $\log_2(\lambda) \in [-6, 0]$ with the grid space (in $\log_2$-scale) 0.1. The Monte Carlo median RMSE and its mean absolute deviation across the regularization paths are recorded for 50 replicates.

\paragraph*{Results} Figure \ref{fig:rmse_final} illustrates the median RMSE across the regularization paths, and the minimum median RMSE over the path and the median absolute deviation (MAD)
at the minimizing $\lambda$ are reported in Table \ref{tab:rmse_final}. CGLASSO-sc2 has the lowest minimum RMSE across all combinations of DGP and $p$, followed by CGLASSO-sc1 which performs comparably to CGLASSO-sc2, with a marginal increase in RMSE. Both CGLASSO-sc1 and CGLASSO-sc2 have lower RMSE across substantial parts of the regularization path containing the corresponding minimum values. Hence this experiment supports that joint regularization and optimization of the real and imaginary parts of the spectral precision matrix yields more accurate estimates than separate penalization and optimization.

\subsection{Full likelihood optimization improves model selection over NWR}\label{subsec:model_selection_improvement}

\begin{figure}[t]
    \centering
    \subfloat{\includegraphics[width=0.25\linewidth]{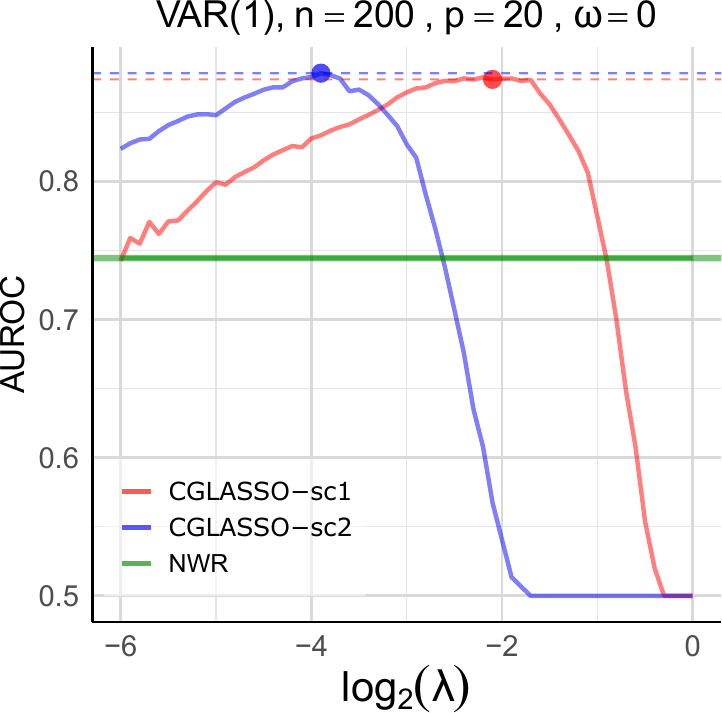}}
    \subfloat{\includegraphics[width=0.25\linewidth]{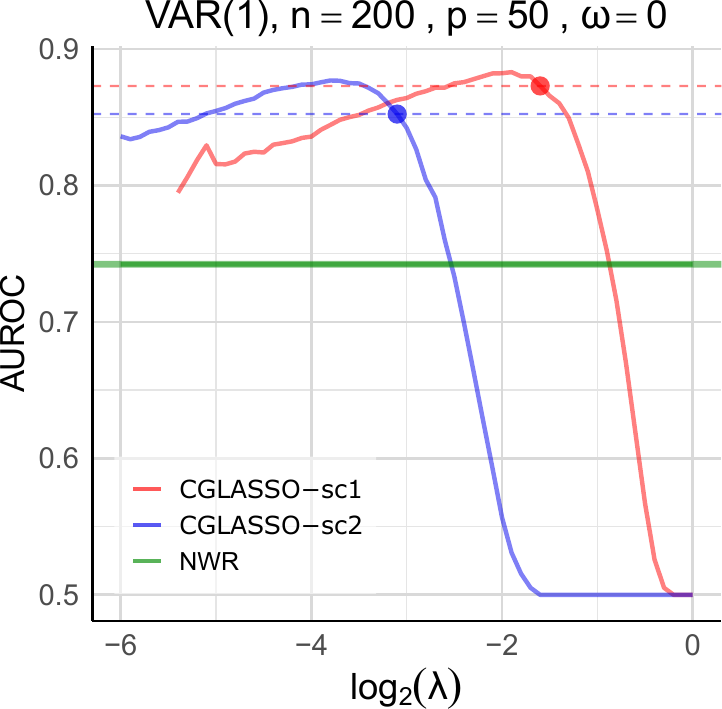}}
    \subfloat{\includegraphics[width=0.25\linewidth]{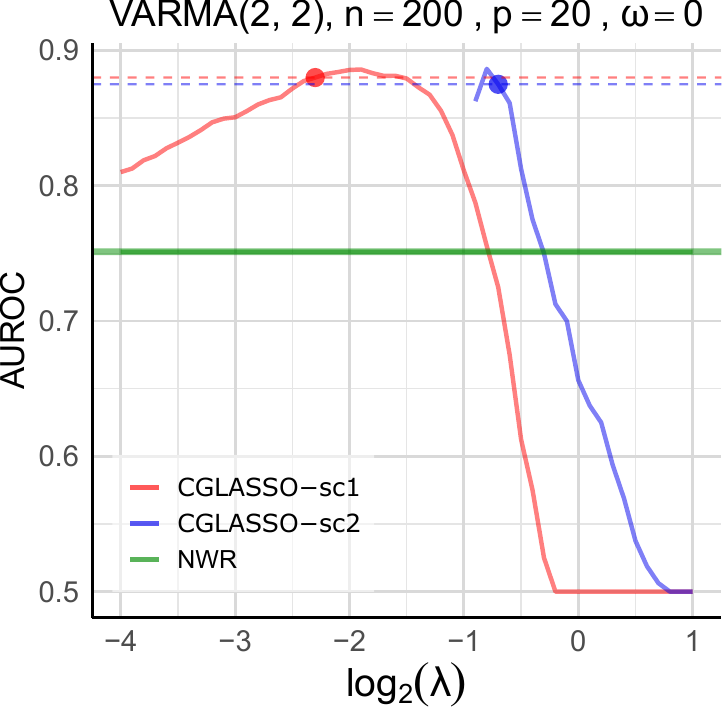}}
    \subfloat{\includegraphics[width=0.25\linewidth]{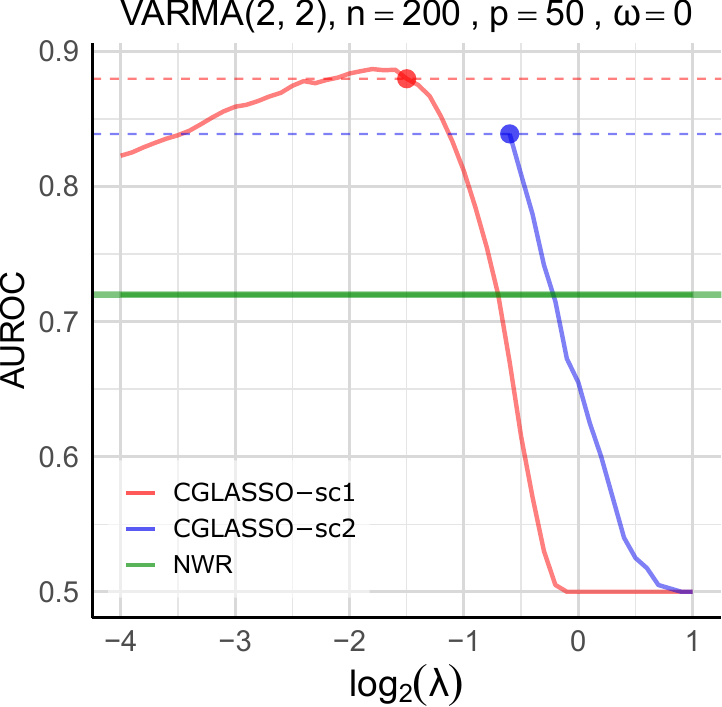}}
    \caption{\textbf{Area under ROC curve (AUROC, median over 50 replicates) of CGLASSO-sc1 (red) and CGLASSO-sc2 (blue) across different values of $\lambda$.} The green lines denote the AUROC corresponding to the NWR estimator. CGLASSO-sc1 and CGLASSO-sc2 have substantially higher AUROC (marked by dots on the paths) than NWR for different choices of $p$.}
    \label{fig:auroc_final}
\end{figure}

\begin{table}[t]
\centering
\caption{\textbf{AUROC (\%) selected by BIC for CGLASSO-sc1, CGLASSO-sc2 and NWR.} The underlying DGP are VAR(1) and VARMA(2,2) models with $n = 200$, $p \in \{10, 20, 50\}$, and $\omega = 0$. Higher AUROC (in bold) means better model selection.}
\label{tab:auroc_final}
\small
\setlength{\tabcolsep}{2pt}
\begin{tabular}{llccc@{\hspace{10pt}}ccc}
\toprule
 & & \multicolumn{3}{c@{\hspace{10pt}}}{VAR(1)} & \multicolumn{3}{c}{VARMA(2,2)} \\
\cmidrule(lr){3-5} \cmidrule(lr){6-8}
$p$ &  & CGLASSO-sc1 & CGLASSO-sc2 & NWR & CGLASSO-sc1 & CGLASSO-sc2 & NWR \\
\midrule
10 &  & 81.83 (4.15) & \textbf{84.45} (4.10) & 73.53 (4.41) & 84.70 (3.65) & \textbf{95.00} (0.00) & 74.30 (3.25) \\
20 &  & 87.38 (2.19) & \textbf{87.83} (2.04) & 74.44 (4.27) & \textbf{88.00} (2.04) & 87.50 (4.53) & 75.12 (2.45) \\
50 &  & \textbf{87.29} (1.37) & 85.23 (1.53) & 74.22 (1.75) & \textbf{87.97} (1.18) & 83.89 (0.67) & 71.99 (1.73) \\
\bottomrule
\end{tabular}
\end{table}
CGLASSO algorithms solve penalized likelihood optimization problems where the likelihood is a function of the DFTs across all the nearby frequencies in a bandwidth as well as all dimensional components. In contrast, nodewise regression (NWR) involves regressing one column of the DFT matrix on the others (Section \ref{subsubsec:method-nodewise-reg}), and hence recovers one column of the spectral precision matrix at a time. Therefore, NWR does not incorporate the full likelihood information as CGLASSO, and requires solving $p$-regression problems in order to fully recover $\thetajstar$. In this experiment, we show that CGLASSO outperforms NWR in model selection consistency due to full likelihood implementation.

To measure the model selection performance, we calculate AUROC that assesses the tradeoff between \textit{sensitivity} and \textit{specificity} \citep[Section 9.2.5]{hastie2009elements}. We compare the AUROC values across the regularization path and the AUROC corresponding to $\lambda$ selected with highest BIC. The Monte Carlo median AUROC and its MAD at the $\lambda$ across the regularization paths are calculated for 50 replicates of all experiments. In addition, we record the BIC values for each regularization path and calculate the median BIC across the paths. The median AUROC of CGLASSO-sc1 and CGLASSO-sc2 at $\lambda$ that minimizes the corresponding median BIC are compared against the median (across 50 Monte Carlo trials) AUROC of NWR. 

\paragraph*{Experiment Setup} Two DGPs are considered in this setting:
\begin{itemize}
    \item[(a)] The VAR(1) model in Section \ref{subsec:rmse_improvement}.
    
    \item[(b)] VARMA(2,2) model $X_t = A_1 X_{t-1} + A_2 X_{t-2} + \varepsilon_t + B_1 \varepsilon_{t-1} + B_2 \varepsilon_{t-2}$. We choose $A_1 = 0.4 I_p,\ A_2 = 0.2 I_p,\ B_1 = I_{p/5} \otimes M_1,\ B_2 = I_{p/5} \otimes M_2$. $M_1$ has 0 as upper-diagonal entries, diagonal entries 3 and off-diagonal entries 1.5. Similarly $M_2$ has 0 as upper-diagonal entries, 1.5 as diagonal entries and lower-diagonal entries 0.75. The errors $\varepsilon_t$ has i.i.d. components and $\varepsilon_{i,t}\sim \nor(0, (0.1)^2)$.
\end{itemize}
For DGP (b), the simulation  is run with $X_0 = \mathbf{0}_{1\times p}$ with a burn-in period 1000 and $\log_2(\lambda) \in [-4, 1]$ with the grid space (in $\log_2$-scale) 0.1.

\paragraph*{Results} The median AUROC across the regularization paths are illustrated in Figure \ref{fig:auroc_final}. CGLASSO has higher AUROC than the baseline AUROC of NWR for most of the regularization paths for all choices of DGP and $p$. The BIC-selected median AUROC for CGLASSO-sc1 and CGLASSO-sc2 are reported in Table \ref{tab:auroc_final} and the MAD of AUROC at the BIC chosen $\lambda$ are reported in parentheses. Both CGLASSO-sc1 and CGLASSO-sc2 have higher AUROC than NWR at the selected $\lambda$ and all choices of $p$. As value of $p$ increases for both VAR(1) and VARMA(2,2) generated data, CGLASSO-sc1 marginally outperforms CGLASSO-sc2. Therefore, the experiment reinforces that incorporating the full likelihood boosts the model selection performance.

\subsection{Adaptive penalization improves estimation accuracy}\label{subsec:adaptation_improvement}

\begin{figure}[!t]
    \centering
    \subfloat[]{\includegraphics[width=0.25\linewidth]{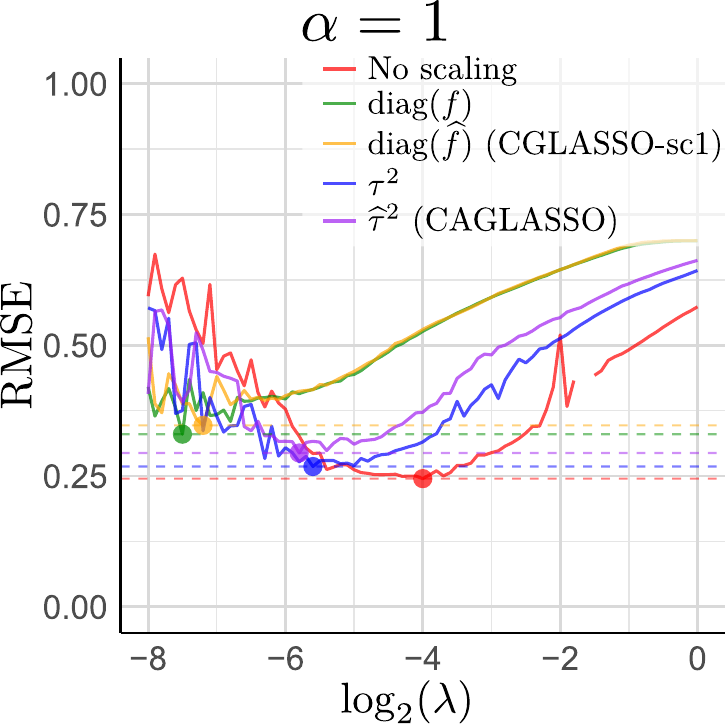}}
    \subfloat[]{\includegraphics[width=0.25\linewidth]{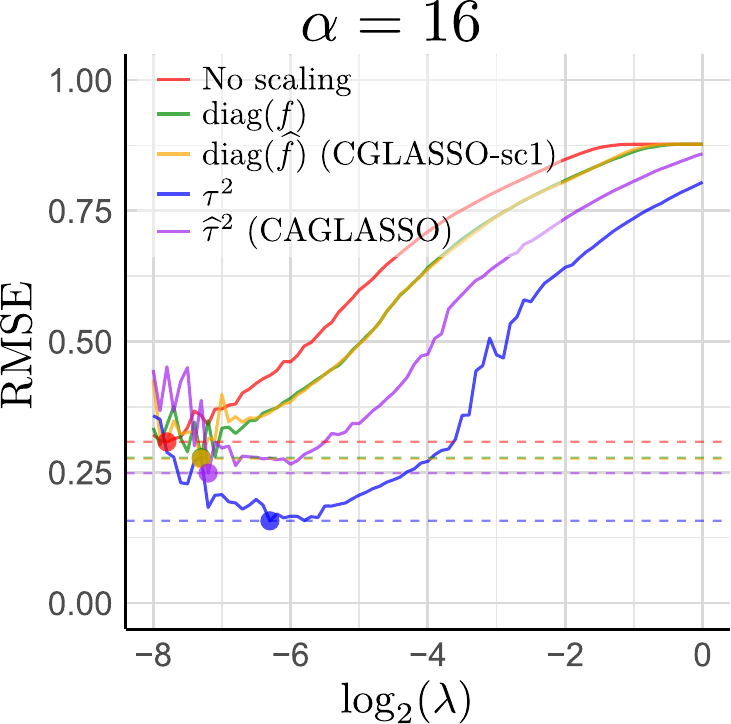}}
    \subfloat[]
    {\includegraphics[width=0.25\linewidth]{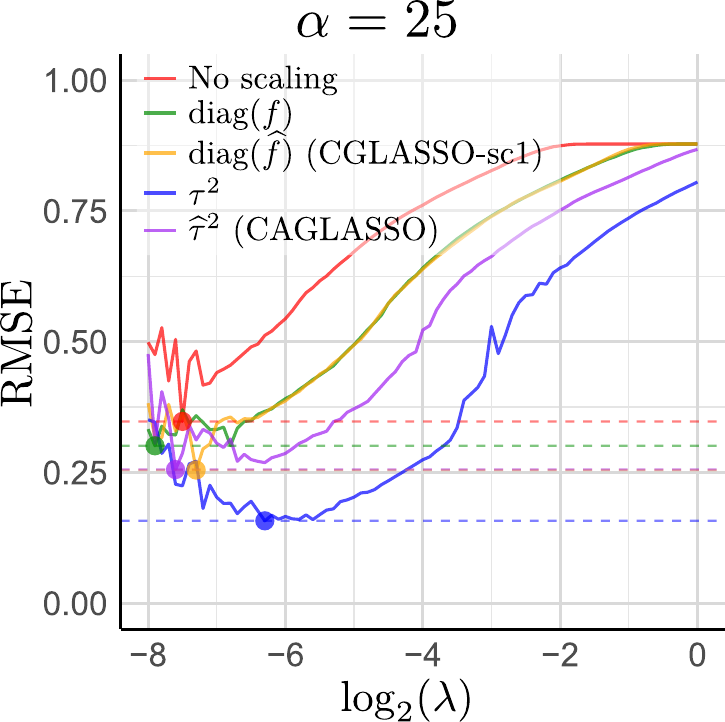}}
    \subfloat[]{\includegraphics[width=0.25\linewidth]{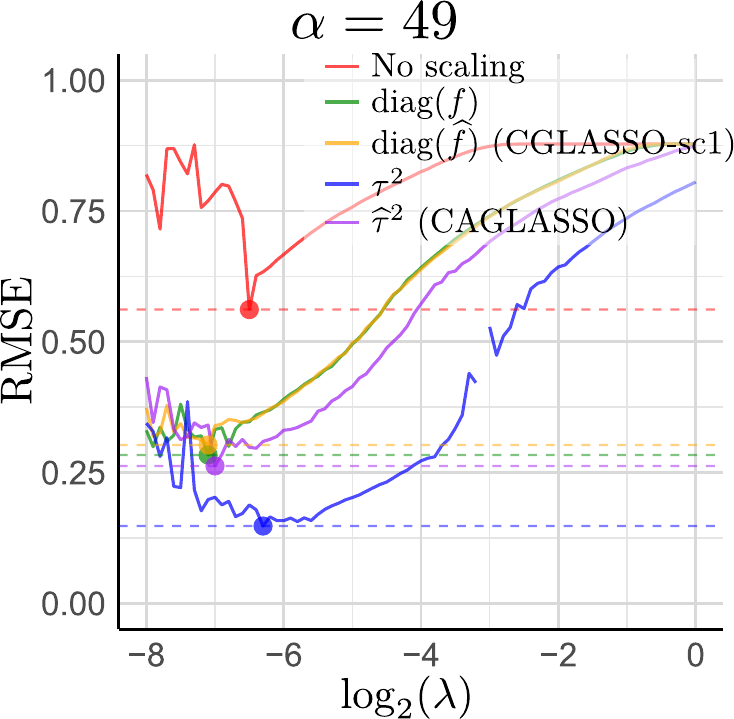}}
    \caption{\textbf{RMSE of CGLASSO with uniform penalty (red), CGLASSO scaled with $\dg(f)$ (green), CGLASSO-sc1 (orange), CGLASSO scaled with $\tau_a^2$'s (blue) and CAGLASSO}. For $\alpha = 1$ in panel (a), CGLASSO with uniform penalty yields the lowest RMSE (red dot). For larger values of $\alpha$ in panel (b), (c) and (d), CGLASSO scaled with $\tau_a^2$'s yields the lowest median RMSE across the regularization paths.}
    \label{fig:adaptive_path}
\end{figure}

\begin{table}[!t]
\centering
\caption{RMSE (\%) under different adaptive penalties for varying $\alpha$ that controls the scaling disparity. The columns imply the choices for the scaling matrix $D^2$. Lower RMSE (in bold) means more accurate estimation.}
\label{tab:rmse_adaptive}
\small
\setlength{\tabcolsep}{8pt}
\begin{tabular}{r@{\hspace{15pt}}cccc|c}
\toprule
$\alpha$ & No Scaling & $\dg(f)$ & $\dg(\hat f)$ (CGLASSO-sc1) & $\hat\tau^2$ (CAGLASSO) & $\tau^2$ (\text{Oracle}) \\
\midrule
1  & \textbf{24.50} & 33.01 & 34.68 & 29.41 & 26.82 \\
16 & 30.82 & 27.80 & 27.55 & 24.84 & \textbf{15.72} \\
25 & 34.74 & 30.08 & 25.44 & 25.53 & \textbf{15.73} \\
49 & 56.14 & 28.33 & 30.25 & 26.25 & \textbf{14.76} \\
\bottomrule
\end{tabular}
\end{table}

We compare the CGLASSO algorithm with entry-wise penalties being uniform, penalties adapted to estimated marginal variances, and penalties adapted to estimated partial variances (CAGLASSO) when the entries of $\Theta(\omega)$ are highly heterogeneous. We use Monte Carlo median RMSE across 50 trials to measure the estimation accuracy.

\paragraph*{Experiment Setup} We consider a $p$-dimensional white noise process $\text{WN}(\Sigma)$, $\Omega := \Sigma^{-1}$ is a block-diagonal matrix with two $p/2$-dimensional Toeplitz structured blocks $A$ and $B$ with $A_{i,j} := (0.2)^{|i-j|}$, $B_{i,j} := (0.8)^{|i-j|}$ for $i,j\in [p]$. We consider a parameter $\alpha$ that accounts for the scaling disparity between two blocks. $\alpha$ is set $\alpha \in \{1, 16, 25, 49\}$. The spectral precision matrix is $\Theta(\omega_j) = 2\pi \Omega$ for all $j \in F_n$.
We generate the data with $n = 1000$ and $p = 20$, and fix $\omega_j = 0$. $\hat f(0)$ is calculated and scaled to $\hat R = D^{-1} \hat f(0) D^{-1}$ with the diagonal matrix $D$. Then $\hat K$, i.e. CGLASSO output with $\hat R$, is rescaled to $\hat \Theta = D^{-1} \hat K D^{-1}$. The choices for the entries of $D$ are:
\begin{itemize}
    \item[(a)] $D_{a, a}^2 = 1$ (entry-wise uniform penalty),
    \item[(b)]  $D_{a,a}^2 = [f(0)]_{a,a}$ (diagonals of true spectral density),
    \item[(c)] $D_{a,a}^2 = [\hat f(0)]_{a,a}$ (diagonals of estimated averaged periodogram),
    \item[(d)] $D_{a,a}^2 = 1/(\thetajstar)_{a, a}$ (true scales of partial variances for NWR), and
    \item[(e)] $D_{a,a}^2 = \hat \tau_a^2 $ as in CAGLASSO. 
\end{itemize}

\paragraph*{Results} Figure \ref{fig:adaptive_path} shows the median RMSE (across 50 trials) across the regularization paths, and the minimum of median RMSE are reported in Table \ref{tab:rmse_adaptive}. For $\alpha=1$ i.e. the diagonals of $\Theta$ are all 1, uniform penalization (a) exhibits the minimum RMSE across all the competing methods. If $\alpha$ is increased, uniform penalization yields higher minimum RMSE. On the other hand, the scaled variants (b)-(e) tend to produce lower RMSE than uniform penalization. The penalization method (d) with the oracle partial variances yields the lowest minimum RMSE as well as uniformly lowest RMSE across the regularization path for larger $\alpha$. With estimated partial variance as penalty weights, the RMSE for CAGLASSO is marginally higher than the previous case. Scaling with true spectral density as in (b), and averaged periodogram estimator in (c) yield similar RMSE across the regularization path. Thus for a given $\lambda$, the empirically most accurate method in terms of RMSE corresponds to scaling with oracle partial variances, followed by CAGLASSO under higher heterogeneity i.e. larger $\alpha$-- supporting the choice of weights for estimating the spectral precision matrix.

\section{fMRI Data Analysis}\label{sec:FC_fMRI}
We implement CGLASSO to estimate functional connectivity (FC) among human brain regions. The data set used here is a publicly available, high-resolution, preprocessed magnetic resonance imaging data from the Human Connectome Project (HCP)—Young Adult S1200 release \citep{van2013wu, dhamala2020sex}. We consider the time series data of 1003 healthy adults observed over same time horizon across four complete resting state fMRI scans. The details of data preprocessing are deferred to Appendix \ref{sec:fmri_pre}.

\begin{figure}[t]
    \centering
    \subfloat{\includegraphics[width=0.33\textwidth]{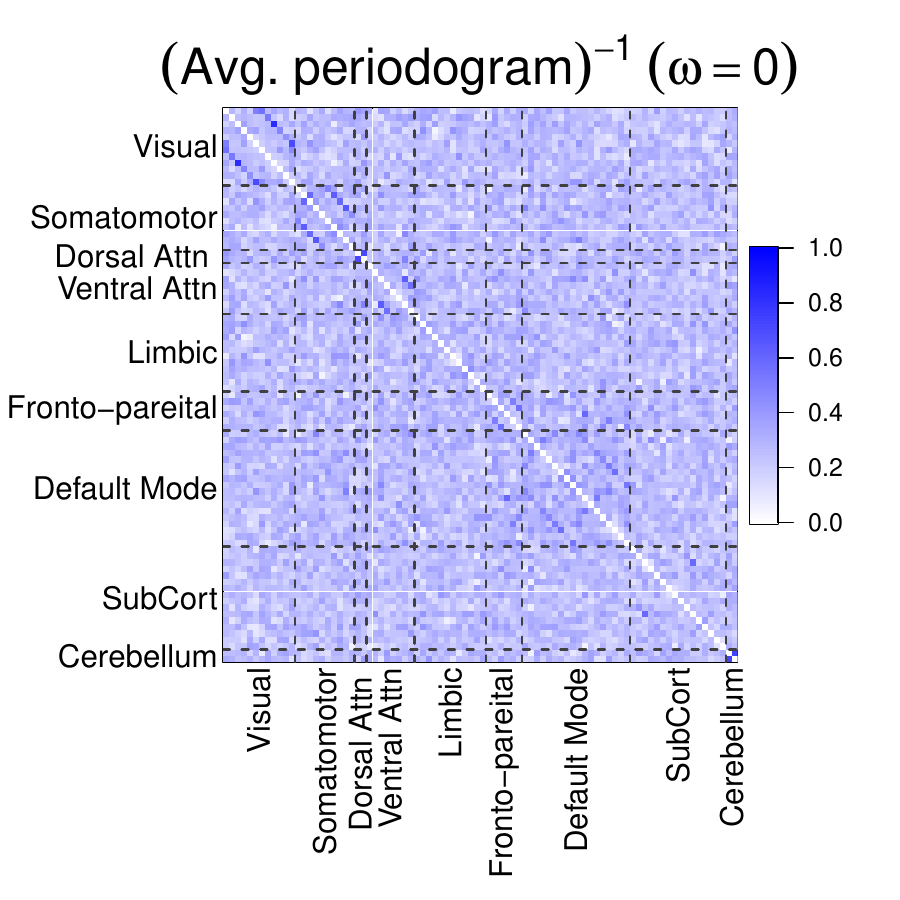}}
    \subfloat{\includegraphics[width=0.33\textwidth]{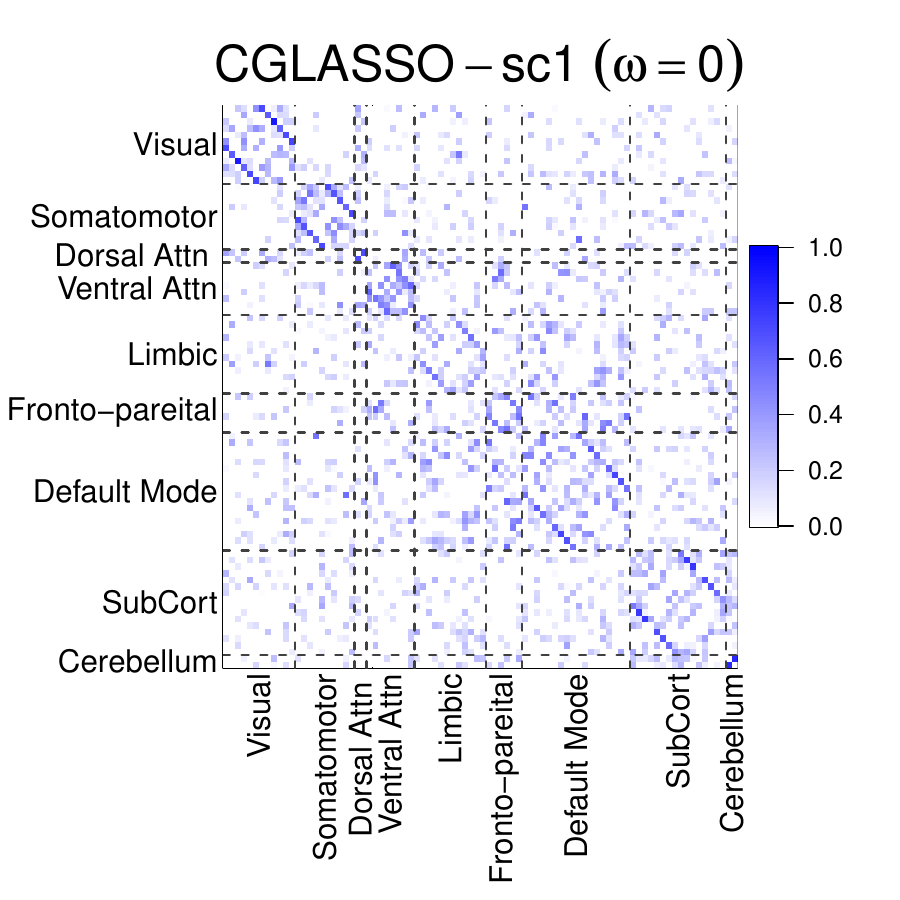}}
    \subfloat{\includegraphics[width=0.33\textwidth]{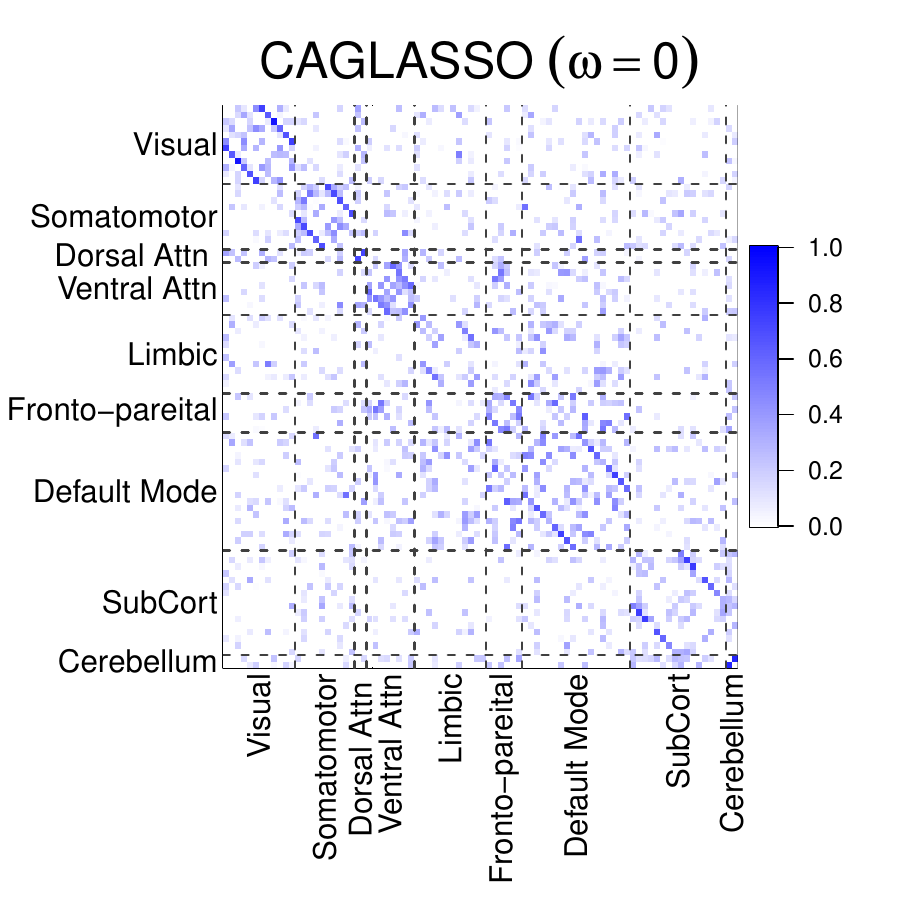}}
    \\
    \vspace{-10pt}
    \subfloat{\includegraphics[width=0.33\textwidth]{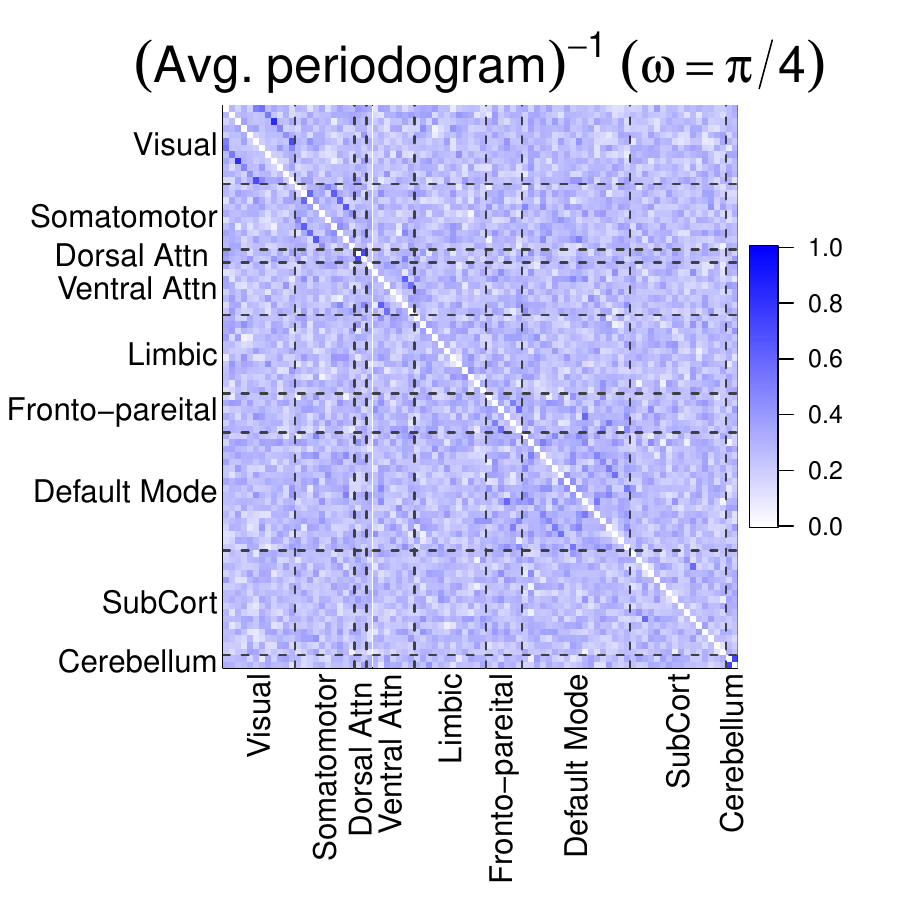}}
    \subfloat{\includegraphics[width=0.33\textwidth]{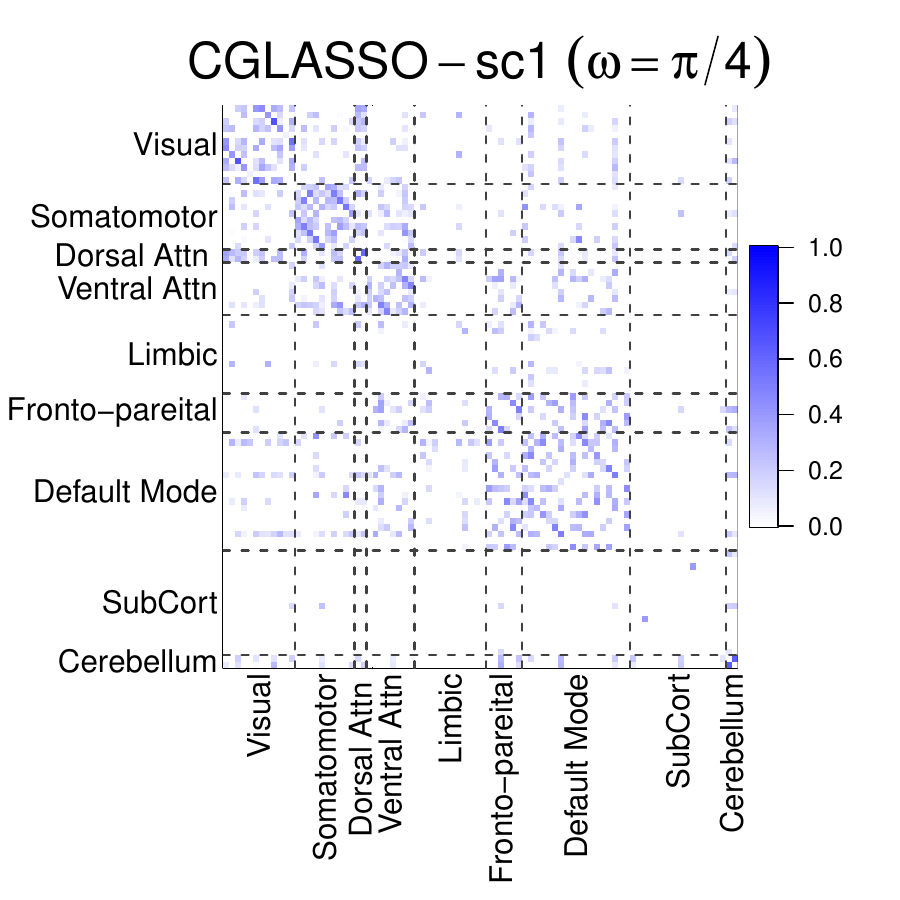}}
    \subfloat{\includegraphics[width=0.33\textwidth]{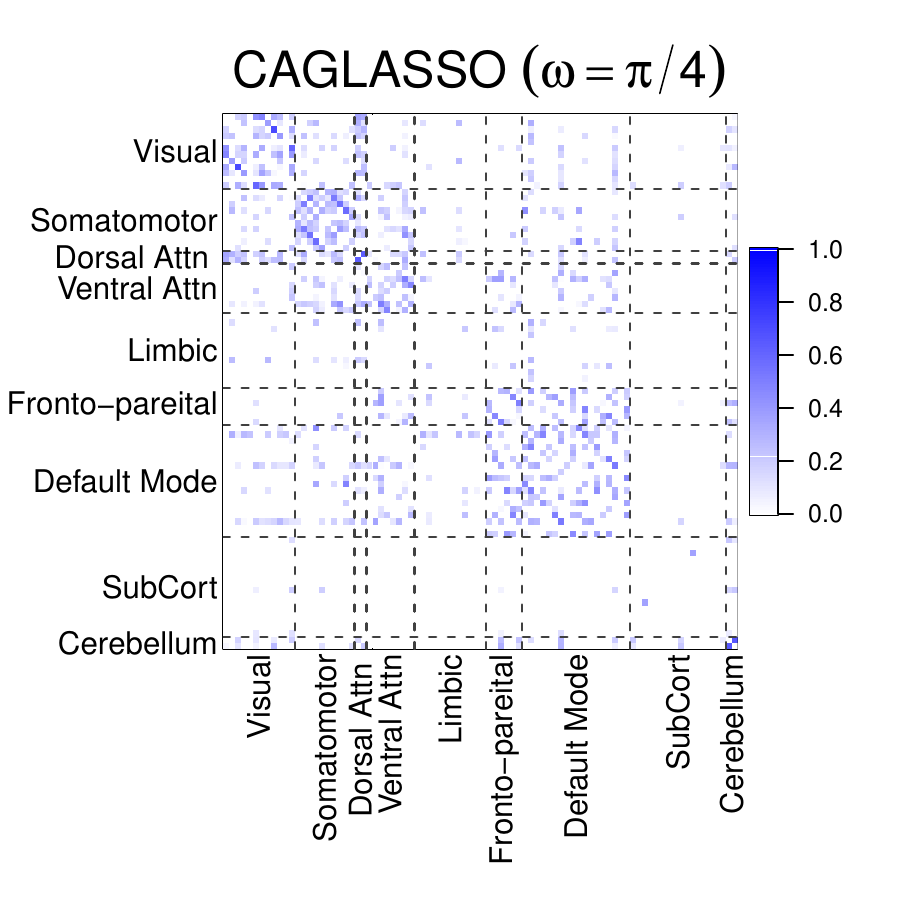}}
    \caption{Heatmaps of partial coherence (square-root-transformed) for averaged periodogram inversion, CGLASSO-sc1 and CAGLASSO on fMRI data of one subject. }
    \label{fig:hcp_heatmaps}
\end{figure}

\paragraph*{Data Analysis} Each scan has $n = 1200$ time points and $p = 86$ brain regions. The brain regions are further parcellated in 9 groups viz. visual, somatomotor, dorsal attention, ventral attention, limbic, default mode, fronto-parietal, subcortical and cerebellum functional networks. For $m = \lfloor\sqrt{n}\rfloor$ \citep{bohm2009shrinkage}, the effective sample size is $2m+1 = 69 < p$. Therefore $\hat f(\omega_j)^{(r)}$ i.e. the averaged periodogram for $r$\textsuperscript{th} scan is not invertible at any $\omega_j$. We take a further average of $\hat f(\omega_j)^{(r)}$ across the four scans i.e. $\hat f(\omega_j) := \sum_{r=1}^4 \hat f(\omega_j)^{(r)}/4$. $\hat f(\omega_j)$ is used as a proxy of averaged periodogram to obtain the classical spectral precision estimator $\hat\Theta_j = [\hat f(\omega_j)]^{-1}$. Additionally $\hat f(\omega_j)$ is used as the input of CGLASSO-sc1 and CAGLASSO algorithms. We illustrate our analysis with $\omega_j = 0, \pi/4$. For invertible $\hat f(\omega_j)$, the classical estimator of partial coherence is the matrix with $(a,b)$ entry: $|(\hat f(\omega_j)^{-1})_{a,b}|\big/\sqrt{|(\hat f(\omega_j)^{-1})_{a,a}|\cdot |(\hat f(\omega_j)^{-1})_{b,b}|}$. For CGLASSO-sc1 and CAGLASSO, the analogous partial coherence matrices are calculated by replacing $\hat f(\omega_j)^{-1}$ with $\hat\Theta_j$ obtained from the underlying CGLASSO problem. By construction, the partial coherence matrices have all diagonals 1. Therefore we set the diagonals of the partial coherence matrices to zero for improved visualization of the off-diagonal entries.

\paragraph*{Results} Figure~\ref{fig:hcp_heatmaps} presents an example of functional connectivity (FC) coherence matrices for a representative subject from our sample. The proposed CGLASSO-sc1 and CAGLASSO algorithms effectively captures the underlying sparsity structure, retaining physiologically relevant connections while suppressing spurious noise-induced entries in the averaged periodogram. In particular, the CGLASSO estimated partial coherence matrices prominently recover known \textit{bilateral functional connections}, especially between homologous regions in the visual and Somatomotor systems---consistently reported in the literature~\citep{zuo2010growing, sun2018large}. Notably, functional links in the Limbic and sub-Cortical regions with other regions are stronger at frequency~0 and diminish at~$\pi/4$, whereas Ventral and Dorsal attention-somatomotor links are more prominent at~$\pi/4$. CAGLASSO produces marginally sparser estimates than CGLASSO-sc1, attributed to its adaptive penalization using NWR partial variances which more effectively attenuates weak non-zero entries.

\section{Discussion}\label{sec:discussion}

In this work, we propose an estimator of the spectral precision matrix namely complex graphical lasso (CGLASSO)-- tailored to recover frequency-specific conditional graphical models. We further propose the method called complex adaptive graphical lasso (CAGLASSO) that adapts the entry-wise penalties to the scales of estimated partial variances obtained from nodewise regression (NWR) of the DFTs. We develop a fast pathwise coordinate descent algorithm for complex lasso (CLASSO) and subsequently use that to develop algorithms for CGLASSO and CAGLASSO. We conduct a non-asymptotic analysis of our proposed methods and show that consistent estimation is possible in high-dimensional regimes under sparsity assumptions and sufficient tail control on the underlying time series. We investigate that both on simulated data sets and a real fMRI data, the performance of our proposed estimators is substantially better than the benchmark methods in appropriate performance metric.

We did not delve into a number of questions related to our method. First, extending the algorithm from a single frequency to all the frequencies is required for recovering the true graphical model for stationary time series. Such an extension needs additional work on both algorithmic and theoretical fronts. A fast and scalable pathwise algorithm for jointly optimizing over different frequencies and different tuning parameters is left as a future work. 

Another important question is the selection of bandwidth $m$. For simplicity, we choose $m = \lfloor\sqrt{n}\rfloor$ following the custom in the literature of high dimensional spectral density matrix estimation \citep{bohm2009shrinkage}. However, a data driven choice of $m$ is more appropriate in practice \citep{ombao2001simple}. Allowing for different bandwidth choices for each off-diagonal entry is ideal for accurate estimation, however more challenging to analyze and compute with. We leave these directions for future research. Finally, a number of recent works have investigated the inferential properties of frequency-domain estimators \citep{chang2025statistical, krampe2025frequency} and delved into the nonstationary time series in high-dimension \citep{zhang2021convergence, basu2023graphical}. Extending CGLASSO and CLASSO to these broader frameworks are also left as avenues for future research.

\section*{Acknowledgments}
ND, AK and SB acknowledge partial support from  NIH award R21NS120227. In addition, SB  acknowledges partial support from NSF awards DMS-1812128, DMS-2210675, DMS-2239102 and NIH award R01GM135926. ND also thanks Dr. Younghoon Kim for maintaining the package containing implementation of our methods, valuable input and insightful discussions.

\bibliographystyle{abbrvnat}
\bibliography{00-0-all}
\appendix
\etocdepthtag.toc{appendix}
\begin{center}\bfseries\large Table of Contents for the Appendix\end{center}
\etocsettocstyle{}{}
{\etocsettagdepth{appendix}{subsection}
\etocsettagdepth{main}{none}
\tableofcontents}

\section{Supplement for the theory of DFT-NWR}\label{sec:supp_dft_nwr}

\subsection{Restricted eigenvalue condition}\label{subsubsec:re}
For linear regression in high dimensional regime, lasso estimated coefficients are consistent under a  \textit{restricted eigenvalue} (RE) condition \citep{bickel2009simultaneous, van2009conditions}. In words, RE condition ensures that the prediction error is small only when the estimation error of $\bjahat$ is small and restricts the eigenvalues of the covariance-like matrix i.e. the averaged periodogram  $\hat f(\omega_j)$ away from zero in the direction of the estimation error vector. For formally characterizing the RE condition, we define a cone set as
$$ \Cscr_{p}(S, t) := \{v\in \CC^p: \|v_{S\cmplt}\|_1 \le t \|v_S\|_1 \},\quad S \subseteq [p],\ t > 0. $$
Similar to \citet[Section 6.2.3]{buhlmann2011statistics}, we can show that the error vector of NWR $\bjahat - \bjastar$ lie in a cone $\Cscr_{p-1}(S_a, 3)$ if $\lambda_a \ge 2\| {\Zma}^\dagger \ea/(2m+1) \|_\infty$. Hence, RE condition is not very stringent even though $\Zma$ is singular as long as the deviation $\| {\Zma}^\dagger \ea/(2m+1) \|_\infty$ is sufficiently small (see Proposition \ref{prop:deviation}). Our next result states that the RE condition holds with high probability if the sample size and the bandwidth are sufficiently large.

\begin{proposition}[Restricted eigenvalue (RE)]\label{prop:re}
 Consider the DFT-NWR setup of \eqref{eq:nodewise-reg} with $p > 1$ satisfying Assumption \ref{asn:summable}. Then there exist universal constants $c_i > 0$ such that for the sample size $n$ satisfying $n \succsim \M \lf \max\{1, \rho^2\} s_a\min\{s_a\log p, \log (21 ep/s_a)\}$ and $\tnf \le 1/(4\M)$, and the bandwidth satisfying $\max\{1, \rho^2\} s_a\min\{\log p, \log (21 ep/s_a)\} \precsim m \le n/(4\M \lf)$, the following holds with probability greater than $1 - c_0 \exp[-c_1(2m+1)\min\{1, \rho^{-2}\} ]$
\begin{equation}\label{eq:re_condition}
    \inf_{v\in \Cscr_{p-1}(S_a, 3)\setminus\{0\}} \frac{\|\Zma v\|_2^2}{(2m+1)\|v\|_2^2} \ge \are,
\end{equation}
where $\are := \M^{-1}$, $\rho := c_2\M^2$.
\end{proposition}

The proof of Proposition \ref{prop:re} is deferred to Appendix \ref{pf:re}. \eqref{eq:re_condition} is referred as the \textit{RE condition} with the parameter $\are$. Assumption \ref{asn:summable} ensures bounded spectrum i.e. $\M < \infty $, resulting to the RE condition to meaningfully hold with $\are > 0$. For $s_a > c_0$, $s_a \log(c_0 p/s_a)$ is smaller than $s_a \log p$ that is the bandwidth requirement for consistent estimation of $\bjahat$ as in Theorem \ref{thm:consistency_dft_nwr}. Additionally, $\tau = c_3\M^2$ in the bandwidth complexity in Proposition \ref{prop:re} implies that less stable processes require larger bandwidth, hence larger sample size is needed for ensuring that RE conditions hold. The conditions $m \le n/(4\M \lf)$ and $\tnf \precsim 1/\M$ limit the truncation bias due to the finite sample DFT and the smoothing bias across nearby frequencies. Less stable processes with larger $\M$ admit tighter constraint on the truncation error $\tnf$ and a tighter upper bound on $m$.

\subsection{Deviation condition}\label{subsubsec:deviation_condition}

A sufficient condition for the consistency of DFT-NWR is that the empirical process term $ {\Zma}^\dagger \ea/(2m+1)$ concentrates around zero with high probability. We formally state this deviation condition as follows.
\begin{proposition}[Deviation condition]\label{prop:deviation}
    Consider the DFT-NWR setup \eqref{eq:nwr_model} satisfying Assumption \ref{asn:summable}. Assume that the sample size satisfies $\tnf\le 1/(4\M), ~~\text{and}~~ n \ge 4\M \lf$. Then for any $a \in [p]$ and $A > 0$, and the bandwidth $m \succsim \log p$, there exists a constant $c > 0$ such that with probability greater than $1 - 4 \exp(-(cA - 1) \log p)$,
    \begin{equation}\label{eq:deviation_ineq}
        \left\|\frac{1}{2m+1} {\Zma}^\dagger \ea \right\|_\infty \le \M^2 \sqrt{\frac{3A\log p}{2m+1}}.
    \end{equation}
    
\end{proposition}

The proof of Proposition \ref{prop:deviation} is deferred to Appendix \ref{pf:deviation}, and relies on concentration bounds for quadratic forms of Gaussian time series. The coordinates of the empirical process term concentrates around zero with error bound $\M^2 \sqrt{ \log p /m}$ -- with $\M^2$ capturing the effect of temporal dependence, and $\sqrt{\log p / m}$ being similar to the deviation bound associated with supremum of the empirical process term in lasso \citep[Example 7.14]{wainwright2019high}. The DFTs being asymptotically distributed as independent complex valued Gaussian random variables, a similar rate shows up in our analysis.

The bandwidth complexity $m \succsim \log p$ shares similarity with to the sample size requirement $n \succsim \log p$ for deviation bound in regression with i.i.d samples \citep[Example 7.14]{wainwright2019high}. The finite sample truncation and smoothing of DFTs impose additional sample size restrictions.

\subsection{Basic inequality}\label{subsec:basic_ineq}
We derive the basic inequality for DFT-NWR estimator obtained from lasso. We use \eqref{eq:nwr_model_mat} and the optimization problem \eqref{eq:nodewise-reg} and get,

\begin{align*}
    & \frac{1}{2m+1}\|\Za - \Zma\bjahat \|_2^2 + \lambda_a \|\bjahat\|_1 \le \frac{1}{2m+1}\|\Za - \Zma\beta \|_2^2 + \lambda_a \|\beta\|_1\\
    \implies & \frac{1}{2m+1}\|\ea + \muja - \Zma\bjahat \|_2^2 + \lambda_a \|\bjahat\|_1 \le \frac{1}{2m+1}\|\ea + \muja - \Zma\beta \|_2^2\\
    & \hspace{300pt}+ \lambda_a \|\beta\|_1\\
    \implies & \frac{1}{2m+1}\|\Zma \bjahat - \muja\|_2^2 - \frac{2}{2m+1}{\ea}^\dagger (\Zma \bjahat - \muja) + \lambda_a \|\bjahat\|_1 \\
    & \hspace{30pt} \le \frac{1}{2m+1}\|\Zma \beta - \muja\|_2^2 - \frac{2}{2m+1}{\ea}^\dagger (\Zma \beta - \muja) + \lambda_a \|\beta\|_1\\
    \implies & \frac{1}{2m+1}\|\Zma\bjahat - \muja\|_2^2 + \lambda_a \|\bjahat\|_1 \le \frac{2}{2m+1}{\ea}^\dagger \Zma (\bjahat - \beta) + \\
    & \hspace{150pt} + \frac{1}{2m+1}\|\Zma\beta - \muja\|_2^2 + \lambda_a \|\beta\|_1,
\end{align*}
hence the proof of the basic inequality is complete. \pfend

\section{Proofs of the propositions}\label{sec:appendix_proofs_prop}

\subsection{\texorpdfstring{Proof of Proposition \ref{prop:phi1}: Field isomorphism of $\varphi$}{Proof of Proposition phi1: Field isomorphism of phi}}\label{pf:phi1}

We first show that the map $\varphi:\CC\rightarrow\Mb$ is a field isomorphism. Clearly $\CC$ is a field. Also, $\Mb$ is a field with the following additive identity (zero) and multiplicative identity (unit)
$$ \mathbf{0}_{\Mb} = \begin{bmatrix}  
0 & 0\\0 & 0
\end{bmatrix},~~~ 
\mathbf{1}_{\Mb} = \begin{bmatrix}
    1 & 0\\
    0 & 1
\end{bmatrix}
$$
and for any $A = \begin{bmatrix}
    a & -b\\ b & a
\end{bmatrix}$, where both $a$ and $b$ are not zero together, the inverse element of the field is
$$ A^{-1} = \frac{1}{a^2 + b^2}\begin{bmatrix}
    a & b\\-b & a
\end{bmatrix} \in \mathcal{M} $$
Next we consider the map $\varphi$. For any $z, z' \in \CC$, $\re(z + z') = \re(z) + \re(z')$, $\im(z + z') = \im(z) + \im(z')$. Hence,
$$ \varphi(z + z') = \begin{bmatrix}
    \re(z + z') & -\im(z + z')\\
    \im(z + z') & \re(z + z')
\end{bmatrix} = \varphi(z) + \varphi(z') $$
Let $z = x+\i y$ and $z' = x' + \i y'$. Therefore,
\begin{align*}
    \varphi(zz') & = \varphi((x+\i y)(x' + \i y'))\\
    & = \varphi((xx' - yy') + \i(xy' + x'y))\\
    & = \begin{bmatrix}
        xx' - yy' & -xy' - x'y\\
        xy' + x'y & xx' - yy'
    \end{bmatrix}\\
    & = \begin{bmatrix}
        x & -y\\ y & x
    \end{bmatrix} \ 
    \begin{bmatrix}
        x' & -y'\\ y' & x'
    \end{bmatrix}\\
    & = \varphi(z) \varphi(z')
\end{align*}
Hence, the map $\varphi$ is an isomorphism. Also note that $\varphi(0) = \mathbf{0}_{\Mb}$ and $\varphi(1) = \mathbf{1}_{\Mb}$. In addition for any $z\ne 0$, $\varphi(z^{-1}) = [\varphi(z)]^{-1}$. Hence, $\varphi$ is a field isomorphism between $\CC$ and $\Mb$.

For any $z = x+\i y\in \CC$, $\varphi_1(z) = (x, y)^\top$ and $\varphi_2(z) = (-y, x)^\top$. Hence,

$$ \varphi_1(z)^\top \varphi_2(z) = x(-y) + yx = 0 $$
Also
$$ \|\varphi_1(z)\|^2 = \|\varphi_2(z)\|^2 = x^2 + y^2 = |z|^2$$
$$ \varphi(z^\dagger) = \varphi(x - \i y) = \begin{bmatrix}
    x & y\\
    -y & x
\end{bmatrix} = \varphi(z)^\top $$
Hence properties (a)-(c) are proved. $\hfil\square$

\subsection{Proof of Proposition \ref{prop:phi2}: Realification of complex vectors}\label{pf:phi2}

It can be shown that both $\CC^n$ and $\RR^{2n\times 2}$ are rings under addition `$+$' and entry-wise Hadamard product `$\odot$' of $n$-dimensional complex-valued vectors and $2n\times 2$-dimensional real matrices respectively, for any positive integer $n$ \citep{herstein1991topics}. We consider the extended map $\varphi:\CC^n\rightarrow \RR^{2n\times 2}$ as described in \eqref{eq:phi-extn-1}. For any $\zb, \zb' \in \CC^n$ such that $\zb = (z_1, \ldots, z_n)^\top, \zb' = (z_1', \ldots, z_n')^\top$, and $i \in [n]$
$$ \varphi(\zb + \zb)_i = \varphi(z_i + z_i') = \varphi(z_i) + \varphi(z_i') = (\varphi(\zb) + \varphi(\zb'))_i, $$
and
$$ \varphi(\zb \odot \zb')_i = \varphi(z_i z_i') = \varphi(z_i)\varphi(z_i') = (\varphi(\zb) \odot \varphi(\zb'))_i. $$

Therefore the extended $\varphi:\CC^n\rightarrow \RR^{2n\times 2}$ of \eqref{eq:phi-extn-1} is a ring homomorphism. By \citet[Proposition 5, Section 7.3]{dummit2004abstract}, $\mathrm{Image}(\varphi)$ is a subring of $\RR^{2n\times 2}$, and the kernel of $\varphi$, denoted by $\text{ker}(\varphi) = \{\zb: \varphi(\zb) = \mathbf{0}_{2n\times 2}\}$, is a subring of $\CC^n$. In this case, $\text{ker}(\varphi)$ consists of all those complex numbers $z$ such that $\re(\zb) = 0,\ \im(\zb) = 0$, i.e. $\zb = 0$. Hence $\varphi$ has a trivial kernel and hence it is a one-to-one map. Thus $\varphi$ in \eqref{eq:phi-extn-1} is a ring isomorphism.

Also we use Proposition \ref{prop:phi1} to get
$$ \varphi_1(\zb)^\top \varphi_2(\zb') = \sum_{i = 1}^n \varphi_1(z_i) \varphi_1(z_i') = 0. $$
Additionally we have
$$ \|\varphi_1(\zb)\|_2^2 = \sum_{i=1}^n \|\varphi_1(z_i)\|_2^2 = \sum_{i=1}^n \frac{1}{2}\|\varphi(z_i)\|_F^2 = \frac{1}{2}\sum_{i=1}^n \|\varphi(z_i)\|_F^2 = \frac{1}{2}\|\varphi(\zb)\|_F^2,  $$
that proves (a). Additionally,
$$ \|\varphi_1(\zb)\|_2^2 = \sum_{i=1}^n \|\varphi_1(z_i)\|_2^2 = \sum_{i=1}^n |z_i|^2 = \|\zb\|_2^2.  $$
Similarly we can show the properties for $\varphi_2$ and the proof of (b) is complete. \pfend

\subsection{Proof of Proposition \ref{prop:phi3}: Realification of complex matrices}\label{pf:phi3}

Since $\CC^{m\times n}$ is a ring under $(+, \odot)$, the proof of the fact that $\varphi$ in \eqref{eq:phi-extn-2} is a ring isomorphism, is similar to Proposition \ref{prop:phi2}. To show (a)-(e), we use . For $Z \in \CC^{m \times n}$, we have $\varphi_1(Z_{\cdot, j})^\top \varphi_2(Z_{\cdot, j}) = 0$ by Proposition \ref{prop:phi2}(a). Additionally, by Proposition \ref{prop:phi2}(b)
$$ \|\varphi(Z)\|_F^2 = \sum_{j = 1}^n \|\varphi(Z_{\cdot, j})\|_F^2 = 2\sum_{j=1}^n \|Z_{\cdot, j}\|_2^2 = 2\|Z\|_F^2. $$
For proving (c), we consider any unit vector $v \in \CC^n$. By Proposition \ref{prop:phi2}(b), $\|\varphi_1(v)\|_2 = \|v\|_2 = 1$ and we can write
\begin{align*}
    & \varphi(Z) \varphi_1(v) = \varphi_1(Zv.\\
    \implies~ &~ \|\varphi(Z) \varphi_1(v)\|_2 = \|\varphi_1(Zv)\|_2 = \|Zv\|_2.
\end{align*}
Taking suprema on both sides with respect to $v \in \CC^n$, we prove (c). Next by Proposition \ref{prop:phi1}, we have
$$ \varphi(Z^\dagger) = (\varphi(\overline{Z}_{i, j}))_{i,j} = (\varphi(Z_{j,i})^\top)_{i,j} = \varphi(Z)^\top. $$
Since $\text{GL}_n(\CC)$ is a field, by the homomorphic properties of $\varphi$ and the field properties of $\text{GL}_n(\CC)$, $\mathrm{Image}(\varphi)$ is also a field and $\varphi$ is a field isomorphism. Thus for any $Z\in GL_n(\CC)$, 
$$\varphi(Z) \varphi(Z^{-1}) = \varphi(Z Z^{-1}) = \varphi(I_n) = I_{2n}.$$ 
Therefore, $\varphi(Z^{-1}) = [\varphi(Z)]^{-1}$, and the proof of (e) is complete. \pfend

\subsection{Proof of Proposition \ref{prop:nwr_approx_oracle}: Linear approximation with the oracle}\label{pf:nwr_approx_oracle}

We denote the following $(p-1)$-dimensional vector: 
\begin{equation}\label{eq:bja_tilde}
    \bjatilde := \left( - \frac{(\thetajstar)_{1,a}}{(\thetajstar)_{a,a}},\ldots, -\frac{(\thetajstar)_{a-1, a}}{(\thetajstar)_{a,a}},  -\frac{(\thetajstar)_{a+1,a}}{(\thetajstar)_{a,a}},\ldots, -\frac{(\thetajstar)_{p,a}}{(\thetajstar)_{a,a}} \right)^\top.
\end{equation}
Then $(\bjatilde)_{S_a\cmplt} = 0$, and $|\{b \in [p]\setminus \{a\}: (\bjatilde)_a \ne 0\}| \le s_a$. From \eqref{eq:oracle_beta} we have   
\begin{align}
    \frac{\|\Zma \bjastar - \muja\|_2^2}{2m+1}~ &~ \le \frac{\|\Zma \bjatilde - \muja\|_2^2}{2m+1} \nonumber \\
    &~ = \sum_{k \in \wmj} \frac{|(\dkma)^\dagger (\bjatilde - \bka)|^2}{2m+1}. \label{eq:oracle_pred_upper_bound}
\end{align}
We denote $ \vka:= \bjatilde - \bka$. Each summand in \eqref{eq:oracle_pred_upper_bound} is $|{\dkma}^\dagger \vka|^2$. We denote $I^{-(a)}$ as the $(p-1)\times p$ dimensional matrix that has every row of $I_{p\times p}$ except the $a$\textsuperscript{th} row. An immediate consequence is that $\dkma = I^{-(a)} d_k$. Therefore we have,
$$ {\dkma}^\dagger \vka = d_k^\dagger {I^{-(a)}}^\top \vka, $$
and 
$$ \frac{\|\Zma \bjastar - \muja\|_2^2}{2m+1} \le \sum_{k \in \wmj} \frac{|d_k^\dagger {I^{-(a)}}^\top \vka |^2}{2m+1}. $$
We denote $\delta_{m,n} := 6\M^5\vmnf$. By Lemma \ref{lem:onk_theta_approx}, in the given regime of $m, n$,
$$\|\vka\|_2^2 \le \delta_{m,n}^2 ~~\implies~~ \|{I^{-(a)}}^\top \vka\|_2^2 \le \delta_{m,n}^2. $$
We apply Lemma \ref{lem:dk_quad_concentration}. For any $R>0$, there exists a constant $c > 0$ such that with probability greater than $1 - 4\exp(-c(2m+1)\min\{A, A^2/2\})$,
$$ \frac{\|\Zma \bjastar - \muja\|_2^2}{2m+1} \le \delta_{m,n}^2\left(\M(A+1) + \tnf  + \frac{1}{n}\lf\right). $$
Since $\tnf \le \frac{1}{4\M}$ and $\lf/n \le \frac{1}{4m\M}\le \frac{1}{4\M}$, we have
$$ \frac{\|\Zma \bjastar - \muja\|_2^2}{2m+1} \le \left(A + \frac{3}{2}\right)\M\delta_{m,n}^2 $$
Hence by choice of $\delta_{m,n}$, the proof follows.
\pfend

We state the lemmata used for proving Proposition \ref{prop:nwr_approx_oracle}.

\begin{lemma}[Concentration of quadratic forms of DFT]\label{lem:dk_quad_concentration}
    Consider a sequence of vectors $\{v_k\}_{k \in \wmj}$ in $\CC^p$ such that $\|v_k\|_2 \le \delta_{m,n}$ for some $\delta_{m,n} > 0$. Then for $R>0$, there exists a universal constant $c > 0$ such that with probability greater than $1 - 4 \exp\left(- c(2m+1) \min\{A, A^2\}\right)$,
    $$ \sum_{k \in \wmj} \frac{|d_k^\dagger v_k|^2}{2m+1} \le \delta_{m,n}^2\left(\M(A + 1) + \tnf + \frac{1}{n}\lf\right). $$
\end{lemma}

\begin{lemma}[Approximation of $\bka$ with $\bjatilde$]\label{lem:onk_theta_approx}
    Consider the DFT-NWR setup \eqref{eq:nodewise-reg}. For any $a \in [p]$, denote $\bjatilde$ as in \eqref{eq:bja_tilde}, and $\bka$ as in \eqref{eq:nwr_true_beta} for $k \in \wmj$. Then for $\tnf \le \frac{1}{4
    \M}$ and $m \le n/(4\M\lf)$, the following holds,
    $$ \|\bjatilde - \bka\|_2 \le 6 \M^5 \vmnf. $$
\end{lemma}

We now provide with the proofs of the necessary lemmata for Proposition \ref{prop:nwr_approx_oracle}. In the proofs, we use an alternative representation of the DFT as 
\begin{equation}\label{eq:dft_sin_cosine}
    d_j \equiv \frac{1}{\sqrt{2\pi}} \X^\top\Ec_k,
\end{equation}
where $\Ec_k := C_k - \i S_k$ and
\begin{align}
    C_k := & \frac{1}{\sqrt{n}} (1, \cos \omega_k, \ldots, \cos(n-1)\omega_k)^\top,~~ \label{eq:Ck}\\
    S_k := & \frac{1}{\sqrt{n}} (1, \sin \omega_k, \ldots, \sin(n-1)\omega_k)^\top. \label{eq:Sk}
\end{align}

\subsubsection{Proof of Lemma \ref{lem:dk_quad_concentration}: Concentration of quadratic forms of DFT}

Throughout this proof, we use the following notation: for any two matrices $A$ and $B$ of same dimensions, we write $A \peq B$ if they are equal up to permutation of rows and columns.

We simplify \eqref{eq:dft_sin_cosine} to obtain the following
\begin{align*}
    d_k^\dagger v_k =~ & \frac{1}{\sqrt{2\pi}}(C_k + \i S_k)^\top \X (\vkr + \i \vki)\\ 
    =~ & \frac{1}{\sqrt{2\pi}} (C_k^\top \X \vkr - S_k^\top \X \vki) + \i(C_k^\top \X \vki + S_k^\top \X \vkr).
\end{align*}
Therefore,
$$ |d_k^\dagger v_k|^2 = \frac{1}{2\pi}(C_k^\top \X \vkr - S_k^\top \X \vki)^2 + (C_k^\top \X \vki + S_k^\top \X \vkr)^2. $$
Using the identity $\vecop{ABC} = (C^\top \otimes A) \vecop{B}$, we get the following:
\begin{align*}
    C_k^\top \X \vkr - S_k^\top \X \vki = & \left(\vkr \otimes C_k - \vki \otimes S_k \right)^\top \vecop{\X}\\
    C_k^\top \X \vki + S_k^\top \X \vkr = & \left(\vki \otimes C_k + \vkr \otimes S_k \right)^\top \vecop{\X}\\
\end{align*}
Therefore, $|d_k^\dagger v_k|^2 = \frac{1}{2\pi} \vecop{\X}^\top \left(\xi_k \xi_k^\top\right) \vecop{\X}$, where 
\begin{align*}
    \xi_k := & \left[\vkr \otimes C_k - \vki \otimes S_k ~~~~ \vki \otimes C_k + \vkr \otimes S_k \right]\\
    = & \left[ (I_p\otimes C_k) \vkr - (I_p \otimes S_k) \vki ~~~~ (I_p\otimes C_k) \vki + (I_p\otimes S_k) \vkr \right]\\
    = & \left[ I_p\otimes C_k \quad I_p\otimes S_k \right] \begin{bmatrix}
        \vkr & \vki\\
        -\vki & \vkr
    \end{bmatrix}\\
    ~\peq~ & \left( I_p \otimes \left[ C_k ~~~ S_k \right] \right)~ \varphi\left(\overline{v_k}\right)
\end{align*}
And the underlying quadratic form is
$$\sum_{k \in \wmj} |d_k^\dagger v_k|^2 = \frac{1}{2\pi} \vecop{\X}^\top \left(E_{j,m} E_{j,m}^\top\right) \vecop{\X},$$ 
where 
$$E_{j,m}^\top := \left[ \xi_{j-m}^\top: \ldots :\xi_{j+m}^\top  \right]_{np \times (4m+2))}.$$
Since $E_{j,m}$ is a submatrix of $E_n$ defined in \eqref{eq:E_n}, and thus by Lemma \eqref{lem:kronecker_norm_boud}, 
$$\|E_{j,m} E_{j,m}^\top\|\le \|E_{j,m}\|^2 \le \|E_n\|^2 \le \delta_{m,n}^2.$$
Also, $\rank(E_{j,m} E_{j,m}^\top) \le \rank(E_{j,m}) \le 4m+2$. Additionally,
$$ \EE[|d_k^\dagger v_k|^2] = v_k^\dagger \Snk v_k \le \|v_k\|_2^2 \|\Snk\| \le \delta_{m,n}^2 \left(\M + \tnf + \frac{1}{n}\lf \right). $$
We use Lemma \ref{lem:hanson_wright_gaussian} with $\eta = A\delta_{m,n}^2$ and $A > 0$. Thus we obtain the following
\begin{align*}
    & \PP\left( \frac{1}{2m+1}\sum_{k \in \wmj} |d_k^\dagger v_k|^2 > 
    \delta_{m,n}^2\left( \M(A+1) + \tnf + \frac{1}{n}\lf \right) \right) \\
    = &\PP\left(\frac{1}{2m+1} \sum_{k \in \wmj} |d_k^\dagger v_k|^2 - \delta_{m,n}^2 \left(\M + \tnf + \frac{1}{n}\lf \right) > A\M \delta_{m,n}^2 \right)\\
    \le &~ \PP\left( \left|\sum_{k \in \wmj} |d_k^\dagger v_k|^2 - \EE\bigg[\sum_{k \in \wmj} |d_k^\dagger v_k|^2\bigg]\right| > \M \eta (2m+1)  \right)\\
    = & ~ \PP\left( \left|\vecop{\X}^\top \left(E_{j,m} E_{j,m}^\top\right) \vecop{\X} - \EE\left[\vecop{\X}^\top \left(E_{j,m} E_{j,m}^\top\right) \vecop{\X}\right]\right| > 2\pi \M \eta (2m+1)) \right)\\
    \le & ~ 4\exp\left[-c \min\left\{\frac{\eta(2m+1)}{\|E_{j,m} E_{j,m}^\top\|},  \frac{\eta^2 (2m+1)^2}{\rank(E_{j,m} E_{j,m}^\top)\|E_{j,m} E_{j,m}^\top\|^2}\right\}\right]\\
    \le & ~ 4\exp \left[ -c \min\left\{ \frac{A\delta_{m,n}^2 (2m+1)}{\delta_{m,n}^2}, \frac{A^2\delta_{m,n}^4 (2m+1)^2}{(4m+2)  \delta_{m,n}^4} \right\}\right]\\
    \le &~ 4\exp\left[ -c'(2m+1)\min\{A, A^2\} \right],
\end{align*}
hence the proof is done.
\pfend

\subsubsection{Proof of Lemma \ref{lem:onk_theta_approx}: Approximation of $\bka$ with $\bjatilde$} \label{pf:onk_theta_approx}

We write the following for any $k \in \wmj$:
\begin{align*}
    \|\bjatilde - \bka\|_2  = & \left\| \frac{(\thetajstar)_{a, \cdot}}{(\thetajstar)_{a, a}} - \frac{(\Onk)_{a, \cdot}}{(\Onk)_{a, a}} \right\|_2.
\end{align*} 
By Lemma \ref{lem:dft_cov_approx}, we have $\|f_j^* - \Snk\| \le \vmnf $. Since $\tnf \le \frac{1}{4\M}$ and $m \le n/(4\M\lf)$, we have $\vmnf \le \frac{1}{2\M}$. We apply \eqref{eq:spectral_difference_inv4} in Lemma \ref{lem:spectral_difference_inv} to get
$$ \|\bjatilde - \bka\|_2 \le 6\M^5 \vmnf,$$
and the proof is complete. \pfend

\subsection{Proof of Proposition \ref{prop:re}: Restricted eigenvalue (RE)}\label{pf:re}

For any $v \in \CC^{p-1}$ and $p > 1$, we denote 
$v' := (v_1,\ldots, v_{a-1}, 0, v_a,\ldots, v_{p-1})^\top$ with $v_a' = 0$. We have $\|v'\|^2 = \|v\|^2$. Additionally, let $S_a = \{i_1,\ldots, i_{s_a}\} \subset [p]$. We define a $j$-augmented set of non-zero entries as $S_a' = \{i_1',\ldots, i_{s_a}'\}$, and for $k \in [s_a]$,
$$ i_k' = \begin{cases}
    i_k & \text{ if } i_k < a,\\
    i_k + 1 & \text{ otherwise}.
\end{cases} $$
Thus, $|S_a'| = |S_a| = s_a$, and for any $v \in \Cscr_{p-1}(S_a,3)$, we have $v_j \in \Cscr_{p}(S_a',3)$. Without loss of generality, we assume $\|v\|_2 = 1$ since $v\in \Cscr_{p-1}(S_a,3)\setminus \{0\}$ if and only if $v/\|v\|_2 \in \Cscr_{p}(S_a,3)\setminus\{0\}$ and $v' \in \Cscr_{p-1}(S_a',3)\setminus\{0\}$. Therefore,
$$ \inf_{\substack{v\in \Cscr_{p-1}(S_a,3),\\ \|v\|_2 = 1}}\|\Zma v\|_2^2 = \inf_{\substack{v'\in \Cscr_p(S_a',3): v_a' = 0,\\ \|v'\|_2=1}} \|\Zcal v'\|_2^2 \ge \inf_{\substack{v\in \Cscr_{p}(S_a',3),\\ \|v\|_2=1}}\|\Zcal v\|_2^2. $$
Hence proving RE condition of $\Zcal$ over $\Cscr_p(S_a', 3)$ is sufficient. We have,
$$ \inf_{\substack{v\in \Cscr_p(S_a', 3), \|v\|_2 =1}} v^\dagger f_j^* v \ge \frac{1}{\M} > 0. $$
It remains to show that $v^\dagger (\hat f_j - f_j^*) v$ is sufficiently small uniformly over $\Cscr_p(S_a', 3)$. From \eqref{eq:avg_pdgram_deviation} in Lemma \ref{lem:single_deviation_bound}, under the conditions of Lemma \ref{prop:re}, the following holds with probability greater than $1 - c'  \exp\{-c(2m+1) \min\{\eta, \eta^2\}\}$:
$$ |v^\dagger (\hat f_j - \EE [\hat f_j]) v| \le \M\eta. $$
For any positive integer $s$, we denote the set of $k$-sparse unit vectors as $\Kscr(s) := \BB_0(s) \cap \BB_2(1)$. From Lemma \ref{lem:cone_to_sparse} and \ref{lem:sparse_set_quad}, weget the following
\begin{align*}
    \sup_{\substack{v\in \Cscr_p(S_a', 3), \|v\|_2 = 1}} |v^\dagger (\hat f_j - \EE [\hat f_j]) v| \le & ~ \sup_{v\in 5~ \text{cl}(\text{conv}\{\Kscr(s_a)\})} |v^\dagger (\hat f_j - \EE [\hat f_j]) v|\\
    \le & ~25\sup_{v\in \text{cl}(\text{conv}\{\Kscr(s_a)\})} |v^\dagger (\hat f_j - \EE [\hat f_j]) v|\\
    \le &~ 75 \sup_{v\in \Kscr(2s_a)} |v^\dagger (\hat f_j - \EE [\hat f_j]) v|.
\end{align*}
In addition, Lemma \ref{lem:conc_sparse_set} implies,
\begin{align}
    & \PP(\sup_{v\in \Kscr(2s_a)} |v^\dagger (\hat f_j - \EE [\hat f_j]) v| > \M\eta)\nonumber\\
    \le~ & c' \exp[-c(2m+1)\min\{\eta, \eta^2\} + 2s_a \min\{ \log p, \log(21 ep/2s_a) \} ]. \label{eq:re_tail_decay}
\end{align}
Therefore \eqref{eq:avg_periodogram_bias} in Lemma \ref{lem:single_deviation_bound} implies,
\begin{equation}\label{eq:re_bias}
    \sup_{\substack{v\in \Cscr_p(S_a', 3), \|v\|_2 = 1}} |v^\dagger(\EE [\hat f_j] - f_j^*) v| \le~ \|\EE[\hat f_j] - f_j^*\|_\infty \le \vmnf.
\end{equation}
Denote $\Delta := 75\M\eta + \vmnf$. Combining \eqref{eq:re_tail_decay} and \eqref{eq:re_bias},
\begin{align*}
    & \PP\bigg(\sup_{\substack{v\in \Cscr_p(S_a', 3)\\ \|v\|_2 = 1}} |v^\dagger (\hat f_j -  f_j^*) v| \ge \Delta\bigg)\\
    \le~ & c' \exp[-c(2m+1)\min\{\eta, \eta^2\} + 2s_a \min\{ \log p, \log(21 ep/2s_a) \}].
\end{align*}
We use $\min\{\eta, \eta^2\} \ge \min\{1,\eta^2\}$ and set $\eta = (150 \M^2)^{-1}$, $\rho = 1/\eta$. With the given regimes
$$\Delta = \frac{1}{2\M} + \vmnf \le \frac{1}{\M},$$
Also we have $m \succsim \max\{1, \rho^2\} \min\{s_a\log (21e p/s_a), s_a\log p)\}$. Therefore,
$$ \PP\bigg(\sup_{\substack{v\in \Cscr_p(S_a', 3)\\ \|v\|_2 = 1}} |v^\dagger (\hat f_j -  f_j^*) v| \ge \M^{-1}\bigg)\le  c'\exp[-c(2m+1)\min\{1, \tau^{-2}\}],$$
and the proof follows. \pfend

We state the necessary lemma for proving Proposition \ref{prop:re}.

\begin{lemma}[Approximation of nearby DFT covariance with spectral density]\label{lem:dft_cov_approx}
The DFT covariance matrix $\Snk$ in \eqref{eq:Snk} satisfies 
\begin{equation}\label{eq:Snk_f_diff}
    \|\Snk - f_k^*\|\le \tnf + \frac{1}{4\pi n}\lf.
\end{equation}
Under Assumption \ref{asn:summable}, 
\begin{equation}\label{eq:truncation_bias}
    \|\Snk - f_j^*\| \le \vmnf.
\end{equation}
\end{lemma}

Lemma \ref{lem:dft_cov_approx} is essential in quantifying the covariance matrix of the DFTs with the spectral density at the central frequency $\omega_j$. Under Assumption \ref{asn:summable} the spectral density is bounded and attains smoothness properties, hence the finite sample truncation and averaging bias can be controlled.

\begin{lemma}[Element-wise deviation bound on averaged periodograms for Gaussian processes]\label{lem:single_deviation_bound}
Let $\{X_t\}_{t\in[n]}$ be observations from a stable Gaussian centered time series satisfying Assumption \ref{asn:summable}. Then the bias of averaged periodogram is controlled as
\begin{equation}\label{eq:avg_periodogram_bias}
    \| \EE[\hat f_j] - f_j^* \|_\infty \le \vmnf.
\end{equation}
For $\|v\|_2 \le 1$, there exists $\eta > 0$ such that $\hat f_j$ in \eqref{eq:avg_periodogram} satisfies
\begin{equation}\label{eq:avg_pdgram_deviation}
    \PP\left(|v^\dagger (\hat f_j -\EE[\hat f_j]) v |\geq \M \eta\right)\leq c' \exp\{-c(2m+1)\min\{\eta,\eta^2\}\}.
\end{equation}
Furthermore, for any $a, b \in [p]$, the following holds
\begin{equation}\label{eq:avg_pdgram_deviation_entrywise}
    \PP\left(|(\hat f_j)_{a,b}-\EE[(\hat f_j)_{a,b}] |\geq \M \eta\right)\leq c' \exp\{-c(2m+1)\min\{\eta,\eta^2\}\}.
\end{equation}
In addition, assume that the sample size satisfies $$n \succsim \lf \M^3 \log p, \quad \text{and}~~ \tnf \le 1/(4\M).$$
Then for bandwidth $m$ satisfying $m \succsim\M^2 \log p$ and $m \le n/(4\M\lf)$, $A>0$ and $\thres$ in \eqref{eq:threshold}, there exist constants $c, c' > 0$ such that

\begin{equation}\label{eq:single_deviation_max}
    \PP\left(\big\|\hat f_j - f_j^* \big\|_\infty \ge \thres\right) \le c' \exp(-c\log p)
\end{equation}
\end{lemma}

Lemma \ref{lem:single_deviation_bound} is a combined statement of \citet[Proposition 3.3 and Proposition 3.5]{sun2018large} and guarantees a non-asymptotic upper bound on on the entry-wise difference between the averaged periodogram estimator and the true spectral density. The single deviation upper bound is a key result for proving the main results viz. Proposition \ref{prop:re}, Theorem \ref{thm:consistency_cglasso} and Theorem \ref{thm:caglasso_consistency}. The error bounds in Lemma \ref{lem:single_deviation_bound} has two components: (a) $\M \sqrt{\log p/m}$ is analogous to the rate of convergence for covariance estimation with i.i.d samples \citep{bickel2008regularized, ravikumar2011high}, and (b) $\vmnf$ captures the approximation bias induced in the averaged periodograms.

\subsubsection{Proof of Lemma \ref{lem:dft_cov_approx}: Approximation of nearby DFT covariance with spectral density} \label{pf:dft_cov_approx}

From \eqref{eq:Snk}, we have
\begin{align*}
     \|f_k^* - \Snk\| \le~ & \left\|\frac{1}{2\pi}\sum_{h \in \ZZ} \Gamma(h) e^{-\i h \omega_k} + \frac{1}{2\pi} \sum_{|h| < n}\left( 1 - \frac{|h|}{n} \right) \Gamma(h) e^{-\i h \omega_k} \right\|\nonumber\\
     \le~ & \frac{1}{2\pi} \sum_{|h|> n}\|\Gamma(h)\| + \frac{1}{2\pi} \sum_{|h|\le n} \frac{|h|}{n}\|\Gamma(h)\|\\
     \le~ & \frac{1}{2\pi} \sum_{|h|> n}\|\Gamma(h)\| + \frac{1}{4\pi n}\cdot 2 \sum_{h=0}^n h\|\Gamma(h)\|\\
     \le~ & \tnf + \frac{1}{4\pi n}\lf.
\end{align*}
Additionally by Lemma \ref{lem:lip_smooth}, we get the following
\begin{align*}
    \|f_j^* - \Snk\| \le & \|f^*_k - \Snk\| + \|f_j^* - f^*_k\| \\
    \le & \tnf + \frac{1}{4\pi n} \lf + \lf\frac{|k-j|}{2n} \\
    \le & \tnf + \lf \frac{m+\frac{1}{2\pi}}{2n} \le \vmnf.
\end{align*}
Hence the proof is complete. \pfend

\subsubsection{Proof of Lemma \ref{lem:single_deviation_bound}: Element-wise deviation bound on averaged periodograms for Gaussian processes}\label{pf:pre-main}

Similar to \citet[Proposition 3.3]{sun2018large}, the bound on the bias term is
\begin{align*}
    \| \EE[\hat f_j] - f_j^* \|_\infty ~
    \le~ & \left\|\frac{1}{2m+1} \sum_{k \in \wmj} \EE [I(\omega_k)] - f_j^*\right\|_\infty\\
    \le ~ & \frac{1}{2m+1} \sum_{k \in \wmj} \|\Snk - f_j^*\|\\
    \le~ & \vmnf,
\end{align*}
where the last step follows from Lemma \ref{lem:dft_cov_approx}. Hence \eqref{eq:avg_periodogram_bias} is proved.

By \citet[Prop 3.5]{sun2018large}, there exist constants $c,c'>0$ such that \eqref{eq:avg_pdgram_deviation} and \eqref{eq:avg_pdgram_deviation_entrywise} hold for any $\eta>0$, $\|v\|_2 \le 1$ and $a, b\in [p]$. We set $\eta = \sqrt{A\log p/m}$ for $A > 0$. In the regime  $m \succsim \M^2 \log p$, we have $\eta \ge \eta^2$. Taking a union bound over $a, b\in [p]$, we obtain the following:
\begin{align}
    &\PP\left(\|\hat f_j - \EE [\hat f_j]\|_\infty \geq \M \sqrt{\frac{A\log p}{m}}\right)\nonumber \\ & \leq p^2 \cdot  c' \exp\left\{ -c(2m+1)\frac{A\log p}{m} \right\}\nonumber\\
    & \le p^2c' \exp\{-2cA\log p\}
    = c'\exp[-(2cA - 2)\log p] \label{eq:taildecay_fhat}.
\end{align}
Combining \eqref{eq:taildecay_fhat} with \eqref{eq:avg_periodogram_bias} in the given regime, we complete the proof of \eqref{eq:single_deviation_max}. \pfend

\subsection{Proof of Proposition \ref{prop:deviation}: Deviation condition}\label{pf:deviation}

By \eqref{eq:nwr_model},
\begin{align*}
    \eka = \overline{\dka} - \big(\dkma\big)^\dagger \bka = \sum_{b = 1}^p \frac{(\Onk)_{a,b}}{(\Onk)_{a,a}} \overline{\dkb} = d_k^\dagger u_{k,a},
\end{align*}
where $u_{k,a}:= (\Onk)_{a, \cdot} /(\Onk)_{a,a}$. From \eqref{eq:dft_sin_cosine},
$$ \eka = u_{k,a}^\top \overline{d_k} = \frac{1}{\sqrt{2\pi}}u_{k,a}^\top \X^\top \Ec_k = \frac{1}{\sqrt{2\pi}}(u_{k,a}^\top \otimes \Ec_k^\dagger) \vecop{\X}. $$
In vector form, $\ea = (\varepsilon_{j-m}^{(a)}, \ldots, \varepsilon_{j+m}^{(a)})^\top = \frac{1}{\sqrt{2\pi}} J_1 \vecop{\X}$, where
\begin{equation}\label{eq:J1}
    J_1 := \begin{bmatrix}
    u_{j-m,a}^\top \otimes \Ec_{j-m}^\dagger\\
    \vdots\\
    u_{j+m,a}^\top \otimes \Ec_{j+m}^\dagger
\end{bmatrix}_{(2m+1) \times np}.
\end{equation}
Similarly for any $k \in \wmj$, $\dkma = \frac{1}{\sqrt{2\pi}}{I^{-(a)}}^\top \X^\top \Ec_k$, and for $b \ne a$
$$ \dkb = \frac{1}{\sqrt{2\pi}}\e_b^\top \X^\top \Ec_k =\frac{1}{\sqrt{2\pi}} \vecop{\X}^\top (\e_b \otimes \Ec_k). $$
Thus in matrix notation,
\begin{equation}\label{eq:dkb_vec}
    \begin{bmatrix}
    d_{j-m}^{(b)} &: \ldots :& d_{j+m}^{(b)}
    \end{bmatrix} = \frac{1}{\sqrt{2\pi}} \vecop{\X}^\top \begin{bmatrix}
    \e_b \otimes \Ec_{j-m} & :\ldots: & \e_b \otimes \Ec_{j+m}
\end{bmatrix}.
\end{equation}
We fix any $v \in \CC^{p-1}$ with $\|v\|_2 \le 1$. \eqref{eq:dkb_vec} implies, $ v^\dagger {\Zma}^\dagger = \frac{1}{\sqrt{2\pi}}\vecop{\X}^\top J_2$ where
$$ J_2 = \begin{bmatrix}
\overline{v^{(a)}} \otimes \Ec_{j-m} & :\ldots: & \overline{v^{(a)}} \otimes \Ec_{j+m}
\end{bmatrix}_{np \times (2m+1)}, $$
and $(v^{(a)})_r := (v_1, \ldots, v_{a-1}, 0, v_a,\ldots, v_{p-1})^\top$. Hence,
$$ v^\dagger {\Zma}^\dagger \ea = \frac{1}{2\pi}\vecop{\X}^\top (J_2 J_1) \vecop{\X}. $$
\eqref{eq:exp_varnwr_err} implies $\EE[v^\dagger {\Zma}^\dagger \ea] = 0$. In addition, $\rank(J_2 J_1) \le 2m+1$ and
$$ J_1 J_2 = \dg(u_{j-m, a}^\top \overline{v^{(a)}}, \ldots, u_{j+m, a}^\top \overline{v^{(a)}}).$$
Thus,
\begin{align*}
    \|J_2 J_1\| = \|J_1 J_2\| \le~ & \max_{k\in \wmj} |u_{k, a}^\top \overline{v^{(a)}}|\\
    \le ~ & \max_{k \in \wmj} |{\bka}^\top v|\\
    \le~ & \max_{k \in \wmj}\|\bka\|_2 \le \sqrt{3}\M,
\end{align*}
where we use Lemma \ref{lem:bjatilde_bounded}. We apply Lemma \ref{lem:hanson_wright_gaussian} with $\eta>0$ as follows:
\begin{align*}
    &~~ \PP\left( \left|\frac{1}{2m+1} v^\dagger {\Zma}^\dagger \ea \right| > \M\eta\times  \sqrt{3}\M \right)\\ 
    \le & ~~\PP\left( \left| \vecop{\X}^\top J_2 J_1 \vecop{\X} - \EE[\vecop{\X}^\top J_2 J_1 \vecop{\X}] \right| > 2\pi\M \times \sqrt{3}\M (2m+1) \eta \right)\\
    \le & ~~ 4\exp\left[ - c \min\left\{ \frac{(2m+1) \sqrt{3}\M \eta}{\sqrt{3}\M}, \frac{(2m+1)^2 (\sqrt{3}\M)^2 \eta^2 }{(2m+1) (\sqrt{3}\M)^2} \right\} \right]\\
    \le & ~~ 4 \exp\left[ - c(2m+1) \min\{\eta, \eta^2\} \right].
\end{align*}
We set $v = \e_b$ for $b\in [p-1]$, and take union bound over $b$. We set $\eta = \sqrt{A\log p/(2m+1)}$ for $R > 0$. Since $m \succsim \log p$, we have $\eta ^2 \le \eta$ and 
$$ \PP\left( \left\|\frac{1}{2m+1} {\Zma}^\dagger \ea \right\|_\infty >  \M^2 \sqrt{\frac{3A\log p}{2m+1}} \right) \le 4\exp(- (cA - 1) \log p ). $$
Hence the proof is complete. \pfend

We use the following lemma in the proof of the deviation condition.

\begin{lemma}[Boundedness of NWR coefficients]\label{lem:bjatilde_bounded}
    Consider the DFT-NWR setup \ref{subsubsec:nwr_pop} in population with $\bjatilde$ as in \eqref{eq:bja_tilde} and $\bka$ as in \eqref{eq:nwr_true_beta}. Then under $\tnf \le \frac{1}{4\M}$ and $n \ge 2\M \lf$, (i) $\|\bjatilde\|_2 \le \M$, and (ii) $\|\bka\|_2 \le \sqrt{3}\M$.
\end{lemma}

If we form approximate coefficient vectors from the columns of $\thetajstar$, then that vector is bounded. Similarly, Lemma \ref{lem:bjatilde_bounded} estbalishes boundedness of the nodewise regression coefficients. The proofs of this necessary lemma is the following.

\subsubsection{Proof of Lemma \ref{lem:bjatilde_bounded}: Boundedness of NWR coefficients} \label{pf:bjatilde_bounded}

We use the shorthand $\Theta^* \equiv \thetajstar$ throughout the proof. We consider the partition of $\Theta$ \wrt\ $a\in [p]$ as follows:
$$ \Theta^* = \begin{bmatrix}
    \Theta^*_{a,a} & \Theta^*_{a, -a}\\
    \Theta^*_{-a, a} & \Theta^*_{-a, -a}
\end{bmatrix}. $$
Since $\Theta^*$ is positive definite and Hermitian, by the formula of Schur complement of $\Theta^*_{a,a}$, the matrix $\Theta^*_{-a, -a} - {\Theta^*_{a,a}}^{-1} \Theta^*_{-a, a}{\Theta^*_{-a, a}}^\dagger $ is positive definite. Thus,
$$ \|\Theta^*_{-a, a}\|_2^2 < \|\Theta^*_{-a, -a}\| \cdot \Theta^*_{a,a} \le \|\Theta^*\| \cdot \Theta^*_{a,a}.  $$
Therefore we have,
$$ \|\bjatilde\|_2^2 \le \frac{\|\Theta^*_{-a, a}\|^2}{{\Theta^*_{a,a}}^2} < \frac{\|\Theta^*\|}{\Theta^*_{a,a}} \le \M^2. $$
Hence part (i) is complete. By Lemma \ref{lem:spectral_difference_inv} part (i), \eqref{eq:Snk_f_diff} and the condition $\tnf \le \frac{1}{4\M}$ and $n \ge 2\M \lf$, we have
$$ \| \Onk \| \le \frac{1}{\frac{1}{\M} - \left(\tnf + \frac{1}{2n}\lf\right)} \le 2\M, $$
and $$ (\Onk)_{a,a} \ge \frac{1}{\|\Snk\|} \ge \frac{1}{\M + \left(\tnf + \frac{1}{2n}\lf\right)} \ge \frac{2}{3\M}. $$
Therefore a similar approach to part (i) implies,
$$ \|\bka\|_2^2 < \|\Onk\|/(\Onk)_{a,a} \le 3\M^2, $$
and the proof is complete.

\pfend

\section{Proofs of the theorems}\label{sec:appendix_proofs_thms}

\subsection{Proof of Theorem \ref{thm:consistency_cglasso}: Consistency of CGLASSO for Gaussian time series}\label{pf:consistency_cglasso}

We condition on $\mathscr{E}:= \left\{ \big\|\hat f_j^* -f_j^*\big\|_\infty \le \thres \right\}$ and denote 
$$r :=2\kt\left(\big\|\hat f_j^* -f_j^*\big\|_\infty + \lambda\right).$$
From the assumptions of Theorem \ref{thm:consistency_cglasso}
$$ r\le 2\kt\left(\thres + \frac{8}{\alpha}\thres\right)  = 2 \kt C_{\alpha} \thres.$$
Since $C_{\alpha} > 1$, and $\thres d \le 1/(6\M^3 \kt^2 C_{\alpha}^2)$, hence $\thres d \ \le 1/(6\M^3 \kt^2)$. Additionally, $\M, \kt \ge 1$ according to the assumptions of Theorem \ref{thm:consistency_cglasso}. By Lemma \ref{lem:dual_witness_bound}, we have
$$\|G\|_\infty = \|\tilde \Theta_j - \thetajstar\|_\infty \le r,$$
where $G := \tilde\Theta_j - \thetajstar$ and $\tilde\Theta_j$ is the primal-dual witness of the restricted problem in \eqref{eq:graphical-lasso-restricted}. The remainder of the first order Taylor expansion of $(\thetajstar + G)^{-1}$ around $\thetajstar$ is denoted by 
$${\cal R}(G) := (\thetajstar + G)^{-1} - \thetajstar + {\thetajstar}^{-1}G {\thetajstar}^{-1}.$$
Using Lemma \ref{lem:dual_witness_bound} and \ref{lem:remainder_bound}, we have
\begin{align*}
    \|{\cal R}(G)\|_\infty \le \frac{3}{2}d\M^3 r^2 \le & \frac{3}{2}d\M^3 \cdot 4 \kt^2 C_\alpha^2 \thres^2\\
    \le & 6 d \M^3 \kt^2 C_\alpha^2\thres \cdot \frac{\alpha \lambda}{8}\\ \le & \frac{\alpha \lambda}{8}.
\end{align*}
Therefore we have
$$ \max\{\big\|\hat f_j^* -f_j^*\big\|_\infty, \|{\cal R}(G)\|_\infty\} \le \frac{\alpha \lambda}{8}. $$
Lemma \ref{lem:strict_duality} implies successful construction of the primal-dual witness with $\tilde \Theta_j = \hat\Theta_j$, ensuring
$$\|\hat \Theta_j - \thetajstar\|_\infty = \|\tilde \Theta_j - \thetajstar \|_\infty \le r \le 2\kt C_{\alpha}\thres.$$
Lemma \ref{lem:single_deviation_bound} implies that under the conditions on $m$, $\PP(\mathscr{E})\ge 1 - c \exp[-(c'A-2)\log p]$. Hence the proof of (a) follows.

Primal-dual witness method and part (a) implies $(\hat\Theta_j)_{S\cmplt} = (\tilde\Theta_j)_{S\cmplt} = 0$. Thus $E(\hat{\Theta}_j) = E(\tilde{\Theta}_j)$. Also for $(a,b)\in S$ such that $\left|(\thetajstar)_{a,b}\right| > 2C_{\alpha} \kt \thres$, we use triangle inequality
\begin{align*}
    |(\hat \Theta_j)_{a,b}| & = |(\hat \Theta_j)_{a,b} - (\thetajstar)_{a,b} + (\thetajstar)_{a,b}|
    \ge |(\thetajstar)_{a,b}| - |(\hat \Theta_j)_{a,b} - (\thetajstar)_{a,b}| > 0.
\end{align*}
Therefore $E(\hat\Theta)$ contains all edges $(a,b)$ with $|(\thetajstar)_{a,b}| > 2C_{\alpha} \kt \thres$. Hence (b) is proved. 

Similar to \citet[Corollary 3]{ravikumar2011high},
\begin{align*}
    \|\hat \Theta - \Theta^*\|_F &\le \Big(\sum_{a=1}^p|\hat\Theta_{a,a} - \Theta^*_{a,a}|^2~~ + \sum_{(a,b)\in E(\Theta^*)} |\hat\Theta_{a,b} - \Theta^*_{a,b}|^2\Big)^{1/2}\\
    & \le 2\kt C_{\alpha} \thres \sqrt{s+p},
\end{align*}
and
$$ \|\hat\Theta_j - \thetajstar\|_2 \le d \|\hat\Theta_j - \thetajstar\|_{\infty} \le 2(d+1)\kt C_{\alpha}\thres, $$
where we use the fact that the rows of $\hat\Theta_j$ and $\thetajstar$ both have at most $d$ non-zero entries. Next using the fact that Frobenius norm dominates $\ell_2$ norm, the proof of part (c) follows. \pfend

\subsubsection{Construction of the primal-dual witness}\label{subsubsec:primal_dual_witness}

The penalized log-determinant program \eqref{eq:graphical-lasso} restricted to the augmented edge set $S$ is:
\begin{equation}\label{eq:graphical-lasso-restricted}
    \tilde\Theta_j = \argmin_{\Theta\in \pdhermit:\ \Theta_{S\cmplt} = 0} - \log \det \Theta + \trace(\hat f_j^*\Theta) + \lambda \|\Theta^-\|_1.
\end{equation}
By construction, $\tilde\Theta$ is positive-definite Hermitian as well as $(\tilde\Theta_j)_{S\cmplt} = 0$. Let $\tilde \Psi \in \partial \|(\cdot)^-\|_1(\tilde\Theta)$. For $(a, b)\in S\cmplt$, we replace $(\tilde \Psi)_{a, b}$ with 
\begin{equation*}
    \tilde \Psi_{a, b} = \frac{1}{\lambda}\left[-(\hat f_j^*)_{a, b} + (\tilde \Theta_j^{-1})_{a, b} \right].
\end{equation*}
The pair $(\tilde\Theta_j, \tilde \Psi)$ satisfies the equation \eqref{eq:kkt} replacing $(\hat\Theta_j, \hat \Psi)$, and hence are called the \textit{primal and dual witnesses} respectively. We note that a necessary condition for the witness $(\tilde\Theta_j, \tilde\Psi)$ to be a feasible solution is that the \textit{strict dual feasibility condition} is satisfied, i.e. $|\tilde \Psi_{a, b}| < 1 \text{ for all } (a, b) \in S\cmplt$. Primal-dual witness methods are popular and useful tools for sparsistency guarantees for $\ell_1$-penalized convex programs \citep{wainwright2009sharp, meinshausen2006high, ravikumar2011high}. For the proof of Theorem \ref{thm:consistency_cglasso}, the first step shows that the strict dual feasibility condition holds with $(\tilde \Theta_j)_S = (\hat \Theta_j)_S$. The next step is to prove that the primal witness $\tilde \Theta_j$ exhibits sparsistency properties.

We now state the lemmata used in the proof \ref{pf:consistency_cglasso} and provide the proof sketches.

\begin{lemma}[Existence and uniqueness]\label{lem:existence_uniqueness}
The penalized log-determinant program \ref{eq:graphical-lasso} has a unique solution $\hat\Theta$ characterized by the KKT condition 
\begin{equation}\label{eq:kkt}
    \hat f_j^* - {\hat\Theta_j}^{-1} + \lambda \hat \Psi = 0,
\end{equation}
where $\hat \Psi$ is an element in the set of  subdifferentials of $\partial \|(\cdot)^{-}\|_1$ at $\hat\Theta_j$ as in \eqref{eq:sign}.
\end{lemma}

\begin{lemma}[Strict duality]\label{lem:strict_duality}
Suppose that the following holds:
$$\max\left\{\big\|\hat f_j^* -f_j^*\big\|_\infty, \|{\cal R}(G)\|_\infty \right\} \leq \frac{\alpha}{8}\lambda.$$
Then the vector $\tilde \Psi$ constructed in the primal dual witness construction \ref{eq:graphical-lasso-restricted} satisfies $\|\tilde\Psi_{S\cmplt}\|_{\infty} < 1$. Hence strict duality holds, and $\tilde \Theta_j = \hat \Theta_j$.
\end{lemma}

\begin{lemma}[Bound on the primal-dual witness]\label{lem:dual_witness_bound}
Suppose that
$$r = 2\kt\left(\big\|\hat f_j^* -f_j^*\big\|_\infty + \lambda\right) \le \min\left\{ \frac{1}{3\M d}, \frac{1}{3\M^3 \kt d} \right\}$$ 
holds. Then $\|G\|_\infty = \|\tilde\Theta_j - \thetajstar\|_\infty \le r $.
\end{lemma}

\begin{lemma}[Bound on the remainder]\label{lem:remainder_bound}
$\|G\|_\infty\le \frac{1}{3\M d}$ implies 
$$ \|{\cal R}(G)\|_\infty\le \frac32 d \|G\|_\infty^2 \M^3. $$
\end{lemma}

Next, we briefly sketch the proofs of the necessary lemmata.

\subsubsection{Proof of Lemma \ref{lem:existence_uniqueness}: Existence and uniqueness}\label{pf_existence_uniqueness}

The proof is inspired from the proof of \citet[Lemma 3]{ravikumar2011high}, with $\hat \Sigma$ replcaed by $\hat f_j^*$. \eqref{eq:graphical-lasso} is a strict convex program due to convexity of the off-diagonal $\ell_1$-penalty and the log-determinant barrier \citep[Section 3.1.5]{boyd2004convex}. Hadamard's inequality for Hermitian positive definite matrices \citep[Theorem 7.8.1]{horn2012matrix} imply $\log \det \Theta \le \sum_{a = 1}^p \log \Theta_{a, a}$, and 
$$ \sum_{a=1}^p \Theta_{a, a}(\hat f_j)_{a, a} - \log\det \Theta  \ge \sum_{a=1}^p \left(\Theta_{a, a}(\hat f_j)_{a, a} - \log\Theta_{kk} \right). $$
The right hand side is coercive in $(\Theta_{1,1}, \ldots, \Theta_{p,p})$. Hence the level sets of the objective function are compact, and continuity implies that a minimum is attained. In addition, strict convexity guarantees the uniqueness of the minimum.

A sufficient condition for $\hat{\Theta}_j$ to be a solution of \eqref{eq:graphical-lasso} is that the KKT condition 
$$0 \in \partial\left(- \log \det \Theta + \trace(\Theta\hat f_j^*) + \lambda \|\Theta^-\|_1\right)\big|_{\Theta = \hat\Theta_j}$$
is satisfied. Using \citet[Theorem 2.9, 2.10]{weylandt2020computational}, the KKT conditions to \eqref{eq:graphical-lasso} is the same as \eqref{eq:kkt}. \pfend

\subsubsection{Proof of Lemma \ref{lem:strict_duality}: Strict duality}\label{pf:strict_duality}

The proof is a straight-forward extension of \citet[Lemma 4]{ravikumar2011high} with the linear algebraic computations and norm inequalities performed in the space of complex Hermitian matrices.

\subsubsection{Proof of Lemma \ref{lem:dual_witness_bound}: Bound on the primal-dual witness}\label{pf:dual_witness_bound}

The proof follows the route of \citet[Lemma 6]{ravikumar2011high} and involves linear algebra of complex matrices. By Lemma \ref{lem:strict_duality}, the restricted problem \eqref{eq:graphical-lasso-restricted} has a unique solution. Upon taking partial derivative \wrt the restricted element, it vanishes at the unique optimum
\begin{equation}\label{eq:zero_gradient_condition}
H_1(\tilde \Theta_S)= 0~~ \text{ where, }H_1(\Theta_S) := [\hat f_j^*]_S - [\Theta^{-1}]_S + \lambda \tilde\Psi_S  ,  
\end{equation}

and $\Theta_S$ denotes the submatrix of $\Theta$ restricted to the augmented edge set $S$. This restricted zero gradient condition is necessary and sufficient for the optimal solution, hence the solution $\tilde \Theta_S$ is unique.

We define $\overline{G}$ as the vectorized form of $G$, and the map $H_2:\CC^{|S|}\rightarrow \CC^{|S|}$ as follows
\begin{equation*}\label{eq:defn_h2}
    H_2(\overline{G}_S) := -(\Upsilon_{SS}^*)^{-1}\ \overline{H}_1(\Theta_S^* + G_S) + \overline{G}_S.
\end{equation*}
From construction, $H_2(\overline{G}_S) = \overline{G}_S$, i.e. $H_1(\Theta_S^* + G_S) = H_1 (\tilde\Theta_S^*) = 0$. Following \citet[Equation 72 to 74]{ravikumar2011high}, 
$$H_2(\BB_\infty(r)) \subset \BB_\infty(r).$$ 
Since $H_2$ is a continuous function on $\CC^{|S|}$ and $\BB_\infty(r)$ is convex and compact in $\CC^{|S|}$, by \textit{Brouwere's fixed point theorem} \citep{kellogg1976constructive}, existence of a fixed point $\overline{G}_S\in \BB_\infty(r)$ of the function $H_2$ is guaranteed. Also uniqueness of the zero gradient condition \eqref{eq:zero_gradient_condition} implies the uniqueness of the fixed points of $H_2$. Therefore, we can conclude that $\|(\tilde\Theta_j)_S - (\thetajstar)_S\|_\infty \le r$, and hence the claim is proved. \pfend

\subsubsection{Proof of Lemma \ref{lem:remainder_bound}: Bound on the remainder}\label{pf:remainder_bound}

The proof follows from \citet[Lemma 5]{ravikumar2011high}, and involves a a power series representation of complex matrix-valued function $(\thetajstar + G)^{-1}$ as
$$(\thetajstar + G)^{-1} = \sum_{k = 0}^\infty ({\thetajstar}^{-1} G)^k {\thetajstar}^{-1} $$
The power series expansion can be verified as $\|G\|_\infty \le \frac{1}{3\M d}$ implies $\vertiii{{\thetajstar}^{-1}G}_\infty < 1$ (the radius of convergence condition, \citet[Equation 68, Appendix B]{ravikumar2011high}). From the results in holomorphic functional calculus involving the general power series expansion of an operator on Banach space \citep{dudley2011concrete, haase2018lectures}, the power series representation is valid. Rest of the proof is similar to \citet[Lemma 5]{ravikumar2011high}.

\subsection{Proof of Theorem \ref{thm:consistency_dft_nwr}: Consistency of DFT-NWR}\label{pf:consistency_dft_nwr}

We define two events: 
\begin{align}
    \Sscr_1 := & \left\{\mathrm{RE~ condition}~\eqref{eq:re_condition} ~ \mathrm{holds} \right\},\\
    \Sscr_2 := & \left\{\mathrm{Deviation~ condition}~ \eqref{eq:deviation_ineq} ~\mathrm{holds} \right\}.
\end{align}
On $\Sscr_1$ and $\Sscr_2$, we apply \citet[Theorem 6.2]{van2014asymptotically} to obtain
\begin{equation}\label{eq:consistency_conditioned}
    \frac{\|\Zma \bjahat - \muja\|_2^2}{2m+1} + \lambda_a \|\bjahat - \bjastar\|_1 \le \frac{48 \lambda_a^2 s_a}{\are} + \frac{3\|\Zma\bjastar - \muja\|_2^2}{2m+1}.
\end{equation}
The second term on the right side of \eqref{eq:consistency_conditioned} is associated with the linear approximation with the oracle. Proposition \ref{prop:nwr_approx_oracle} with $A = 1$ yields that the oracle approximation term is bounded by $270 \M^{11}\vmnf^2$ with probability greater than $1- 4 \exp(-c_0 (2m+1))$. Under the conditions of Theorem \ref{thm:consistency_dft_nwr} and by Proposition \ref{prop:re}, $\Sscr_1$ holds with probability greater than $1 - c_1 \exp[-c_2 (2m+1)/\M^4]$. Similarly, under the conditions of Theorem \ref{thm:consistency_dft_nwr} and by Proposition \ref{prop:deviation}, $\Sscr_2$ holds with probability greater than $1 - c_3 \exp(-(c_4 A - 1) \log p)$. Since $m \succsim \M^4 \log p$, \eqref{eq:consistency_conditioned} holds with probability greater than $1 - c_5 \exp(-(c_4 A - 1) \log p)$ and can be written as
$$ \frac{\|\Zma \bjahat - \muja\|_2^2}{2m+1} + \lambda_a \|\bjahat - \bjastar\|_1 \le 48 \M s_a\lambda_a^2 + 270 \M^{11}\vmnf^2. $$
Using the expressions of $\threse$ and $\thresa$ in \eqref{eq:threshold_nwr},
the proof is complete. \pfend

\subsection{Proof of Theorem \ref{thm:caglasso_consistency}: Estimation consistency of CAGLASSO} \label{pf:caglasso_consistency}

We combine the results from Theorem \ref{thm:consistency_cglasso} with Lemma \ref{lem:scaled_avg_periodogram}. Denote $\Dt := D_\tau^{-1} \otimes D_\tau^{-1}$. Similar to \eqref{eq:hessian}, the Hessian associated with $K_\tau$ in \eqref{eq:caglasso} is
$$ \Upsilon^*_\tau := R_\tau \otimes R_\tau = D_\tau^{-1} f_j^* D_\tau^{-1} \otimes D_\tau^{-1} f_j^* D_\tau^{-1} = \Dt \Upsilon^* \Dt. $$
Thus $\kt = \vertiii{(\Upsilon^*)_{S, S}}$ is translated for $\Upsilon_\tau^*$ to
$$ \kt' := \vertiii{(\Upsilon^*_\tau)_{S, S}}_\infty = \|(\Dt)_{S, S}\|_\infty^2 \vertiii{(\Upsilon^*)_{S, S}}_\infty \le \M^2 \kt,  $$
Similarly we obtain
$$ \kf' := \vertiii{D_\tau^{-1} f_j^* D_\tau^{-1}}_\infty \le \M \kf, $$
and we also have
\begin{align*}
    \max_{e\in S\cmplt}\|(\Upsilon_\tau^*)_{e, S} [(\Upsilon_\tau^*)_{S, S}]^{-1}\|_1 =~ & \max_{e\in S\cmplt} \|(\Dt)_{e, e} \Upsilon_{e, S}^* (\Upsilon_{S, S}^*)^{-1} (\Dt^{-1})_{S, S}\|_1\\
    \le~ & \max_{e\in S\cmplt} (\Dt)_{e, e} \|(\Dt^{-1})_{S, S}\|_\infty \|\Upsilon_{e, S}^* (\Upsilon_{S, S}^*)^{-1}\|_1 \\ \le~ & r_S (1-\alpha).
\end{align*}
Hence the incoherence assumption \ref{assumption:incoherence} is satisfied by $K_\tau$ for $\alpha \in (1 - r_S^{-1}, 1)$ and incoherence constant $\alpha' := 1 - r_S(1-\alpha)$. Additionally we can use Lemma \ref{lem:scaled_avg_periodogram} to bound the error of the scaled averaged periodogram i.e. $\|\hat R_\tau - R_\tau\|_\infty$. Hence similar to the proof of Theorem \ref{thm:consistency_cglasso} in Section \ref{pf:consistency_cglasso}, we condition on the event $\mathscr{E} := \{\|\hat R_\tau - R_\tau\|_\infty \le 4\M^{7/2}\thresr\}$ and get
$$ \|\hat K_\tau - K_\tau\|_\infty \le 2\kt' C_\alpha' (4\M^{7/2}\thresr) \le 8 \M^{11/2} \kt C_\alpha' \thresr. $$
From \eqref{eq:rtau_bound} in Lemma \ref{lem:scaled_avg_periodogram}, this error bound holds with probability greater than $1 - c_0 \exp(-(c_1A - 2) \log p)$. The error bound on $\thetajhattau$ is
\begin{align*}
    \|\hat \Theta_j^{(\tau)} - \thetajstar\|_\infty = & \|\hat D_\tau^{-1} \hat K_\tau \hat D_\tau^{-1} - D_\tau^{-1} K_\tau D_\tau^{-1} \|_\infty\\
    \le & \|\hat D_\tau^{-1} - D_\tau^{-1} \|_\infty \|K_\tau D_\tau^{-1}\|_\infty + 2\|D_\tau^{-1}\|_\infty^2 \|\hat K_\tau - K_\tau\|_\infty + \text{product terms}.
\end{align*}
Since $\|K_\tau D_\tau^{-1}\|_\infty = \max_{a,b\in [p]}\frac{(\thetajstar)_{a,b}}{\sqrt{(\thetajstar)_{a, a}}} \le \M\sqrt{\M}$ and $\|D_\tau^{-1}\|_\infty^2 \le \M$, combining these bounds with \eqref{eq:dtauinv_bound} in Lemma \ref{lem:scaled_avg_periodogram},
\begin{align*}
    \|\hat \Theta_j^{(\tau)} - \thetajstar\|_\infty \le & \M\sqrt{\M}\times \M^2 \thresr + \M \times 8 \M^{11/2} \kt C_\alpha' \thresr + \text{ product terms}\\
    \le & 10 \kt\M^{13/2} C_\alpha' \thresr.
\end{align*}
Since the product terms are negligible in order \wrt\ $\M_S \kt\M^{9/2} C_\alpha' \thresr$. The rest of the proof follows from the proof of parts (b), (c) and (d) of Theorem \ref{thm:consistency_cglasso}. \pfend

\begin{lemma}[Estimation error of scaled averaged periodogram]\label{lem:scaled_avg_periodogram}
Under the conditions of Theorem \ref{thm:caglasso_consistency}, there exists constants $c_i > 0$ such that the following holds,
\begin{equation}\label{eq:dtauinv_bound}
\PP\left(\|\hat D_\tau^{-1} - D_\tau^{-1}\|_\infty \ge \M^2 \thresr \right) \le c_0 \exp(-(c_1 A - 2) \log p).
\end{equation}
Furthermore, we have
\begin{equation}\label{eq:rtau_bound}
\PP\left(\|\hat R_\tau - R_\tau\|_\infty\ge 4\M^{7/2} \thresr \right) \le c_2 \exp(-(c_3 A - 2) \log p).
\end{equation}
\end{lemma}

Next, we sketch the proofs of the aforementioned lemmata.

\subsubsection{Proof of Lemma \ref{lem:scaled_avg_periodogram}: Estimation error of scaled averaged periodogram} \label{pf:scaled_avg_periodogram}

We have $\|D_\tau^{-1}\|_\infty = \max_{a\in [p]} \sqrt{|(\thetajstar)_{a,a}}| \le \sqrt{\M}$. By Corollary \ref{cor:consistency_partial_variance}, under the conditions of Theorem \ref{thm:caglasso_consistency},the following holds with probability greater than $1 - c_0 \exp(-(c_1A - 1) \log p)$,
$$ \tauahatsq \ge \frac{1}{(\thetajstar)_{a,a}} - \sqrt{\M}\thresr \ge \frac{1}{2\M} $$
and the following holds with the same probability,
\begin{align*}
    (\hat D_\tau^{-1} - D_\tau^{-1})_{a,a} = \left| \frac{1}{\hat \tau_a} - \sqrt{(\thetajstar)_{a,a}} \right|
    = \frac{\sqrt{(\thetajstar)_{a,a}}}{\hat\tau_a}\cdot \frac{\left| \tauahatsq - 1/(\thetajstar)_{a,a} \right|}{\left(\hat\tau_a + 1\big/\sqrt{(\thetajstar)_{a,a}}\right)} \le \M^2 \thresr.
\end{align*}
Union bound over $a\in [p]$ implies that
$$\|\hat D_\tau^{-1} - D_\tau^{-1}\|_\infty \le \M^2 \thresr. $$
with probability greater than $1 - c_0 \exp(-(c_1A - 2) \log p)$. In order to show that $\{\|\hat R_\tau - R_\tau \|_\infty \le \lambda_0\}$ holds with high probability, we write the following decomposition:
\begin{align*}
    & \hat R_\tau - R_\tau\\
    = & \hat D_\tau^{-1} \hat f_j \hat D_\tau^{-1} - D_\tau^{-1} f_j^* D_\tau^{-1}\\
    = & (\hat D_\tau^{-1} - D_\tau^{-1}) f_j^* D_\tau^{-1} + D_\tau^{-1}(\hat f_j - f_j^*)D_\tau^{-1} + D_\tau^{-1} f_j^* (\hat D_\tau^{-1} - D_\tau^{-1}) + U,
\end{align*}
where $U$ consists of higher order negligible terms containing products of $\hat D_\tau^{-1} - D_\tau^{-1}$ and $\hat f_j - f_j^*$. We use the following matrix properties: (i) for any two matrices $A, B$ of appropriate orders, $\|AB\|_\infty\le \vertiii{A}_\infty \|B\|_\infty$ and $\|AB\|_\infty\le \|A\|_\infty \vertiii{B}_1$, (ii) for any Hermitian matrix $A$, $\vertiii{A}_1 = \vertiii{A}_\infty$, and (iii) For any diagonal matrix $D$, $\vertiii{D}_1 = \vertiii{D}_\infty = \|D\|_\infty$. We have,
\begin{align*}
    & \|\hat R_\tau - R_\tau\|_\infty\\
    \le & \|D_\tau^{-1}\|_\infty^2 \|\hat f_j - f_j^*\|_\infty + 2 \|f_j^*\|_\infty \|D_\tau^{-1}\|_\infty \|\hat D_\tau^{-1} - D_\tau^{-1}\|_\infty + \|U\|_\infty
\end{align*}
By Lemma \ref{lem:single_deviation_bound} and under the conditions of Theorem \ref{thm:caglasso_consistency}, the following holds with probability $1 - c_2 \exp(-(c_3 A - 1 )\log p)$
$$\|\hat f_j - f_j^*\|_\infty \le \M \sqrt{\frac{A \log p}{2m+1}} + \vmnf.$$
With probability greater than $1 - c_4 \exp(-(c_5A - 2) \log p)$, we have,
$$ \|\hat R_\tau - R_\tau\|_\infty \le \M\thres + 2\M \sqrt{\M}\cdot \M^2 \thresr + \|U\|_\infty \le 4\M^{7/2} \thresr. $$
By choice of $\lambda_0$, the proof follows. 
\pfend

\section{Proof of corollaries}\label{sec:pf_cor}
\subsection{Proof of Corollary \ref{cor:consistency_partial_variance}: Consistency of DFT-NWR residual variance} \label{pf:consistency_partial_variance}

By \eqref{eq:nwr_partial_variance2},
\begin{align*}
    & \hat\tau_a^2 - \frac{\|\ea\|_2^2}{2m+1}\\ =~ & \frac{1}{2m+1}(\Za - \Zma \bjahat)^\dagger \Za - \frac{1}{2m+1}{\ea}^\dagger \ea\\
     =~ & \frac{1}{2m+1}(\muja + \ea - \Zma \bjahat)^\dagger (\muja + \ea) - \frac{1}{2m+1} {\ea}^\dagger \ea\\
     =~ & \underbrace{-\frac{1}{2m+1}{\ea}^\dagger \muja}_{=:V_1} + \underbrace{\frac{1}{2m+1}(\Zma \bjahat - \muja)^\dagger \Za}_{=:V_2}
\end{align*}
We now separately bound $V_1$ and $V_2$. The first term is
\begin{align*}
    & \left|\frac{1}{2m+1} {\ea}^\dagger \muja\right| = \left|\frac{1}{2m+1}\sum_{k \in \wmj} \overline{\eka} {\dkma}^\dagger \bka\right|\\ 
    \le & \max_{k\in \wmj}\left|\left(\frac{1}{2m+1} \sum_{k \in \wmj} \overline{\eka} {\dkma}^\dagger \right)\right| \|\bjatilde\|_1 + \left|\sum_{k \in \wmj}\frac{1}{2m+1} \eka {\dkma}^\dagger (\bka - \bjatilde) \right|\\
    \le & \underbrace{\|\bjatilde\|_1 \left\|\frac{1}{2m+1} {\Zma}^\dagger \ea\right\|_\infty}_{=:V_{1,1}} + \underbrace{\frac{1}{2m+1}\sum_{k \in \wmj} |\eka| |{\dkma}^\dagger(\bka - \bjatilde)|}_{=:V_{1,2}}
\end{align*}
We first give a bound on $V_{1,1}$. By Lemma \ref{lem:bjatilde_bounded}, $\| \bjatilde\|_1 \le \sqrt{s_a} \M$. Thus by Proposition \ref{prop:deviation}, the following holds with probability greater then $1 - c_0 \exp(-(c_1 A - 1) \log p)$
\begin{equation}\label{eq:v11_bound}
    |V_{1,1}| \le \sqrt{s_a}\M \lambda_a = \sqrt{s_a}\threse.
\end{equation}
By Proposition \ref{prop:nwr_approx_oracle}, Lemma \ref{lem:bounded_dft_error} part (ii) and applying \textit{Cauchy-Schwarz} inequality, the following holds with probability greater than $1 - c_2 \exp(-c_3 (2m+1))$,
$$|V_{1,2}| \le c_4 \sqrt{\M}\thresa.$$
Next we bound $V_2$. By Theorem \ref{thm:consistency_dft_nwr}, Lemma \ref{lem:bounded_dft_error} part (i) and applying \textit{Cauchy-Schwarz} inequality again, the following holds probability greater than $1 - c_5 \exp(-(c_1A - 1) \log p)$,
\begin{align*}
   |V_2| \le & \left( \frac{\|\Za\|_2^2}{2m+1} \right)^{1/2} \left(\frac{\|\Zma \bjahat - \muja\|_2^2}{2m+1}\right)^{1/2}\\ 
   \le & \sqrt{\M} \left( 48 s_a {\threse}^2 + 270{\thresa}^2 \right)^{1/2} \le \sqrt{270\M}\left(\sqrt{s_a}\threse + \thresa\right)
\end{align*}
Hence we obtain
\begin{equation}\label{eq:variance_bound_1}
    \left|\hat \tau_a^2 - \frac{\|\ea\|_2^2}{2m+1} \right| \le V_{1,1} + V_{1,2} + V_2 \le c_6\sqrt{\M}\left(\sqrt{s_a}\threse + \thresa\right)
\end{equation}
Additionally we apply \eqref{eq:spectral_difference_inv2} in Lemma \ref{lem:spectral_difference_inv}. We get 
\begin{align*}
    & \left| \frac{1}{(\thetajstar)_{a,a}} - \frac{1}{2m+1} \sum_{k \in \wmj} \frac{1}{(\Onk)_{a,a}} \right|\\
    \le~ & \frac{1}{2m+1} \sum_{k \in \wmj} \left|  \frac{1}{(\thetajstar)_{a,a}} - \frac{1}{(\Onk)_{a,a}} \right| \le 3 \M^4 \tnf.
\end{align*}
Hence by Lemma \ref{lem:bounded_dft_error} part (ii), with probability greater than $1 - c_{7} \exp(-c_{8}\log p)$,
\begin{equation}\label{eq:variance_bound_2}
   \left|\frac{\|\ea\|_2^2}{2m+1} - \frac{1}{(\thetajstar)_{a,a}} \right| \le 4 \M \lambda_a + 3 \M^4 \tnf 
\end{equation}
By $\sqrt{\M s_a}\threse \succsim \M \lambda_a$ and $\sqrt{\M}\thresa \succsim \M^4 \tnf$, and combining \eqref{eq:variance_bound_1} and \eqref{eq:variance_bound_2}, we complete the proof.  
\pfend

We now state the necessary Lemmata for proving Corollary \ref{cor:consistency_partial_variance}.

\begin{lemma}[Boundedness of DFT and residual]\label{lem:bounded_dft_error}
Consider the setup of Corollary \ref{cor:consistency_partial_variance}. Then there exist constants $c_i > 0$ such that the following two hold:
\begin{itemize}
\item[(i)] with probability greater than $1 - 2\exp(-c_0(2m+1))$, $$\dfrac{\|\Za\|^2}{2m+1} \le \frac{5\M}{2},$$
\item[(ii)] with probability greater than 
$1 - 2\exp(-c_2 \log p)$,  
$$\frac{\|\ea\|^2}{2m+1} \le c\M,~~ \text{and}~~ \left| \frac{\|\ea\|_2^2}{2m+1} - \frac{1}{2m+1} \sum_{k \in \wmj} \frac{1}{(\Onk)_{a,a}} \right|\le 4 \M^3 \sqrt{\frac{\log p}{m}}.$$
\end{itemize}
\end{lemma}

\subsubsection{Proof of Lemma \ref{lem:bounded_dft_error}: Boundedness of DFT and residual}

By \eqref{eq:dft_sin_cosine}, we have the following for $a\in [p]$ and $k \in \wmj$:
$$ \dka = \frac{1}{\sqrt{2\pi}} \e_a^\top \X^\top \Ec_k = \frac{1}{\sqrt{2\pi}} (\e_a^\top \otimes \Ec_k^\top) \vecop{\X}.$$
Therefore in matrix notation, $\Za = \frac{1}{\sqrt{2\pi}} J_0 \vecop{\X}$, where 
$$ J_0 := \begin{bmatrix}
    {\e_a}^\top \otimes {\Ec}_{j-m}^\dagger\\
    \vdots\\
    {\e_a}^\top \otimes {\Ec_{j+m}}^\dagger
\end{bmatrix} \peq {\e_a}^\top \otimes \begin{bmatrix}
    {\Ec_{j-m}}^\dagger\\
    \vdots\\
    {\Ec_{j+m}}^\dagger
\end{bmatrix}_{(2m+1)\times np}. $$
We note that $\rank(J_0^\dagger J_0) \le \rank(J_0) \le 2m+1$, and $\|J_0^\dagger J_0\| = \|J_0 J_0^\dagger\| = 1$. We apply Lemma \ref{lem:hanson_wright_gaussian} with $\eta = 1$ as follows:
\begin{align*}
    & \PP\left( \left| \frac{{\Za}^\dagger \Za}{2m+1} - \EE\left[\frac{{\Za}^\dagger \Za}{2m+1}\right] \right| > \M \eta \right)\\
    \le~ &~ \PP\left( \left| \vecop{\X}^\top J_1^\dagger J_1 \vecop{\X} - \EE\left[ \vecop{\X}^\top J_0^\dagger J_0 \vecop{\X} \right] \right| > 2\pi \M (2m+1) \right)\\
    \le~ &~ 4 \exp\left[-c_0 (2m+1)\right].
\end{align*}
Additionally by \eqref{eq:Snk_f_diff} and the condition $\tnf \le 1/(2\M)$,
$$ \EE\left[\frac{{\Za}^\dagger \Za}{2m+1}\right] = \frac{1}{2m+1}\sum_{k \in \wmj} (\Snk)_{a,a} \le 3\M/2. $$
Thus the following holds with probability greater than $1 - 2\exp[-c_0(2m+1)]$,
$$ \left|\frac{{\Za}^\dagger \Za}{2m+1}\right| \le \frac{5\M}{2}, $$
completing the proof of (i).

We now prove (ii). By \eqref{eq:J1}, $\ea = J_1 \vecop{\X}$ and $ {\ea}^\dagger \ea = \vecop{\X}^\top J_1^\dagger J_1 \vecop{\X} $.  We have $\rank(J_1^\dagger J_1) \le 2m+1$. In addition, 
\begin{align*}
    \|J_1^\dagger J_1\| \le \max_{k \in \wmj} \|u_{k,a}\|_2^2 = & \max_{k \in \wmj} \frac{\|(\Onk)_{a\cdot}\|_2^2}{|(\Onk)_{a,a}|^2}\\ 
    = & 1 + \max_{k \in \wmj}\|\bka\|_2^2\\
    < & 4\M^2,
\end{align*}
where the last step follows from Lemma \ref{lem:bjatilde_bounded} part (ii). We now apply Lemma \ref{lem:hanson_wright_gaussian} to obtain the following
\begin{align}
    & \PP\left( \left| \frac{{\ea}^\dagger \ea}{2m+1} - \EE\left[ \frac{{\ea}^\dagger \ea}{2m+1} \right] \right| > \M \eta \right) \nonumber\\
    \le~ &~ \PP\left( \left| \vecop{\X}^\top J_1^\dagger J_1 \vecop{\X} - \EE\left[ \vecop{\X}^\top J_1^\dagger J_1 \vecop{\X} \right] \right| > 2\pi \M \eta (2m+1) \right) \nonumber\\
    \le~ &~ 4 \exp\left[-c_2 (2m+1) \min\left\{\frac{\eta}{4\M^2}, \frac{\eta^2}{(4\M^2)^2}\right\}\right]. \label{eq:ea_hanson_wright}
\end{align}
We choose $\eta = 4\M^2 \sqrt{\log p/m}$. By $m \succsim \log p$, we get
$$ \PP\left( \left| \frac{{\ea}^\dagger \ea}{2m+1} - \EE\left[ \frac{{\ea}^\dagger \ea}{2m+1} \right] \right| > 4\M^3 \sqrt{\frac{\log p}{m}} \right) \le 4 \exp(-c_2 \log p), $$
that proves the first part of (ii). Additionally,
$$ \EE\left[ \frac{{\ea}^\dagger \ea}{2m+1} \right] = \frac{1}{2m+1} \sum_{k \in \wmj} \frac{1}{(\Onk)_{a,a}} \le \frac{3\M}{2}.$$
Here we use the fact that under the conditions of Corollary \ref{cor:consistency_partial_variance}, $(\Onk)_{a,a} \ge 2/(3\M)$. Thus by $m \succsim \M^4 \log p$, the following holds with probability greater than $1 - 4\exp(-c_2 \log p)$,
$$ \left| \frac{{\ea}^\dagger \ea}{2m+1}\right|\le \EE\left[\frac{{\ea}^\dagger \ea}{2m+1}\right] + 4 \M^3 \sqrt{\frac{\log p}{m}} \le c\M, $$
and the proof of (ii) is done.
\pfend

\section{Consistency results for general linear processes}\label{sec:consistency_linear}

We extend our theoretical results for general linear processes with not necessarily Gaussian errors. We start with an assumption that generalizes Assumption \ref{asn:summable}.

\begin{assumption}[Summable linear processes] \label{asn:summable_general}
$\{X_t\}$ is a stable linear process as $X_t = \sum_{h = 0}^\infty B_h e_{t-h}$, where $B_h \in \RR^{p \times p}$ and $e_t \in \RR^p$ are i.i.d. Furthermore,
\begin{equation}\label{eq:summable_general}
    \sum_{h = 0}^\infty|h| \|B_h\| < \infty.
\end{equation}
\end{assumption}

Assumption \ref{asn:summable_general} is sufficient to ensure that (a) the process $X_t$ is stationary \citep{rosenblatt2012stationary}, (b) the autocovariance function $\Gamma(\cdot)$ is well-defined (can be shown similar to \citet[Lemma C.7]{sun2018large}), and (c) the condition \eqref{eq:summable} is satisfied. We show the last implication in the next lemma.

\begin{lemma}\label{lem:summable_general}
    Assumption \ref{lem:summable_general} implies that \eqref{eq:summable} holds.
\end{lemma}

\begin{proof}
\begin{align*}
    \sum_{\ell = 0}^\infty \ell \|\Gamma(\ell)\| = \sum_{\ell = 0}^\infty \ell \left\| \sum_{t = 0}^\infty B_{t+\ell} B_{t}^\top \right\|
    = &  \sum_{s = 0}^\infty \sum_{t = s}^\infty (t-s) \|B_t B_s^\top\|\\
    = & \frac{1}{2} \sum_{s = 0}^\infty \sum_{t = 0}^\infty |t-s| \|B_t\| \|B_s\| \\
    \le & \frac{1}{2}\left( \sum_{s, t = 0}^\infty (|t| + |s|) \|B_t\| \|B_s\| \right)\\
    \le & \left(\sum_{s=0}^\infty \|B_s\|\right) \left(\sum_{t=0}^\infty t\|B_t\|\right),
\end{align*}
hence we are done with the proof. 
\end{proof}

Next we extend the assumption of Gaussian time series in Assumption \ref{asn:summable} to linear processes accommodating more general distribution families for the errors.

\begin{assumption}[Error distribution]\label{asn:error_dist}
The errors $e_t = (e_{t, 1},\ldots, e_{t, p})^\top$ in Assumption \ref{asn:summable_general} come form one of the following three distribution families:

\begin{enumerate}[label=\textbf{(F\arabic*)}]
\normalfont{\item \label{f1} Sub-Gaussian}:~~ $\PP(|e_{t,a}|>x)\le 2\exp\left(-\frac{x^2}{2\sigma^2}\right),$ for all $x>0$ and some $\sigma > 0$.

\item \label{f2} Generalized sub-exponential:~~ $\PP(|e_{t,a}|>x^{\nu})\le \theta\exp(-\gamma x)$, for all $x> 0$, and some positive constants $\theta$ and $\gamma$.

\item \label{f3} Distribution with finite 4\textsuperscript{th} moment:~~ $\EE(e_{t,a}^4)\le K < \infty$ for some $K > 0$.
\end{enumerate}
\end{assumption}

For generalizing the CGLASSO results for linear processes, we modify the threshold quantity in \eqref{eq:threshold} with $A > 0$ and $g(p, A)$ in Table \ref{tab:notation_general} as
\begin{equation}\label{eq:threshold_general}
 \thres := \M \cdot \frac{g(p, A)}{\sqrt{m}} + \vmnf.
\end{equation}

\begin{table}[t!]
\centering
\renewcommand{\arraystretch}{1.4}
\begin{tabular}{@{} lccc @{}}
\toprule
\textbf{Object} & \textbf{\ref{f1}} & \textbf{\ref{f2}} & \textbf{\ref{f3}} \\
\midrule

$\underline{m}_0$ 
& $\M^4 s_a \min \left\{ \log p, \log \left( \frac{21 ep}{s_a} \right) \right\}$ 
& $\M^4 \left( s_a \min \left\{ \log p, \log \left( \frac{21 ep}{s_a} \right) \right\} \right)^{4(1+\nu)}$ 
& $\M^4 \min\left\{ p^{s_a}, \left( \frac{21ep}{s_a} \right)^{s_a} \right\}$ \\

$\underline{m}_1(x)$ 
& $\M^4 x\log p$ 
& $\M^4 (x\log p)^{4(1+\nu)}$ 
& $\M^4p^{x}$ \\

$\underline{m}_2$ 
& $\M^2 \log p$ 
& $\M^2 (\log p)^{4(1+\nu)}$ 
& $\M^2 p^2$ \\

$g(p,A)$ 
& $(A\log p)^{1/2}$ 
& $(A\log p)^{2(1+\nu)}$ 
& $p^{A}$ \\

$\Bcal_1(p, A)$ 
& $c \exp\{-(c' A - 2)\log p\}$ 
& $c \exp\{-(c' A - 2)\log p\}$ 
& $c \exp\{-2A \log p\}$ \\

$\Bcal_2(m, A)$ 
& $4 \exp\!\left[-c (2m+1) \min\{A, A^2\}\right]$ 
& $c \exp\!\left[-c' (\sqrt{2m+1}A)^{\tfrac{1}{2(1+\nu)}}\right]$ 
& $\tfrac{c}{(2m+1)A^2}$ \\

$\Bcal_3(p, A)$ 
& $c \exp\{-(c'A - 1)\log p\}$ 
& $c \exp\{-(c'A - 1)\log p\}$ 
& $c \exp\{-2A \log p\}$ \\

$\tscr$ 
& $4\exp\!\left[-c\min\!\left\{\tfrac{\eta}{\|M\|}, \tfrac{\eta^2}{\rank(M)\|M\|^2}\right\}\right]$ 
& $c \exp\!\left[-c' \left(\tfrac{\eta}{\sqrt{\rank(M)}\|M\|} \right)^{\tfrac{1}{2(1+\nu)}}\right]$ 

& $\tfrac{c\ \rank(M)\|M\|^2}{\eta^2}$\\
\bottomrule
\end{tabular}
\caption{Notation for the bandwidth lower bounds $\underline{m}_0$, $\underline{m}_1(x)$ and $\underline{m}_2$, the tail control parameters $g(p, A)$, the tail probabilities $\Bcal_1(m, A), \Bcal_2(p, A), \Bcal_3(p, A)$ and $\tscr$ under the distribution families \ref{f1}, \ref{f2}, and \ref{f3}. Here $A, \eta$ are fixed positive number determining the tail probabilities and the rates of convergence, and $c, c'>0$ denote universal constants and are allowed to change values for every pair of the defined object and distribution family.}
\label{tab:notation_general}
\end{table}

\begin{theorem}[Consistency of CGLASSO for general linear process]\label{thm:consistency_cglasso_gen}
Let $\{X_t\}_{t\in [n]}$ be observations from a stationary time series satisfying Assumptions \ref{asn:summable_general}, \ref{asn:error_dist} and \ref{assumption:incoherence}. Assume that $\kf, \kt \ge 1$ and the sample size $n$ satisfies $n \succsim \lf \M^3 g(p, R)$ and $\tnf \le 1/(4\M)$. Then for any $A > 0$, $\thres$ as in \eqref{eq:threshold_general} satisfying $d \thres \le [6\kt^2 \kf^3 C_\alpha]$, the choice of penalty $\lambda = (8/\alpha)\thres$ and the bandwidth $m \succsim\M^2 g(p, R)$ and $m \le n/(4\M \lf)$, $\hat\Theta_j$ satisfies (a), (b) and (c) in Theorem \ref{thm:consistency_cglasso} with probability greater than $1 - \Bcal_1(p, A)$.
\end{theorem}

We now generalize Theorem \ref{thm:consistency_dft_nwr} for distribution families in Assumption \ref{asn:error_dist}, and then state the general versions of the CAGLASSO consistency results in Section \ref{subsec:caglasso}. The proofs for general linear process are similar to that of Gaussian time series, with the concentration bounds replaced.

\begin{proposition}[Restricted eigenvalue for general linear process]\label{prop:re_gen}
Consider the DFT-NWR setup of \eqref{eq:nodewise-reg} with $p > 1$ satisfying Assumption \ref{asn:summable_general} and \ref{asn:error_dist}. Then for the sample size $n \succsim \M \lf \underline{m}_0$ and $\tnf \le 1/(4\M)$, and the bandwidth $m \succsim\underline{m}_0$ and $m \le n/(4\M \lf)$, the RE condition \eqref{eq:re_condition} holds with probability greater than $1 - \Bcal_2(m, \M^{-2})$.
\end{proposition}

\begin{proposition}[Linear approximation with the oracle for general linear process]\label{prop:nwr_approx_oracle_gen}
    Consider the DFT-NWR setup \eqref{eq:nwr_model} satisfying Assumption \ref{asn:summable_general} and \ref{asn:error_dist}. Assume that the sample size satisfies $\tnf \le 1/(4\M)$. Then for $a \in [p]$, and the bandwidth $m \le n / (4\M \lf)$, there exist constants $c_i > 0$ such that \eqref{eq:nwr_approx_oracle} holds with probability greater than $ 1 -  \Bcal_2(m, A)$.  Here $\Bcal_2(m, A)$ for distribution families \ref{f1}, \ref{f2} and \ref{f3} are given in Table \ref{tab:notation_general}.
\end{proposition}

\begin{proposition}[Deviation condition for general linear process]\label{prop:deviation_gen}
    Consider the DFT-NWR setup \eqref{eq:nwr_model} satisfying Assumption \ref{asn:summable_general} and \ref{asn:error_dist}. Assume that the sample size satisfies $\tnf\le 1/(4\M)$ and $n \ge 4\M \lf$. Then for $a \in [p]$ and $A > 0$, the following holds with probability greater than $1 - \Bcal_3(p, A)$:
    $$ \left\|\frac{1}{2m+1} {\Zma}^\dagger \ea \right\|_\infty \le \M^2\frac{\sqrt{3}g(p, A)}{\sqrt{2m+1}}. $$
    Here $\Bcal_3(p, A)$ and $g(p, A)$ for distribution families \ref{f1}, \ref{f2} and \ref{f3} are given in Table \ref{tab:notation_general}.
\end{proposition}

\begin{theorem}[Consistency of DFT-NWR for general linear process]\label{thm:consistency_dft_nwr_gen}
    Consider the DFT-NWR setup \eqref{eq:nwr_model} satisfying Assumption \ref{asn:summable_general} and \ref{asn:error_dist}. Assume $a \in [p]$ and the sample size satisfies $n \succsim \M \lf \underline{m}_1(s_a)$ and $\tnf \le 1/(4\M)$. Then for $A > 0$, $\lambda_a \ge 4\M^2\frac{\sqrt{3}g(p, A)}{\sqrt{2m+1}}$ and the bandwidth 
    $m \succsim \underline{m}_1(s_a)$ and $m \le n/(4\M\lf)$, the solution $\bjahat$ of the sample NWR problem \eqref{eq:nodewise-reg} has the estimation error bound \eqref{eq:estimation_consistency_nwr} around $\bjastar$ and prediction error bound \eqref{eq:prediction_consistency_nwr} around $\muja$ with probability greater than $1 - \Bcal_3(p, A)$. Here $\underline{m}_1(s_a)$, $g(p, A)$ and $\Bcal_3(p, A)$ for distribution families \ref{f1}, \ref{f2} and \ref{f3} are given in Table \ref{tab:notation_general}.
\end{theorem}

\begin{theorem}[Estimation consistency of CAGLASSO for general linear process] \label{thm:caglasso_consistency_gen}
Let $\{X_t\}_{t\in[n]}$ be a stationary time series satisfying Assumptions \ref{asn:summable_general}, \ref{asn:error_dist} and \ref{assumption:incoherence} with $\alpha \in (1 - r_S^{-1}, 1)$. Assume that $n \succsim \M\lf \underline{m}_1(d)$ and $\tnf, \kf, \kt$ are as in Theorem \ref{thm:caglasso_consistency}.  Then for $A > 0$, $\lambda_a \ge 4\M^2 \frac{\sqrt{3} g(p, A)}{\sqrt{2m+1}}$, $\thresr$ and $\lambda$ both satisfying the conditions in Theorem \ref{thm:caglasso_consistency}, and the bandwidth $m \succsim \underline{m}_1(d)$ and $m \le n/(4\M\lf)$, there exist constants $c_i >0 $ such that the solution $\thetajhattau$ of the CAGLASSO problem \eqref{eq:caglasso} satisfies (a), (b) and (c) in Theorem \ref{thm:caglasso_consistency} with probability greater than $1 - \Bcal_1(p, A)$.
\end{theorem}

The proofs of Proposition \ref{prop:re_gen}, \ref{prop:nwr_approx_oracle_gen}, \ref{prop:deviation_gen} and Theorem \ref{thm:consistency_cglasso_gen}, \ref{thm:consistency_dft_nwr_gen}, \ref{thm:caglasso_consistency_gen} follow similar to the proofs of Proposition \ref{prop:re}, \ref{prop:nwr_approx_oracle}, \ref{prop:deviation} and Theorem \ref{thm:consistency_cglasso}, \ref{thm:consistency_dft_nwr}, \ref{thm:caglasso_consistency} respectively-- with the concentration results applied for Gaussian time series (Lemma \ref{lem:hanson_wright_gaussian}) replaced with the same for general linear processes (see Remark \ref{lem:hanson_wright_general} for details). Heavier-tailed error distributions increase the sample size and bandwidth required for consistency of CGLASSO and CAGLASSO. For example, in case of sub-Gaussian errors, consistency holds when $m \succsim \log p$; for sub-exponential errors, the requirement strengthens to $m \succsim (\log p)^{4(1+\nu)}$; and with only finite fourth moments, it further grows to $m \succsim p^{2}$.

\section{Auxiliary Results}\label{sec:aux_results}
This section contains the auxiliary theoretical lemmata used for proving the main theoretical results in Appendix \ref{sec:appendix_proofs_prop} and \ref{sec:appendix_proofs_thms}.

\begin{lemma}[Lipschitz continuity of spectral density]\label{lem:lip_smooth}
    Under Assumption \ref{asn:summable}, the spectral density $f^*(\cdot)$ is a Lipschitz continous function satisfying 
    \begin{equation}\label{eq:lip_smooth}
        \|f^*(\omega) - f^*(\omega')\| \le \frac{\lf}{4\pi} |\omega - \omega'|.
    \end{equation}
\end{lemma}

\begin{lemma}[Hanson-Wright inequality for Gaussian time series] \label{lem:hanson_wright_gaussian}
Consider the data matrix $\X = [X_1:\ldots: X_n]^\top \in \RR^{n\times p}$ with $X_t$ coming from a stable Gaussian time series satisfying Assumption \ref{asn:summable}. Then there exists a universal constant $c>0$ such that for any $\eta > 0$ and $M \in \CC^{np\times np}$,
\begin{align}
    & \PP\left( \left| \vecop{\X}^\top M \vecop{\X} - \EE\left[ \vecop{\X}^\top M \vecop{\X} \right] \right| > 2\pi \M\eta \right)\nonumber\\
    & \hspace{100pt} \le 4\exp\left[-c \min\left\{\frac{\eta}{\|M\|},  \frac{\eta^2}{\rank(M)\|M\|^2}\right\}\right].\label{eq:hanson_wright}
\end{align}
\end{lemma}

\begin{remark}[Hanson-Wright inequality for general linear processes]\label{lem:hanson_wright_general}
Hanson-Wright type concentration inequalities for $M \in \RR^{np \times np}$ and the distribution families in Assumption \ref{asn:error_dist} are provided in \citet[Proposition 4.2]{sun2018large} -- with $\tscr$ in Table \ref{tab:notation_general} appearing in the right side of \eqref{eq:hanson_wright} as the concentration. We only show that the concentration inequalities remain to hold for quadratic forms of $\vecop{\X}$ with complex-valued $M$ only for Gaussian time series since the proof for the general linear processes will hold similarly.
\end{remark}

\begin{lemma}[Perturbation bounds of inverted matrices]\label{lem:spectral_difference_inv}
Consider two matrices $p$-dimensional $A, B \in \pdhermit$ such that $\max\{\|A\|, \|A^{-1}\|\} \le M$ for some $M\ge 1$, and $\|B - A\| \le \delta$ for $0 < \delta \le \frac{1}{2M}$. Then the following holds:
\begin{equation}\label{eq:spectral_difference_inv1}
    \|B^{-1} - A^{-1}\| \le 2M^2 \delta.
\end{equation}
For any $j \in [p]$, the following hold:
\begin{equation}\label{eq:spectral_difference_inv2}
    \left| \frac{1}{(A^{-1})_{j,j}} - \frac{1}{(B^{-1})_{j,j}} \right| \le 3M^4 \delta,
\end{equation}
\begin{equation}\label{eq:spectral_difference_inv3}
    \frac{\|(B^{-1})_{j,\cdot}\|_2}{|(B^{-1})_{j,j}|} \le 4M^2,
\end{equation}
and
\begin{equation}\label{eq:spectral_difference_inv4}
    \left\| \frac{(A^{-1})_{j,\cdot}}{(A^{-1})_{j,j}} - \frac{(B^{-1})_{j,\cdot}}{(B^{-1})_{j,j}} \right\|_2 \le 6M^5 \delta.
\end{equation}
\end{lemma}

\begin{lemma}[Inner product of $C_j$ and $S_k$] \label{lem:inner_product_C_S}
For any $j, k \in F_n$, the inner product between $C_j$ and $S_k$ satisfies the following:
\begin{enumerate}
    \item $C_j^\top S_k = 0$ for every $j,k \in F_n$.
    \item $C_j^\top C_k = 0$ if $|j|\ne |k|$,~~ $C_j^\top C_j = \begin{cases}
        1 & \text{ if } j \in \{0, n/2\},\\
        1/2 & \text{ otherwise}
    \end{cases}$,~~ $C_j^\top C_{-j} = \begin{cases}
        1 & \text{ if } j = 0,\\
        1/2 & \text{ otherwise}.
    \end{cases}$
    \item $S_j^\top S_k =  0$ if $|j| \neq |k|$,~~ $S_j^\top S_j = \begin{cases}
        0 & \text{ if } j \in \{0, n/2\},\\
        1/2 & \text{ otherwise}
    \end{cases}$,~~ $S_j^\top S_{-j} = \begin{cases}
        0 & \text{ if } j = 0,\\
        -1/2 & \text{ otherwise}.
    \end{cases}$
\end{enumerate}
\end{lemma}

\begin{lemma}[Norm bound for structured Kronecker products]\label{lem:kronecker_norm_boud}
Let $\{x_j\}_{j\in F_n}$ be a sequence of vectors in $\CC^p$ such that $\|x_j\|_2 \le \delta$ for every $j \in F_n$. Denote the following
\begin{equation}\label{eq:E_n}
E_n := \begin{bmatrix}
\xi_{-\lfloor\frac{n-1}{2}\rfloor}\\
\vdots\\
\xi_{\lfloor \frac{n}{2} \rfloor}
\end{bmatrix},
\end{equation}
where $\xi_j := (I_p \otimes [C_j ~~ S_j]) \varphi(x_j) \in \RR^{np \times 2}$ and $C_j, S_j,~ j \in F_n$ are defined in \eqref{eq:Ck} and $\eqref{eq:Sk}$ respectively. Then $\|E_n\| \le \delta$.
\end{lemma}

\begin{lemma}[Cone set to sparse set approximation]\label{lem:cone_to_sparse}
For any $S\subset[p]$ with $|S| = s$ and $\kappa > 0$,
$$ \Cscr(S, \kappa) \subseteq \mathbb{B}_1((\kappa + 1)\sqrt{s}) \cap \mathbb{B}_2 (1) \subseteq (\kappa + 2) \mathrm{cl}\{\mathrm{conv}\{\Kscr(s)\}\}. $$
\end{lemma}

\begin{lemma}[Upper bound on quadratic forms on closure of sparse sets]\label{lem:sparse_set_quad}
    For a Hermitian matrix $D \in \CC^{p\times p}$, 
    $$\sup_{v \in \mathrm{cl}\{\mathrm{conv}\{\Kscr(s)\}\}} |v^\dagger D v| \le 3 \sup_{v\in \Kscr(2s)}|v^\dagger D v|.$$
\end{lemma}

\begin{lemma}[Concentration bound on quadratic forms over a sparse set]\label{lem:conc_sparse_set}
    Let $D \in \CC^{p\times p}$ be a Hermitian matrix. If, for any vector $v\in \CC^p$ with $\|v\|_2 \le 1$, and $\eta > 0$,
    $$ \PP(|v^\dagger D v| > C\eta) \le c\exp(\mathcal{T}(\eta)), $$
    then for any integer $s \ge 1$, we have
    $$ \PP\bigg( \sup_{v\in \Kscr(s)} |v^\dagger D v| > C\eta \bigg)\le c\exp\left(\mathcal{T}(\eta) + s\min\{\log p, \log (21ep/s)\}\right). $$
\end{lemma}

The proofs of Lemma \ref{lem:cone_to_sparse}, \ref{lem:sparse_set_quad} and \ref{lem:conc_sparse_set} are available for real matrices in \citet[Lemma F.1, F.3, F.2 respectively]{basu2015regularized} and their generalization for complex-valued matrices are straight-forward. A proof of Lemma \ref{lem:inner_product_C_S} is given in \citet[Lemma C.3]{sun2018large}. We provide proof sketches of rest of the results.

\subsection{Proof of Lemma \ref{lem:lip_smooth}: Lipschitz continuity of spectral density}\label{pf:lip_smooth}

We use the following property: for $x, y \in \RR$,
$$ |e^{\i x} - e^{\i y}| \le |x - y|. $$
For any $\omega, \omega' \in [-\pi, \pi]$ we have
\begin{align*}
    \|f^*(\omega) - f^*(\omega')\| \le & \frac{1}{2\pi} \sum_{h\in \ZZ}\|\Gamma(h)\| |e^{-\i h \omega} - e^{-\i h \omega'}|\\
    \le & \frac{1}{2\pi} \sum_{h \in \ZZ} \|\Gamma(h)\| |h| |\omega - \omega'|\\
    \le & \frac{1}{2\pi}\cdot 2\sum_{h = 0}^\infty \|\Gamma(h)\| h |\omega - \omega'|\\
    = & \frac{\lf}{4\pi} |\omega - \omega'|,
\end{align*}
and hence proved. \pfend

\subsection{Proof of Lemma \ref{lem:hanson_wright_gaussian}: Hanson-Wright inequality for Gaussian time series}

We extend from the proof of \citet[Lemma 3.2]{sun2018large}. For any $p \times p$ complex-valued matrix $M = \re(M) + \i \im(M)$, we can separately apply their result to obtain the following:
\begin{align}
    & \PP\left( \left| \vecop{\X}^\top \re(M) \vecop{\X} - \EE\left[ \vecop{\X}^\top \re(M) \vecop{\X} \right] \right| > 2\pi\M\eta \right)\nonumber\\
    & \hspace{100pt} \le 2\exp\left[-c \min\left\{\frac{\eta}{\|\re(M)\|},  \frac{\eta^2}{\rank(\re(M))\|\re(M)\|^2}\right\}\right], \label{eq:hanson_wright_real0}\\
    & \PP\left( \left| \vecop{\X}^\top \im(M) \vecop{\X} - \EE\left[ \vecop{\X}^\top \im(M) \vecop{\X} \right] \right| > 2\pi\M\eta \right) \nonumber\\
    & \hspace{100pt} \le 2\exp\left[-c \min\left\{\frac{\eta}{\|\im(M)\|},  \frac{\eta^2}{\rank(\im(M))\|\im(M)\|^2}\right\}\right]. \label{eq:hanson_wright_imaginary0}
\end{align}
We denote $r := \rank(M)$. By \textit{rank factorization theorem}, there exist $U \in \CC^{p \times r},\ V \in \CC^{r\times p}$ such that
\begin{align*}
    M = & UV\\ 
    = & (\re(U) + \i \im(U))(\re(V) + \i \im(V))\\
    = & (\re(U)\re(V) - \im(U) \im(V)) + \i(\re(U)\im(V) + \im(U) \re(V)) 
\end{align*}
Thus we have the real part as
\begin{align*}
    \re(M) = & \re(U)\re(V) - \im(U) \im(V)\\
    = & \begin{bmatrix}
        \re(U) & - \im(U)
    \end{bmatrix}_{p \times 2r} 
    \begin{bmatrix}
        \re(V) \\ \im(V)
    \end{bmatrix}_{2r \times p}
\end{align*}
Hence, $\rank(\re(M)) \le 2r$. Additionally we have 
$$\|\re(M)\| = \left\|\frac{1}{2}(M + \overline{M})\right\| \le \frac{1}{2}\left( \|M\| + \|\overline{M}\| \right)\le \|A\|.$$
We now turn to \eqref{eq:hanson_wright_real0} and provide a bound on the concentration involving $\re(M)$ in terms of the rank and norm of $A$. We replace $\eta$ with $\eta/2$ and obtain the following
\begin{align}
    & \PP\left( \left| \vecop{\X}^\top \re(M) \vecop{\X} - \EE\left[ \vecop{\X}^\top \re(M) \vecop{\X} \right] \right| > \pi\M\eta \right)\nonumber\\
    & \hspace{100pt} \le 2\exp\left[-c \min\left\{\frac{\eta}{2\|A\|},  \frac{\eta^2}{8\ \rank(M)\|M\|^2}\right\}\right], \nonumber \\
    & \hspace{100pt} \le 2\exp\left[-\frac{c}{8} \min\left\{\frac{\eta}{\|M\|},  \frac{\eta^2}{\rank(M)\|M\|^2}\right\}\right]. \label{eq:hanson_wright_real1}
\end{align}
Similarly, we can write the bound \eqref{eq:hanson_wright_imaginary0} as 
\begin{align}
    & \PP\left( \left| \vecop{\X}^\top \im(M) \vecop{\X} - \EE\left[ \vecop{\X}^\top \im(M) \vecop{\X} \right] \right| > \pi\M\eta \right)\nonumber\\
    & \hspace{100pt} \le 2\exp\left[-\frac{c}{8} \min\left\{\frac{\eta}{\|M\|},  \frac{\eta^2}{\rank(M)\|M\|^2}\right\}\right]. \label{eq:hanson_wright_imaginary1}
\end{align}
Combining \eqref{eq:hanson_wright_real1} and \eqref{eq:hanson_wright_imaginary1} by union bound, we obtain the desired result. \pfend

\subsection{Proof of Lemma \ref{lem:spectral_difference_inv}: Perturbation bounds of inverted matrices}

We denote the eigenvalues of any $p$-dimensional matrix $X$ in decreasing order by $\{\Lambda_i(X)\}_{i \in [p]}$. Hence $\Lambda_p(A) \ge \frac{1}{\|A^{-1}\|} \ge \frac{1}{M}$. By \textit{Weyl's inequality} we have
$$ |\Lambda_i(B) - \Lambda_i(A)| \le \|A - B\| \le \delta ~~ \implies~~ \frac{1}{M} - \delta \le \Lambda_p(B) \le \Lambda_1 (B) \le M + \delta. $$
Thus $\|B^{-1}\| \le \frac{1}{\Lambda_p(B)} \le \frac{1}{\frac{1}{M} - \delta} $. By the identity $B^{-1} - A^{-1} = -A^{-1}(B - A)B^{-1}$,
$$ \|B^{-1} - A^{-1}\| \le \|A^{-1}\| \|B^{-1}\| \|B - A\| \le \frac{M\delta}{\frac{1}{M} - \delta}. $$
Hence using $\delta \le \frac{1}{2M}$, \eqref{eq:spectral_difference_inv1} follows. From \textit{Rayleigh quotient} formulation of eigenvalues, $|(B^{-1})_{j,j}| \ge \frac{1}{\Lambda_1(B)} \ge \frac{1}{M+\delta}$ for any $j \in [p]$. Thus we get the following
\begin{align*}
    \left|\frac{1}{(B^{-1})_{j,j}} - \frac{1}{(A^{-1})_{j,j}}\right| = & \frac{|(A^{-1})_{j,j} - (B^{-1})_{j,j}|}{|(A^{-1})_{j,j} (B^{-1})_{j,j}|} \\
    \le & \frac{\|A^{-1} - B^{-1}\|}{|(B^{-1})_{j,j}| |(A^{-1})_{j,j}|}\\
    \le & \frac{M^2 (M + \delta)}{\frac{1}{M} - \delta}\delta ~\le~ 3M^4\delta.
\end{align*}
This completes the proof of \eqref{eq:spectral_difference_inv2}. Next, \eqref{eq:spectral_difference_inv3} follows from
$$ \left\| \frac{(B^{-1})_{j,\cdot}}{(B^{-1})_{j,j}} \right\|_2 \le \frac{\|B^{-1}\|}{|(B^{-1})_{j,j}|} \le \frac{1}{\left(\frac{1}{M} - \delta\right)^2} \le 4M^2. $$
And for the final claim,
\begin{align*}
    &~ \left\|\frac{(B^{-1})_{j,\cdot}}{(B^{-1})_{j,j}} - \frac{(A^{-1})_{j,\cdot}}{(A^{-1})_{j,j}}\right\|_2\\
    = ~& \frac{\|(B^{-1})_{j,\cdot} (A^{-1})_{j,j} - (B^{-1})_{j,j} (A^{-1})_{j,\cdot}\|_2}{|(B^{-1})_{j,j} (A^{-1})_{j,j}|} \\
    \le~ & \frac{|(B^{-1})_{j,j} - (A^{-1})_{j,j}| \|(A^{-1})_{j,\cdot}\|_2 + |(A^{-1})_{j, j}| \|(B^{-1} - A^{-1})_{j,\cdot}\|_2}{|(A^{-1})_{j, j}| \cdot | (B^{-1})_{j, j}|}\\
    \le & \frac{\|B^{-1} - A^{-1}\| \|A^{-1}\| + \|A^{-1}\| \|B^{-1}- A^{-1}\|}{|(A^{-1})_{j, j}|  | (B^{-1})_{j, j}|}\\
    \le~ & \frac{M^3 (M + \delta)}{\frac{1}{M} - \delta}2\delta ~\le~ 6M^5 \delta.
\end{align*}
Hence the proof of \eqref{eq:spectral_difference_inv4} is complete. \pfend 

\subsection{Proof of Lemma \ref{lem:kronecker_norm_boud}: Norm bound for structured Kronecker products} 

We note that $E_n^\top E_n$ consists of blocks of the form $B_{j, k} = \xi_j^\top \xi_k$, $j, k \in F_n$. Without loss of generality, we permute the rows of $E_n$ such that $\xi_j$ and $\xi_{-j}$ appear adjacently if both $j, -j \in F_n$. Since operator norm is invariant under row permutation, we apply Lemma \ref{lem:inner_product_C_S} and encounter four possible configuration of blocks in $E_n^\top E_n$:

\begin{enumerate}
    \item If $j = k = 0$, 
    \begin{align*}
        B_{0,0} = \Ec_0^\top \Ec_0 = \varphi(x_0)^\top \left(I_p \otimes\begin{bmatrix}
        1 & 0\\
        0 & 0
    \end{bmatrix}\right) \varphi(x_0) 
    = \begin{bmatrix}
    \|\re(x_0)\|_2^2 & 0\\
    0 & 0
    \end{bmatrix}.
    \end{align*}
    
    Thus we have 
    $$ \|B_{0, 0}\| \le  \|\re(x_0)\|_2^2 \le \delta^2.$$
    \item If $j, k \in F_n$ and $k = -j \ne 0$,
    \begin{align*}
        B_{j, k} = \Ec_j^\top \Ec_k = \varphi(x_k^\top)\left(I_p \otimes \begin{bmatrix}
        C_j^\top C_k & C_j^\top S_k\\
        S_j^\top C_k & S_j^\top S_k
    \end{bmatrix}\right) \varphi(x_k) = \varphi\left(\frac{1}{2}x_j^\dagger x_k\right)
    \end{align*}
    We use Proposition \ref{prop:phi2}(c) and get
    $$ \left\| \begin{bmatrix}
        B_{j,j} & B_{j, k}\\
        B_{k,j} & B_{k,k}
    \end{bmatrix} \right\| \le \frac{1}{2}(\|x_j\|_2^2 + \|x_k\|_2^2) \le \delta^2. $$
    
    Hence the eigenvalues of this block are bounded by the trace $(\|x^j\|_2^2 + \|x^k\|_2^2)/2 \le \delta^2$.
    \item If $j \in F_n$ and $-j \not\in F_n$,
    $B_{j,j} = \varphi\left( \|x^j\|_2^2/2 \right)$, thus $\|B_{j, j}\| \le \delta^2 / 2$.
    \item If $j,k \in F_n$ and $|j| \neq |k|$, $B_{j,k} = 0_{2\times 2}$.
\end{enumerate}
Combining all the cases, $E_n^\top E_n$ is a block diagonal matrix with blocks having operator norm smaller than $\delta^2$. Therefore, $\|E_n\|^2 = \|E_n^\top E_n\| \le \delta^2$ that completes the proof. \pfend

\section{Illustration of frequency-specific graphical model}\label{sec:freq_graphical_model}

\begin{figure}[!h]
    \centering
    \includegraphics[width = 0.9\linewidth]{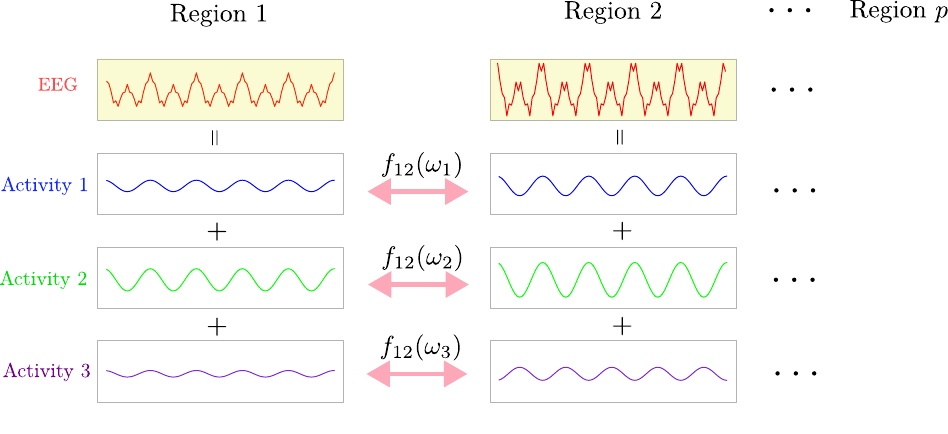}
    \caption{\textbf{Illustration of time series generated from two brain regions as a sum of periodic components.}
    Pink bidirectional edges represent marginal dependencies captured by the spectral density $f(\cdot)$, which can equivalently capture the conditional dependencies encoded in the spectral precision matrix $\Theta(\cdot)$.}
    \label{fig:brain_signal_decomp}
\end{figure}

We illustrate the frequency specific graphical model captured by $f(\omega)$ and the conditional graphical model captured by $\Theta(\omega)$ with a simple example. In the following example and also in Figure \ref{fig:brain_signal_decomp}, the underlying pair of time series $X_t$ and $Y_t$ can be thought of as brain signals associated with two distinct brain regions, and the frequency specific components correspond to physiological activities. Similar examples are available in \citet{ombao2022spectral}.

\begin{example}[Frequncy-decomposition of time series]
We consider two time series $X_t$ and $Y_t$, each having three frequency components, and generate $N = 200$ observations as
\begin{align*}
    & X_{t}^{(j)} = Z_1(\omega_j) \cos \omega_j t, &&  Y_{t}^{(j)} = Z_2(\omega_j) \cos \omega_j t, \\
    & X_t = X_t^{(1)} + X_t^{(2)} + X_t^{(3)}, && Y_t = Y_{t}^{(1)} + Y_{t}^{(2)} + Y_{t}^{(3)},
\end{align*}
 where $j = 1,2,3$, $\omega_1 = 2\pi \times 5/N,\ \omega_2 = 2\pi \times 10 /N,\ \omega_3 = 2\pi \times 40/N$. We generate the amplitude processes from the following distributions
 \begin{align*}
 \begin{bmatrix} Z_1(\omega_1) \\ Z_2(\omega_1) \end{bmatrix} 
& \sim \nor_2\left(\begin{bmatrix} 0 \\ 0 \end{bmatrix}, 
\begin{bmatrix}
    1 & 0.5\\
    0.5 & 1
\end{bmatrix}\right), \quad 
\begin{bmatrix} Z_1(\omega_2) \\ Z_2(\omega_2) \end{bmatrix} 
\sim \nor_2\left(\begin{bmatrix} 0 \\ 0 \end{bmatrix}, 
\begin{bmatrix}
    1 & -0.5\\
    -0.5 & 1
\end{bmatrix}\right), \\
\begin{bmatrix} Z_1(\omega_3) \\ Z_2(\omega_3) \end{bmatrix} 
& \sim \nor_2\left(\begin{bmatrix} 0 \\ 0 \end{bmatrix}, 
\begin{bmatrix}
    2 & 0.1\\
    0.1 & 2
\end{bmatrix}\right).
\end{align*}
$X_t$ and $Y_t$ behave as illustrated in Figure~\ref{fig:brain_signal_decomp}. The components $\cos(\omega_j t)$, $j = 1, 2, 3$, serve as orthogonal basis functions, with $Z_1(\omega_j)$ and $Z_2(\omega_j)$ denoting the corresponding random amplitude coefficients. Each set of coefficients is independent across frequencies $\omega_j$. If the spectral density of the joint process $W_t := (X_t, Y_t)^\top$ is $f_W$, Cram\'{e}r's representation indicates   $[f_W(\omega_j)]_{k,k} = \var(Z_k(\omega_j))$, i.e. $[f_W(\omega_1)]_{1,1} = 1,\ [f_W(\omega_1)]_{2,2} = 1,\ [f_W(\omega_2)]_{1,1} = 1,\ [f_W(\omega_2)]_{2,2} = 1,\ [f_W(\omega_3)]_{1,1} = 2,\ [f_W(\omega_3)]_{2,2} = 2$. For the cross-spectral term, $[f_W(\omega_1)]_{12} = 0.5,\ [f_W(\omega_2)]_{12} = -0.5,\ [f_W(\omega_3)]_{12} = 0.1$.
\end{example}

Therefore $f_W$ fully captures the correlation structure of the same frequency orthogonal components of the joint process $W_t$. Similarly, $\Theta_W(\omega) = [f_W(\omega)]^{-1}$ captures the partial correlation structure of the orthogonal components in the joint process $W_t$.

\section{Supplement for realfying standard statistical optmization problems}\label{sec:realifying_stat_opt_supp}

We provide the details from Section \ref{subsec:realifying_stat_opt} on realifying the standard convex problems of statistical interests.

\subsection{Proof of Lemma \ref{lem:ols}: OLS with complex variables}\label{pf:ols_complex}

For $Y\in \CC^n,\ X \in \CC^{n\times p},\ \beta\in \CC^p$,
\begin{align*}
    \|Y - X\beta\|_2^2 = \frac12\|\varphi(Y - X\beta)\|_F^2 = &\frac12\|\Pi_n^\top \varphi(Y - X\beta)\Pi_1\|_F^2\\
    & \hspace{20pt} \text{[Fr\"{o}benius norm invariant under orthogonal transform]}\\
    & = \frac12\| \Pi_n^\top \varphi(Y) \Pi_1 - \Pi_n^\top \varphi(X\beta)\Pi_1 \|_F^2\\
    & = \frac12\| \Pi_n^\top \varphi(Y) \Pi_1 - (\Pi_n^\top \varphi(X) \Pi_p) (\Pi_p^\top \varphi(\beta)\Pi_1) \|_F^2\\
    & = \frac12 \|\tilde{\tilde{Y}} - \tilde{\tilde{X}} \tilde{\tilde{\beta}} \|_F^2= \frac12 \cdot 2\| (\tilde{\tilde{Y}} - \tilde{\tilde{X}} \tilde{\tilde{\beta}}) \e_1 \|_2^2 = \| \tilde{Y} - \tilde{\tilde{X}} \tilde{\beta} \|_2^2.
\end{align*}
And for the normal equation,
\begin{align*}
    X^\dagger X\beta = X^\dagger Y\ \iff\ &\varphi(X^\dagger X\beta) = \varphi(X^\dagger Y)\\ 
    \iff\ & [\varphi(X)]^\top \varphi(X) \varphi(\beta)\e_1 = [\varphi(X)]^\top \varphi(Y)\e_1\\
    \iff\ & (\Pi_n^\top \varphi(X)\Pi_p)^\top (\Pi_n^\top \varphi(X)\Pi_p) \Pi_p^\top \varphi_1(\beta)\nonumber\\
    &\hspace{2cm} = (\Pi_n^\top\varphi(X)\Pi_p)^\top \Pi_n^\top \varphi_1(Y)\\
    \iff\ & \tilde{\tilde X}^\top \tilde{\tilde X}\tilde\beta = \tilde{\tilde X}^\top \tilde Y.
\end{align*} \pfend

\subsection{Proof of Lemma \ref{lem:log-det}: Complex log-determinant program}\label{pf:glasso_complex}

$Z_1\ldots, Z_n$ are i.i.d samples from a complex normal distribution with mean 0 and covariance $\Theta^{-1}$, $\Theta \in \pdhermit$, and the estimated complex-valued sample Gram matrix is $\hat \Sigma := n^{-1}\sum_{i=1}^n Z_i Z_i^\dagger$. The convex log determinant barrier function is $L_\CC(\Theta) = -\log \det\Theta + \trace(\hat \Sigma\Theta)$ and the real counterpart is $L_\RR(\Theta) = -\log \det\dtilde{\Theta} + \trace\bigg(\dtilde{\hat \Sigma}\dtilde{\Theta}\bigg)$.
If $\Theta = U\Lambda U^\dagger$ is the spectral decomposition of $\Theta$ with $U$ being unitary and $\Lambda$ being a diagonal matrix with all positive real entries, then 
\begin{align*}
    \varphi(\Theta) = \varphi(U \Lambda U^\dagger)
    \varphi(U) \varphi(\Lambda) \varphi(U)^\top
    =  \varphi(U) \Pi_p \begin{bmatrix}
        \Lambda & 0\\ 0 & \Lambda
    \end{bmatrix} \Pi_p^\top \varphi(U)^\top.
\end{align*}
Therefore,
\begin{align*}
    \log \det \varphi(\Theta) & = \log \left(\det(\varphi(U)) \det (\Pi_p) (\det \Lambda)^2 \det(\Pi_p^\top) \det(\varphi(U)^\top)\right)\\
    & = \log\left(\det(\varphi(U)\varphi(U)^\top) (\det \Lambda)^2\right)\\
    & = \log\left(\det(\varphi(UU^\dagger)) (\det \Lambda)^2\right)\\
    & = \log\left((\det \Lambda)^2\right) = 2\log\det \Lambda = 2\log \det \Theta.
\end{align*}
Let $P\Theta = V\Delta V^\dagger$ be the Schur decomposition of $P\Theta$, where $V$ is unitary and $\Delta$ is upper triangular with the eigenvalues of $P\Theta$ on the diagonal. Define the square root of $\Theta$ as $\Theta^{1/2} := U \Lambda^{1/2} U^\dagger$, where $\Lambda^{1/2}$ is the diagonal matrix of square roots of $\Theta$'s eigenvalues. Since $P\Theta$ is similar to the positive definite matrix $\Theta^{1/2} P \Theta^{1/2}$, its eigenvalues—and hence the diagonal entries of $\Delta$—are real and positive. Thus,
\begin{align*}
\trace\!\big(\varphi(P)\varphi(\Theta)\big)
&= \trace\!\big(\varphi(P\Theta)\big)
  = \trace\!\big(\varphi(V\Delta V^\dagger)\big) \\
&= \trace\!\big(\varphi(V)\varphi(\Delta)\varphi(V)^\top\big)
  = \trace\!\big(\varphi(\Delta)\varphi(V)^\top\varphi(V)\big) \\
&= \trace\!\big(\varphi(\Delta)\varphi(V^\dagger V)\big)
  = \trace\!\big(\varphi(\Delta)\big) \\
&= \trace\!\big(\varphi(\Delta)\Pi_p\Pi_p^\top\big)
  = \trace\!\big(\Pi_p^\top\varphi(\Delta)\Pi_p\big) \\
&= \trace\!\big(\text{diag}(\Delta, \Delta)\big)
  = 2\,\trace(\Delta)
  = 2\,\trace(P\Theta).
\end{align*}
Combining the trace terms with the log-barrier terms, we obtain $L_\RR(\Theta) = 2L_\CC(\Theta)$. \pfend

\section{Supplement for CLASSO}\label{sec:classo+supplement}

\subsection{Details of Algorithms \ref{alg:classo} and \ref{alg:classo-cov}}\label{subsec:classo_simplification}

The CLASSO problem \eqref{eq:classo} can be alternatively expressed as an optimization problem with real variables. We use the transformed OLS in Lemma \ref{lem:ols} and the following fact for the $\ell_1$ regularization part of \eqref{eq:classo}
$$ \|\beta\|_1 = \sum_{j = 1}^p |\beta_j| = \sum_{j = 1}^p \|\varphi_1(\beta_j)\|_2 = \sum_{j = 1}^p \|\tilde{\beta_j}\|_2. $$
Therefore, the realification of \eqref{eq:classo} is\eqref{eq:grplasso}, a \textit{group Lasso} problem with $p$ groups, each of size 2. Let the optimizer of \eqref{eq:grplasso} be $(\hat{\tilde{\beta_1}}, \ldots, \hat{\tilde{\beta_p}})^\top$ and $\tilde r^{(j)} = \tilde{Y} - \sum_{k\neq j}\tilde{\tilde{X_k}} \hat{\tilde{\beta_k}}$ be the vector of \textit{$j$\textsuperscript{th} partial residuals}. Following \citet[Equation (4.14)]{hastie2015statistical}, the optimizer $\hat{\tilde{\beta_j}}$ satisfies $\hat{\tilde{\beta_j}} = 0$ if $\left\|\frac1n \tilde{\tilde{X_j}}^\top \tilde r^{(j)}\right\|_2 < \lambda$, and otherwise the following:
\begin{equation}\label{eq:grplasso-update}
    \hat{\tilde{\beta_j}} = \left(\frac1n \tilde{\tilde{X_j}}^\top \tilde{\tilde{X_j}} + \frac{\lambda}{\big\|\hat{\tilde{\beta_j}}\big\|_2 } I_2 \right)^{-1} \frac1n \tilde{\tilde{X_j}}^\top \tilde r^{(j)}.
\end{equation}
The equation \eqref{eq:grplasso-update} does not have a closed form solution in general, but here we exploit the structure of $\tilde{\tilde{X_j}}$. Using Proposition \ref{prop:phi3}(a) and for each $j$, the columns of $\tilde{\tilde{X_j}}$ are orthogonal, implying the following 
$$ \tilde {\tilde {X_j}}^\top \tilde {\tilde {X_j}} = \varphi(X_j)^\top \underbrace{\Pi_n \Pi_n^\top}_{I_{2n}} \varphi(X_j) = \varphi(X_j^\dagger) \varphi(X_j) = \varphi(X_j^\dagger X_j) = \varphi(\|X_j\|_2^2) = \|X_j\|_2^2 I_2. $$
Therefore, the update \eqref{eq:grplasso-update} has a closed form expression of $\hat{\tilde {\beta_j}}$ given by 
\begin{equation}\label{eq:grplasso-update-closed}
    \hat{\tilde {\beta_j}} = \left(1 - \frac{\lambda}{\left\|\frac1n \tilde{\tilde {X_j}}^\top \tilde r^{(j)}\right\|_2}\right)_+ \frac{\frac1n \tilde{\tilde {X_j}}^\top \tilde r^{(j)}}{\frac1n\|X_j\|_2^2}.
\end{equation}

\paragraph*{Soft threshold operator} We define two close variants of the soft threshold operators widely used in Lasso literature. For any $t\in \RR$, let $t_+ = \max\{0, t\}$. For every $z\in \CC$, we define the soft threshold operator $\mathcal{S}_{\lambda}:\CC \rightarrow \CC $ as
\begin{equation}\label{eq:soft_threshold_cplx}
    \mathcal{S}_{\lambda}(z) = (|z| - \lambda)_{+}\frac{z}{|z|},
\end{equation}
and for any $x\in \RR^2$, we similarly define $\tilde{\cal S}_{\lambda}:\RR^2 \rightarrow \RR^2 $ as 
\begin{equation*}\label{eq:soft_threshold_real}
    \tilde{\cal S}_{\lambda}(x) = (\|x\|_2 - \lambda)_+ \frac{x}{\|x\|_2},
\end{equation*}
the \textit{soft threshold operator} in $\RR^2$. The two operators are connected by the map $\varphi$ and the following identity. For any $z\in \CC$, we have 
$$\tilde{\mathcal{S}}_\lambda(\varphi(z)) = (\|\varphi(z)\| - \lambda)_+ \frac{\varphi(z)}{\|\varphi(z)\|_2} = (|z| - \lambda)_{+}\frac{\varphi(z)}{|z|} = \varphi(\mathcal{S}_{\lambda}(z)),$$
where we use the property (iii) from Proposition \ref{prop:phi1}. We now use the soft threshold operators for getting a more convenient form of \eqref{eq:grplasso-update-closed} which is $\hat{\tilde\beta}_j = \tilde S_{\lambda}\left(\tilde{\tilde {X_j}}^\top \tilde r^{(j)}/n\right)\bigg/\left(\frac{1}{n} \|X_j\|_2^2\right)$. If the columns of $X$ are scaled so that $\|X_j\|_2^2 = n$ for every $j = 1,\ldots, n$, then 
\begin{equation}\label{eq:grplasso-update-closed-scaled}
    \hat{\tilde {\beta_j}} = \tilde {\cal S}_{\lambda}\left(\frac1n \tilde{\tilde {X_j}}^\top \tilde r^{(j)}\right).
\end{equation}
Using the group lasso update \citep{hastie2015statistical} and the soft threshold operator notation in \eqref{eq:grplasso-update-closed-scaled}, the group Lasso update in this case is
\begin{equation}\label{eq:soft_threshold_classo}
    \tilde r^{(j)} \leftarrow \tilde r + \tilde{\tilde{X_j}} + \hat{\tilde{\beta_j}};\quad 
    \hat{\tilde{\beta_j}} \leftarrow \tilde{\mathcal{S}}_{\lambda}\left( \frac1n \tilde{\tilde{X_j}}^\top \tilde{r}^{(j)} \right);\quad
    \tilde{r} \leftarrow \tilde r^{(j)} - \tilde{\tilde{X_j}} + \hat{\tilde{\beta_j}}.
\end{equation}
We denote $\hat\beta_j = \hat{\tilde{\beta}}_{j1} + \i \hat{\tilde{\beta}}_{j2}$ i.e. $\hat{\tilde{\beta_j}} = \varphi(\hat\beta_j)$, and $r^{(j)} = \tilde{r}^{(j)}_1 + \i\ \tilde{r}^{(j)}_2$ for $j = 1,\ldots, p$. Therefore, we can write 
\begin{align*} 
\varphi(r^{(j)}) = \tilde r^{(j)}  = \tilde Y - \sum_{k\neq j}\tilde {\tilde {X_k}} \hat{\tilde {\beta_k}}
& =  \varphi(Y) - \sum_{k\ne j}\varphi(X_k) \varphi(\hat\beta_k) \\
& = \varphi(Y - \sum_{k\ne j}X_k\hat\beta_k)\\
& = \varphi(r + X_j\hat\beta_j),
\end{align*}
where $r = Y - \sum_{k=1}^p X_k\hat\beta_k$. Using this representation,  the Lasso update becomes
$$ \varphi(r^{(j)}) \leftarrow \varphi(r + X_j\hat\beta_j);\quad 
    \varphi(\hat\beta_j) \leftarrow \varphi\left({\cal S}_{\lambda}\left(X_j^\dagger r^{(j)}/n\right)\right);\quad
    \varphi(r) \leftarrow \varphi(r^{(j)} - X_j \hat\beta_j). $$
Therefore, $\varphi$ being an isomorphism, we can simply omit it from the Lasso updates above and still we get an equivalent complex-valued update as Algorithm \ref{alg:classo}.

\paragraph*{CLASSO with covariance updates}
We can modify Algorithm \ref{alg:classo} to take the inner product-like objects $X^\dagger X$ and $X^\dagger Y$ as inputs, which is the Lasso step involved in graphical Lasso. In \eqref{eq:soft_threshold_classo}, we can update $X_j^\dagger r^{(j)}$ instead of $r^{(j)}$, i.e. the update for each $j = 1,\ldots, p$ will be 
\begin{align*}
    X^\dagger r^{(j)} = X^\dagger r + X^\dagger X_j \hat\beta_j;\quad
    \hat\beta_j \leftarrow {\cal S}_{\lambda}\left(\frac1n X_j^\dagger r^{(j)}\right);\quad X^\dagger r \leftarrow X^\dagger r^{(j)} - X_j \hat\beta_j.
\end{align*}
Therefore $X^\dagger Y$ and $X^\dagger X$ are sufficient for estimation. The complete algorithm is presented in Algorithm \ref{alg:classo-cov}.

\subsection{Speed comparison of CLASSO with existing algorithms}\label{subsec:speec_comp_exp}

We perform a numerical experiment to compare the computational efficiency of CLASSO with two popular group Lasso \texttt{R} packages: (a) \texttt{gglasso} \citep{yang2015fast} that uses blockwise-majorization-descent (BMD) algorithm, and (b) \texttt{grplasso}\citep{meier2020package} that implements pathwise block-coordinate descent without incorporating the within-group orthogonal predictor structure.    

\paragraph*{Experiment setup} 
We illustrate the speed improvement of CLASSO with data sets from the following DGP: 
$$y = X\beta + \varepsilon,~~ \re(X_{ij}) \sim \nor(0, 2),~~ \im(X_{ij}) \sim \nor(0,1), ~~ \re(\varepsilon_i), \im(\varepsilon_i) \sim \nor(0, 1).$$
$$\beta_j = \begin{cases}
    1 + 1\i~~ \text{ for } 1 \le j \le 5,  ~~~ & -1 + 1\i~~ \text{ for } 6 \le j \le 10\\
    -1~~ \text{  for } 11 \le j \le 20,  ~~~ & 0.5 - 1\i~~ \text{ for } 21 \le j \le 30\\
    0 ~~~\text{otherwise}. & 
\end{cases}$$
The first experiment consists of data sets with sample sizes $n = 100, 200, \ldots, 1000$ and dimensions $p = 100, 200, \ldots, 1000$. For the second experiment we choose two cases, $n = 100, p = 500$, and $n = 70, p = 1500$. We choose the penalty parameter $\log(\lambda)\in [-4, 0]$ with grid space 0.1 in log-scale. The experiment is conducted with 50 trials and the median run times for CLASSO (implemented with Fortran 77), \texttt{gglasso}  and \texttt{grplasso}  are recorded.

\begin{figure}[t]
    \centering
    \subfloat{\includegraphics[width=0.45\linewidth]{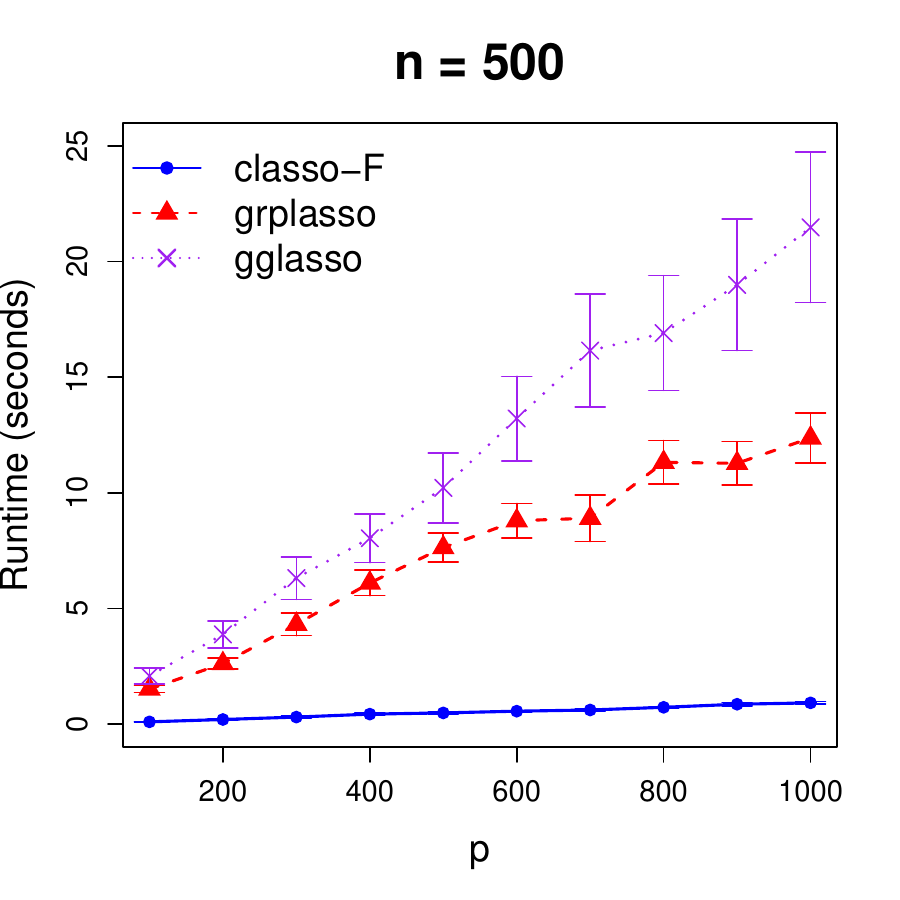}}
    \subfloat{\includegraphics[width=0.45\linewidth]{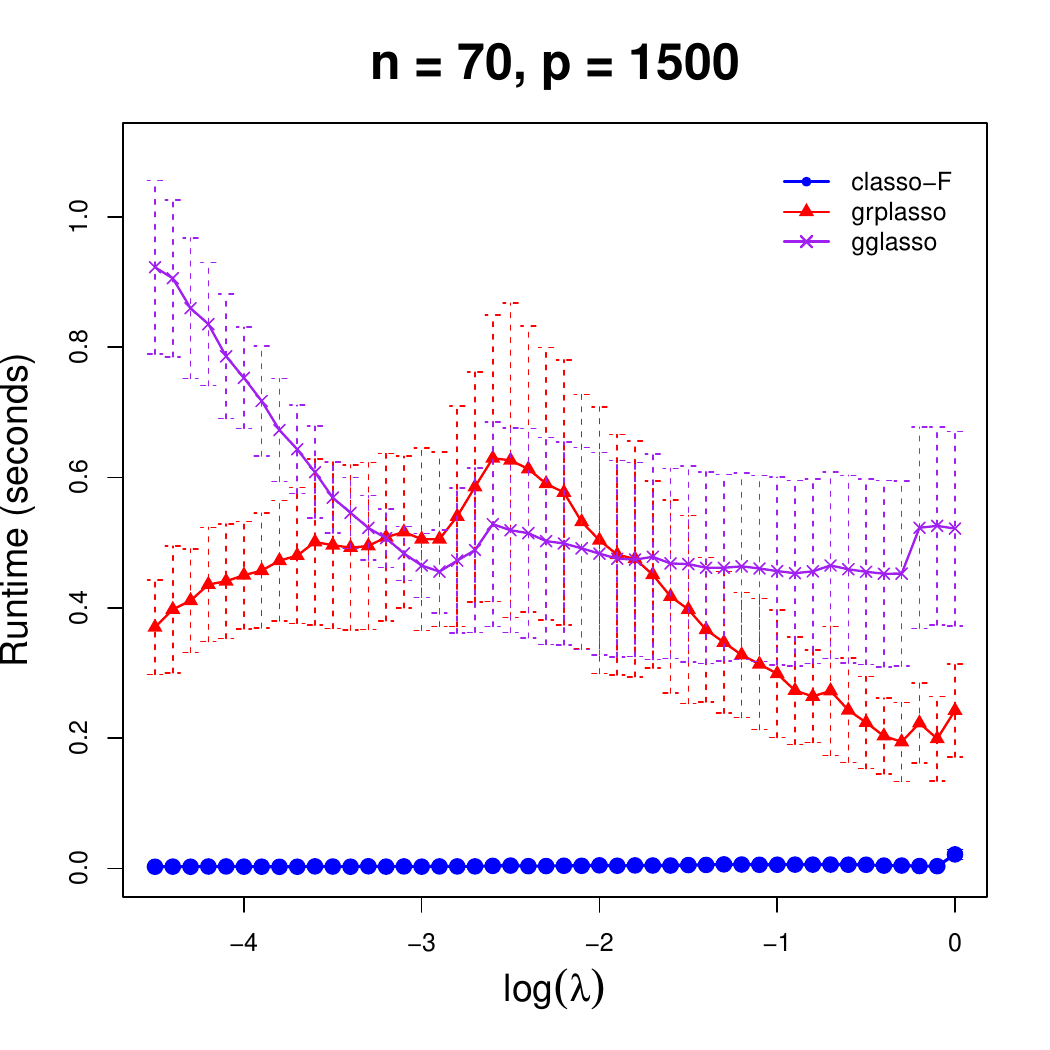}}
    \caption{(a) Total runtime for a regularization path (median across 50 trials) versus dimension $p$ for CLASSO (blue circles), \texttt{grplasso} (red triangles), and \texttt{gglasso} (purple crosses) ($n=500$). (b) Median runtime across 50 trials versus $\log \lambda$ for the same methods ($n = 70, p = 1500$). In panels (a) and (b), vertical bars indicate the median absolute deviation of runtime. CLASSO is substantially faster than the benchmarks across both dimensions and the regularization path.}
    \label{fig:classo_runtime}
\end{figure}

\paragraph*{Results} 
As shown in Figures \ref{fig:classo_runtime} (a) and (b), CLASSO has significantly less median runtime over the total regularization paths for various $p$'s and for a fixed $n$. As illustrated with Figures \ref{fig:classo_runtime} (c) and (d), for fixed $(n,p)$, CLASSO exhibits significant runtime improvement over the two benchmark methods across every $\lambda$ in the regularization path. As $\lambda$ decreases or $p$ increases, the underlying LASSO problem becomes harder to solve and hence the benchmark algorithms start to suffer in runtime. In contrast, CLASSO is scalable to small $\lambda$'s or large $p$'s, and the runtime change is insignificant compared to its benchmarks.

\subsection{Speed improvement strategies of CLASSO}\label{subsec:speed_improvement_strategies}

Since CLASSO algorithm is implemented with a coordinate descent algorithm, it enjoys two speed improvement strategies.

\paragraph*{Warm start} 
Often for practical purposes and tuning the regularization parameter, CLASSO is required to perform on a sequence of $\lambda$ values and consequently produce a regularization path. Such regularization path helps us identify the \textit{best} choice of $\lambda$ in term of certain criterion (e.g. relative mean squared error (RMSE), Bayesian information criterion (BIC) or cross-validation error). A \textit{pathwise} regularization method involves starting with a \textit{reasonably} large value of $\lambda$ (e.g. $\lambda_0  = \max_{j \in [p]}|X_j^\dagger Y_j/n|$) and obtain an all-zero solution $\hat\beta(\lambda_0)$. For each smaller value $\lambda_{t}$ with $t \ge 1$, the solution for $\lambda_{t-1}$ i.e.  $\hat\beta(\lambda_{t-1})$ can be used as an initial value i.e. \textit{warm start} and compute $\hat\beta(\lambda_{t})$. This scales up the efficiency of computing the Lasso solution for all tuning parameters since a small change in the tuning parameter is expected to have a limited influence on the solution. This method is referred to as the \textit{pathwise CD}.

\paragraph*{Active set screening} An active set $\mathcal{A}_t$ is the set of indices for non-zero entries in iteration $t$ of the Lasso algoriTheorem For the vector of current residuals $r^{(t)}$ if $|X_j^\dagger r^{(t)}|/N < \lambda_\ell$ is satisfied, then $j$ is excluded from $\mathcal{A}_{t}$, i.e. $\mathcal{A}_t := \mathcal{A}_{t-1}\setminus \{j\}$. For the indices in $\mathcal{A}_t$, the soft threshold step 4.(b) of Algorithm 1 is performed. The advantage of doing the \textit{active set} selection is that once a index is declared inactive and sparse, the number of soft threshold operation in each step decreases. If after a few iterations $\hat\beta^{(t)}$ starts to become sparse, then the number of soft threshold operations significantly goes down. For implementation of CLASSO, we use the sequential strong screening rule \citep{tibshirani2012strong}, which discards the $j$\textsuperscript{th} predictor from $\mathcal{A}_t$ and sets $\hat \beta_j(\lambda_t) = 0$ if the strong screening condition, i.e. $\left| \frac{1}{n} X_j^\dagger (Y - X\hat \beta(\lambda_{t-1})) \right| \le 2\lambda_t - \lambda_{t-1}$ is satisfied.

\section{Deferred details of fMRI data set in Section \ref{sec:FC_fMRI}}\label{sec:fmri_pre}

\paragraph*{Preprocessing} 
MRIs were acquired on a Siemens Skyra 3 T scanner at Washington University in St. Louis. Each subject underwent four
gradient-echo EPI resting-state functional MRIs (rsfMRI), with TR = 720 ms, TE = 33.1 ms, 2.0 mm isotropic voxels, FoV = 208 × 180 mm\textsuperscript{2}, flip angle = 52$^\circ$,
matrix = 104 × 90, 72 slices. Two 15 min rsfMRI runs were acquired at each of the two sessions. The preprocessed data consists of $n = 1200$ time points for each run, for a total of 4800 time points for each subject over four scans. Each run of each subject's rsfMRI was
preprocessed by the HCP consortium \citep{smith2013resting}. The data were minimally preprocessed \citep{glasser2013minimal} and had artifacts removed using ICA + FIX \citep{griffanti2014ica, salimi2014automatic}. Gray matter was parcellated into 86 cortical ($p = 86$), subcortical and cerebellar regions using FreeSurfer  \citep{Fischl1999CorticalSystem} and regional time series were calculated as the mean of the time series in each voxel of a given region. Each region was further assigned to a Yeo functional network delineating visual, somatomotor, dorsal attention, ventral attention, limbic, default mode, and fronto-parietal networks \citep{Yeo2011TheConnectivity}; we added a subcortical and cerebellar network for whole brain coverage as in previous work \citep{dhamala2020sex}. Yeo parcellation of the regions assembles brain regions based on similar function.

\section{Background on group, ring and field}\label{sec:intro_algebra}

We provide with the basic axiomatic definitions and examples of algebraic structures used in Section \ref{sec:realification}. Readers can find deeper treatment of these topics in \citet{herstein1991topics} and \citet{dummit2004abstract}.

\begin{definition}[Group]
A nonempty set of elements $G$ is a \textbf{group} if in $G$ there is defined a binary operation ` $\cdot$ ', such that
\begin{enumerate}
    \item \textit{Closed}: $a, b \in G\ \implies\ a\cdot b \in G$,
    \item \textit{Associativity}: $a, b, c\in G\ \implies\ a\cdot (b\cdot c) = (a\ \cdot b)\cdot c$,
    \item \textit{Identity}: There exists $e\in G$ such that $a\cdot e = e\cdot a = a$ for all $a \in G$,
    \item \textit{Inverse}: For every $a\in G$, there exists $a^{-1}\in G$ such that $a\cdot a^{-1} = a^{-1}\cdot a = e$.
\end{enumerate}
\end{definition}

\begin{example}[Example of group]
\begin{itemize}
\item[(a)] Set of all integers $\ZZ$ with additive operation `+'.
\item[(b)] Set of all complex numbers $\CC$ with complex addition `+'.
\item[(c)] Set of $2\times 2$ matrices defined in \eqref{eq:matrix-class} under the matrix addition `+'.
\end{itemize}
\end{example}

In our work, we deal with Abelian groups of infinite order, that are groups with infinitely many elements and where the group operation `$\cdot$' is commutative, i.e. $a\cdot b = b\cdot a \in G$ for all $a, b\in G$ . Details can be found in \citet{herstein1991topics}.

\begin{definition}[Ring]\label{def:ring}
    A nonempty set $R$ is said to be an \textbf{associative ring} if in $R$, there are two operations $+$ and $\cdot$ such that for all $a, b, c, \in R$, the following hold:
    \begin{enumerate}
        \item $a + b \in R$,
        \item $a + b = b + a$,
        \item $a + (b + c) = (a + b) + c$,
        \item \textit{Additive identity}: There is an element $0\in R$ such that $a + 0 = 0 + a = 0$ for all $a \in R$,
        \item \textit{Additive inverse}: There is an element $-a\in R$ such that $a + (-a) = 0$,
        \item $a\cdot b \in R$,
        \item $a\cdot (b\cdot c) = (a\cdot b)\cdot c$,
        \item \textit{Distributive law}: $a\cdot (b + c) = a\cdot b + a\cdot c$ and $(b + c)\cdot a = b\cdot a + b\cdot c$.
    \end{enumerate}
\end{definition}

By Definition \ref{def:ring}.1-5, $R$ is an Abelian group under the operation `$+$' (addition). Similarly, Definition \ref{def:ring}.6-7 implies that a ring is closed under associative operation `$\cdot$' (multiplication). In practice, $a\cdot b$ is simply written as $ab$. The two operations are connected by Definition \ref{def:ring}.8. 

\begin{example}[Example of ring]
\begin{itemize}
\item[(a)] The set of real numbers $\RR$ is a commutative ring under the operations $+$ (addition) and $\cdot$ (multuplication). The additive unit is $0$, and $\RR$ also has a multipicative unit denoted by 1.
\item[(b)] Similar to $\RR$, $\CC$ is a ring under $+$ and $\cdot$.
\item[(c)] The set of all matrices characterized by \eqref{eq:matrix-class} is a ring under matrix addition and multiplication.
\end{itemize}
\end{example}

\begin{definition}[Ring homomorphism]
    A map $\phi$ from the ring $R$ into the ring $R'$ is said to be a \textbf{homomorphism} if
    \begin{enumerate}
        \item $\phi(a + b) = \phi(a) + \phi(b)$,
        \item $\phi(ab) = \phi(a) \phi(b)$.
    \end{enumerate}
\end{definition}
The operations $+$ and $\cdot$ on the left side are concerned with the ring $R$, and the ring operations `$+$' and `$\cdot$' appearing on the right are of $R'$.

\begin{example}[Example of ring homomorphism]\label{ex:ring_homomorphism}
\begin{itemize}
\item[(a)] Let $R = R'$ be the set of all real numbers $\RR$, both associated with the ring operations being addition `$+$' and multiplication `$\cdot$'. Then $\phi: R\rightarrow R'$, defined by $\phi(x) = x$ for every $x\in R$, is a ring isomorphism.
\item[(b)] The complex conjugation map $\phi:\CC\rightarrow\CC$, defined as $\phi(z) = \overline{z}$ for all $z\in \CC$, is a ring homomorphism.
\end{itemize}
\end{example}

\begin{definition}[Ring isomorphism]
    A homomorphism of $R$ into $R'$ is said to be an \textbf{isomorphism} if it is a one-to-one mapping. Two rings are said to be isomorphic if there exists an isomorphism of one onto the other.
\end{definition}
\vspace{-0.5cm}

Both cases in Example \ref{ex:ring_homomorphism} for ring homomorphism are one-to-one maps, and hence serve as examples for ring isomorphism as well.

\begin{definition}[Field]
    A ring is said to be a \textbf{division ring} if its non zero elements form a group under multiplication. A \textbf{field} is a commutative division ring.
\end{definition}

\begin{example}[Example of field]
\begin{itemize}
\item[(a)] Some standard widely used fields in mathematics are the rational numbers $\mathbb{Q}$, the set of real numbers $\RR$, complex numbers $\CC$, finite fields $\ZZ_p$ for a prime integer $p$, etc.
\item[(b)] The set of matrices characterized by \eqref{eq:matrix-class} is a field isomorphic to the set of all complex numbers $\CC$ (Proposition \ref{prop:phi1}). 
\end{itemize}
\end{example}

A field is always a ring. Two fields $F$ and $F'$ are said to be \textit{isomorphic} if they are isomorphic as rings. We use the terms ring isomorphism and field isomorphism separately in order to keep the underlying isomorphic objects specified as rings or fields respectivly. A field isomorphism is always a ring isomorphism, however the converse may not be true.

\end{document}